\documentclass[12pt,preprint]{aastex}
\shorttitle{VLASS Analysis, 2015 February 11} \shortauthors{Condon}
\begin{document}

\title{An Analysis of the VLASS Proposal}

\author{Jim~Condon\altaffilmark{1}}
\affil{National Radio Astronomy Observatory,\\ 520 Edgemont Road,
Charlottesville, VA 22903, USA}
\email{jcondon@nrao.edu}

\altaffiltext{1}{The National Radio Astronomy Observatory is a
  facility of the National Science Foundation operated by Associated
  Universities, Inc. 
All views expressed in this article are those of the author and do not
necessarily represent the views of the NRAO.}

\begin{abstract}

``The Jansky--Very Large Array Sky Survey (VLASS)'' 
  \citep{ssg15} comprises two distinct S-band ($2 < \nu < 4
  \mathrm{~GHz}$) surveys: \\ (1) The shallow (rms noise
  $\sigma_\mathrm{n} \approx 69 \,\,\mu \mathrm{Jy~beam}^{-1} \approx 1.5
  \mathrm{~K}$ at $\theta \approx 2\,\farcs5$ resolution) but wide
  (covering all $33,885 \mathrm{~deg}^2$ north of $\delta =
  -40^\circ$) ``All-sky'' 
  and \\(2) the sensitive ($\sigma_\mathrm{n} \approx 1.5 \,\, \mu
  \mathrm{Jy~beam}^{-1} \approx 0.32 \mathrm{~K}$ at $0\,\farcs8$
  resolution or $\approx 0.13 \mathrm{~K}$ at $2\,\farcs 0 \times
  0\,\farcs 8$ resolution) but narrow ($10 \mathrm{~deg}^2$ in three
  patches) ``Deep.''

All-Sky is intended to be a community resource, the JVLA update of
the high-impact FIRST and NVSS VLA surveys made at 1.4~GHz.  FIRST and
NVSS succeeded for two reasons: (1) they are $ > 10 \times$ better in
number of sources detected, sensitivity, resolution, position
accuracy, etc.  than prior ``all-sky'' radio surveys, and (2) they
have not been surpassed for almost two decades, so their high citation
rates have not diminished.

In contrast, (1) the proposed All-Sky is only about $1.5 \times$ more
sensitive than FIRST for point-source populations whose effective
spectral index is $\langle \alpha \rangle \approx -0.7$, and its
sensitivity to extended sources (e.g., many radio galaxies and
quasars, diffuse sources in galaxy clusters, low-redshift star-forming
galaxies) is about $3 \times$ worse than FIRST and about $60 \times$
worse than NVSS because its angular resolution is so high.  The rms
noise $ \sigma_\mathrm{n}$ on survey images has units of apparent brightness
($\mu\mathrm{Jy~beam}^{-1}$ or K), not flux density
($\mu\mathrm{Jy}$), so the high-resolution All-Sky images will miss
$>25$\% of all sources with flux densities $S > 5 \sigma_\mathrm{n}
\approx 350 \,\mu\mathrm{Jy}$, including entire low-brightness
populations such as normal spiral galaxies, whose median S-band
surface brightness is only $\langle T_\mathrm{b} \rangle \approx 0.17
\mathrm{~K} \ll 5 \times 1.5 \mathrm{~ K}$.  (2) All-Sky will probably
be surpassed before its earliest completion date (2023) by the
contemporary ASKAP EMU survey \citep{nor11} covering the whole sky
south of $\delta = +30^\circ$, and by the complementary Westerbork
WODAN survey north of $\delta= +30^\circ$.
 EMU intends to reach
$\sigma_\mathrm{n} = 10 \,\,\mu \mathrm{Jy~beam}^{-1} \approx 0.06
\mathrm{~K}$ at $\theta = 10''$ resolution at L band ($1.1 < \nu <
1.4$~GHz), which is equivalent to $\sigma_\mathrm{n} \approx 6 \,\,\mu
\mathrm{Jy~beam}^{-1} \approx 0.01 \mathrm{~K}$ at S band.  EMU
\emph{will} detect spiral galaxies and generate complete samples of
all sources $~10 \times$ below the All-Sky detection limit for
point sources.  Nearly all multi-wavelength astronomers asking the
question ``Are my favorite objects radio sources?'' will get better
answers from EMU than from All-Sky.  

The uniquely high angular resolution of All-Sky, and hence its poor
surface-brightness sensitivity, was forced by a straw-man optical
identification method that fails to distinguish between extended radio
sources (e.g., core plus lobes and jets) and radio source components
(brightness peaks on images), so it tries to identify individual
clearly resolved radio lobes with unrelated galaxies.  The high resolution
of All-Sky is not necessary for identification reliability (except for
radio stars), and it is counterproductive because it lowers
identification completeness---All-Sky will miss $>25$\% of all sources
with flux densities above its $5 \sigma_\mathrm{n}$ detection limit.
No radio detection, no optical identification.  Incompleteness is
actually worse than unreliability because incompleteness cannot be
corrected, while sources with marginal identifications can be
reobserved.

The high sensitivity and angular resolution of Deep is likely to
remain unique until SKA Phase I is operating.  Deep is qualitatively
like the large PI VLA survey COSMOS, but bigger.  Its top science goals
are (1) a pilot survey for 
constraining dark energy by detecting the effect of weak
gravitational lensing on distant radio sources and (2) studying the
evolution of complete (flux-limited) galaxy samples.  These goals
conflict because resolving most faint sources for goal (1) implies
sample incompleteness for goal (2).  The proposal suggests that Deep
will resolve most sources so it can detect weak lensing, but Deep
won't resolve most sources so its completeness limit in
$\mu\mathrm{Jy~beam}^{-1}$ can be conflated with source flux densities
in $\mu\mathrm{Jy}$.  The Deep proposal must fully address the
angular-size distributions of $\mu$Jy sources before it can be
reviewed responsibly.


The VLASS will cost about 9,000 hours of JVLA observing time plus
20+ FTE-years of Socorro scientific staff effort for development.
User feedback about the VLA time taken by FIRST and NVSS led to the
\citet{bri97} Report, which was ignored by the proposal but should
be required reading.  The front-loaded
VLASS support requirements could divert most of the already-overloaded
Socorro scientific staff from JVLA commissioning and helping observers
for the next few years.  
\vfill\eject

\end{abstract}

\clearpage




\section{Introduction}

The VLASS proposal
originated in the community suggestion
``it was time to think  about a follow-on from NVSS and FIRST''
\citep{ssg15} that would exploit the new capabilities of the JVLA.
The VLASS originally had no specific science goals or technical
specifications, so white papers
were solicited and discussed at a workshop preceding the January
2014 AAS meeting and reported at \newline {\tt
  https://science.nrao.edu/science/surveys/vlass} 
\newline  
See Appendix A ``Motivation and Process'' of \citet{ssg15}
for a longer history.  I was a member of the workshop 
SOC and presented the ``Skeptic's View'' at the
workshop.

The VLASS Survey Science Group (SSG) of all interested
parties (mostly non-NRAO) was formed after the workshop and spent the next
year developing the final VLASS
proposal.  As the SSG ``designated skeptic,'' NVSS PI, and an
NRAO employee, I had potential conflicts of interest.  To
minimize them, I participated in the SSG meetings but did not try to
influence the choice of science goals, except to calculate their
technical implications. After each draft VLASS proposal appeared, I
sent to the SSG my ``VLASS Notes'' analyzing its strengths, weaknesses, and
possible fixes.  The present ``An Analysis of the VLASS Proposal''  update
those notes to
address the ``final'' VLASS proposal
\citep{ssg15}.  

The VLASS is so different from earlier surveys that a careful and
quantitative analysis is needed to show it can meet its many different
and sometimes incompatible science goals.  In particular, high angular
resolution and high completeness of flux-limited source samples don't
mix.  Simple quantities such as survey frequency and source flux
density have to be redefined for observations with large fractional
bandwidths.  Questions like ``How does
optical/IR identification completeness and reliability depend on
survey sensitivity and angular resolution?''~or ``How accurate are
in-band spectral indices?''~need quantitative answers.
Sections~\ref{vlassspecssec}, \ref{fluxbrightnesssec}, and
\ref{surveymetricssec} present the technical calculations.
The capabilities of the VLASS are compared with the proposed
science goals is Section~\ref{sciencesec}.   
  Section~\ref{sec:justification}
reviews VLASS justifications based on technical performance
and predicted science impact.
For the technically inclined, the ongoing debate on the angular
resolution that yields the best optical identifications is
detailed in Appendix~\ref{sec:ids}.

\section{VLASS Specifications}
\label{vlassspecssec}

\begin{deluxetable}{llcccccc}
\tablecolumns{8}
\tablewidth{0pc}
\tablecaption { VLASS Specifications }
\tablehead{
\colhead{ } &
\colhead{Component} & \colhead{Area} & \colhead{$\nu$} & 
\colhead{$\theta$} & \colhead{$\sigma_{\rm n}$} &
\colhead{$\sigma_{\rm n}$} & \colhead{$\sigma_{\rm n}$}  \\
\colhead{Tier} &
\colhead{Name} & \colhead{(deg$^2$)} & \colhead{(GHz)} &
\colhead{(arcsec)} & \colhead{($\mu{\rm Jy/}\Omega_{\rm b}$)} &
\colhead{(${\rm MJy~sr}^{-1}$}) & \colhead{(K)} 
}
\startdata
1 &
All-Sky & $\llap{3}3,885$ & 3.000 &
2.5 & 69 & 0.42 & 1.51  \\
2 & Deep &  \hphantom{00}10 & 3.000 &
0.8 & \hphantom{5}1\rlap{.5} & 0.08{\rlap 8} & 0.32 \\
1 &
All-Sky & $\llap{3}3,885$ & 2.682 &
2.8 & 65 & 0.31 & 1.42  \\
2 & Deep &  \hphantom{00}10 & 2.682 &
0.9 & \hphantom{5}1\rlap{.5} & 0.06{\rlap 9} & 0.32 \\
\enddata
\label{tab:VLASSspecs}
\end{deluxetable}

\begin{figure}
\vskip -1.5in
\epsscale{1.}
\plotone{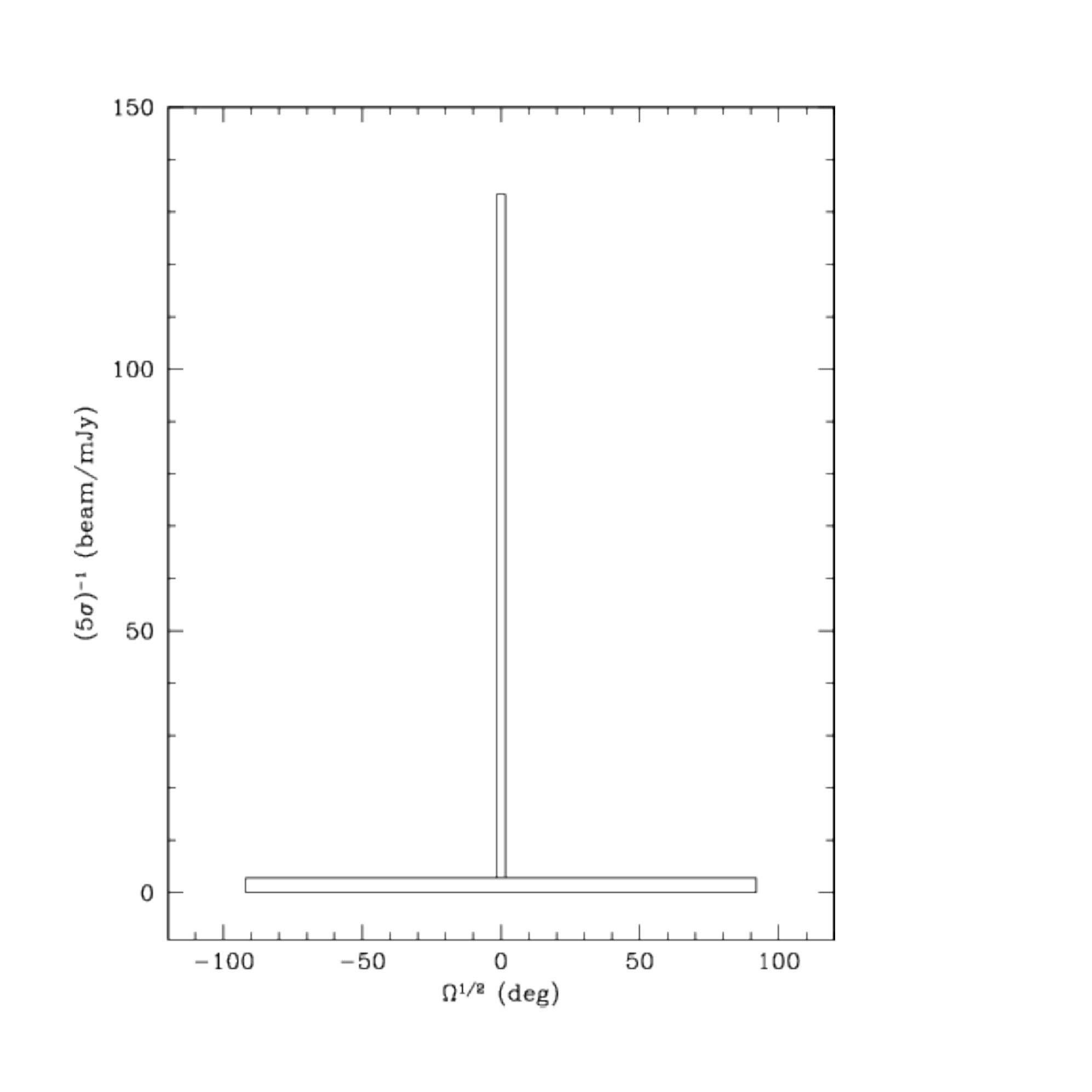}
\vskip -0in
\caption{The VLASS is more like a Prussian hat than a traditional wedding
  cake. The layers are All-Sky (bottom) and Deep (top).  The
  plotted width of each layer is $\Omega^{1/2}$, appropriate for a
  square cake.  The height of each layer is proportional to the
  point-source detection sensitivity.  Abscissa: $\Omega^{1/2}$~(deg)
  Ordinate: Point-source $(5\sigma)^{-1}$ sensitivity
  (beam~mJy$^{-1}$) }
\label{cakefig}
\end{figure}

The ``final'' VLASS proposal \citep{ssg15} describes an S-band survey
with two Tiers (Tier 1 = All-Sky and Tier 2 = Deep) that will use
$\sim 9000$ hours of JVLA observing time in the B and A configurations.
The VLA exposure calculator (for pointed observations with
1.5~GHz input bandwidth, no frequency
weighting, and ``robust'' $(u,v)$ weighting) gave the
Tier specifications at $\nu = 3 \mathrm{~GHz}$ listed in the first two
rows of Table~\ref{tab:VLASSspecs} and
plotted in Figure~\ref{cakefig}.  The first six
columns match Table~1 in \citet{ssg15}, and the final two
columns list their noise levels in brightness units.  (Caveat: most of
the Deep observing time is spent in the ECDFS field at declination
$\delta \approx -28^\circ$, where the 3~GHz beam is a $2\,\farcs0 \times
0\,\farcs8$ ellipse.)  The last two rows are more accurate
for surveys like the
VLASS in which the integration time per position is proportional to
primary beam solid angle, or $\nu^{-2}$
(Section~\ref{instrumentalsensitivitysec}).  Thus the effective
frequency of the VLASS is about 10\% lower, the synthesized beamwidth
is about 10\% larger, and the rms noise is slightly lower than
specified in \citet{ssg15}.


The Executive Summary of the \citet{ssg15}
stresses (original emphases)
``Both components make optimal utilization of the Jansky VLA's unique
capabilities: \emph{high resolution imaging} and
\emph{exquisite point-source
sensitivity}, critical for source identification; \emph{wide bandwidth
coverage}, enabling instantaneous spectral index determination; and
\emph{full polarimetry} with good performance even in lines of sight with
high Faraday depth, enabling instantaneous rotation measure and
Faraday structure determinations.''

The uniquely \emph{high angular resolution} chosen for the All-Sky is
not critical for source identification reliability
(Appendix~\ref{sec:ids}), and it comes at the cost of low completeness
for flux-limited source samples, and hence low identification
completeness.  The high rms \emph{surface brightness} noise levels
(columns 7 and 8 in Table~\ref{tab:VLASSspecs}) of both VLASS Tiers
mean that neither can detect low-brightness sources, normal
star-forming galaxies in particular, no matter how high their total
flux densities.

The \emph{exquisite point-source sensitivity} at 3~GHz of All-Sky
is only $1.5 \times$ better than FIRST for sources with
mean spectral index $\langle \alpha \rangle \equiv
d \ln S / d \ln \nu = -0.7$ and is $10 \times$ worse than
the sensitivity of the planned EMU \citep{nor11}
survey (Table~\ref{tab:FNEspecs}).
The point-source sensitivity of Deep is unique and should remain so
until SKA Phase I exists.

The \emph{wide bandwidth coverage} (2 to $4 \mathrm{~GHz}$) enables
instantaneous spectral-index determination in principle, but in
practice it is useful only for very strong, compact sources because
(1) the in-band $\alpha$ is so vulnerable to noise errors that
a signal-to-noise ratio
SNR $\approx 50$ is required to reach $\sigma_\alpha \approx 0.1$, and 
(2) the
spectral indices of extended sources are biased steep because
the synthesized beam solid angle falls by $4 \times$ across the
band (Section~\ref{alphasec}).

The relatively high VLASS center frequency (about twice the 1.4~GHz of
most competing surveys) allows for higher angular resolution, reduces
Faraday depolarization, and helps to protect the All-Sky snapshot
images from being dynamic-range limited.  The tradeoffs are (1) a
smaller instantaneous field-of-view for discovering transients having
durations shorter than the $\sim 32$~month cadence of All-Sky, (2) a
lower instrumental survey speed
(Section~\ref{instrumentalsurveyspeedsec}), and (3) a lower
point-source detection rate for most radio sources
(Section~\ref{pointsourcesurveyspeedsec}).  If surveys at frequencies
$\nu_1$ and $\nu_2$ detect the same numbers per steradian of radio
sources stronger than $S_1$ and $S_2$, then the effective spectral
index of the radio-source population can be defined as
\begin{equation}
\langle \alpha \rangle \equiv \ln (S_1/S_2) / \ln (\nu_1 / \nu_2)
\end{equation}
Over a wide range of frequencies near
 $\nu \sim 3$~GHz, $\langle \alpha \rangle = -0.7$
\citep{con84} so, on average, sources are $1.58 \times$
weaker at 2.682~GHz
than at 1.4~GHz.  The $\sigma_\mathrm{n} = 65 ~\mu
\mathrm{Jy~beam}^{-1}$ 2.682~GHz VLASS All-Sky survey should detect about
as many point sources per square degree as a $\sigma \approx 102\,\mu
\mathrm{Jy~beam}^{-1}$ 1.4~GHz survey.
Table~\ref{tab:FNEspecs} can be
used with Table~\ref{tab:VLASSspecs} to compare
the point-source and brightness sensitivities of VLASS with
FIRST \citep{bec95}, NVSS \citep{con98}, and
EMU \citep{nor11}.

\begin{deluxetable}{lcccccccc}
\tablecolumns{9}
\tablewidth{0pc}
\tablecaption {FIRST, NVSS, and EMU Specifications at 1.4~GHz
and converted  to 2.682~GHz using the mean source spectral index $\langle
\alpha  \rangle = -0.7$.}
\tablehead{
\colhead{} & \colhead{Area} & 
\colhead{$\theta$} & 
\multicolumn{3}{c}{$\sigma_{\rm n}$ (1.4 GHz)} &
\multicolumn{3}{c}{$\sigma_{\rm n}$ (2.682 GHz, $\alpha = -0.7$)}  \\
\colhead{Name} & \colhead{(deg$^2$)} &
\colhead{(arcsec)} & 
\colhead{{\llap (}$\mu{\rm Jy/}\Omega_{\rm b}$)} & 
\colhead{{\llap (}${\rm MJy~sr}^{-1}$)} & \colhead{(K)} &
\colhead{\hphantom{0}($\mu{\rm Jy/}\Omega_{\rm b}$)} & 
\colhead{{\llap (}${\rm MJy~sr}^{-1}$)} & \colhead{(K)} 
}
\startdata
FIRST & $\sim 10^4$ & 
$5\rlap{.4}$ & 
150 & {\llap 0}.193 & {\llap 3}.21\hphantom{0} &
\hphantom{0}95 & {\llap 0}.122 & 0.56 \\
NVSS & $\llap{3}.4 \times  10^4$ &
45 & 
450 & {\llap 0}.008{\rlap 3} & {\llap 0}.139 &
\hphantom{0}{\llap 2}85 & {\llap 0}.005{\rlap 3} & 0.024\\
EMU & $\llap{3}.1 \times  10^4$ &
10 &
10 & ${\llap 0}.003{\rlap 8}$ & {\llap 0}.062 &
\hphantom{0}6{\rlap {.3}} & {\llap 0}.002{\rlap 4} & 0.011 \\
\enddata
\label{tab:FNEspecs}
\end{deluxetable}

The main difficulty faced by any blind JVLA sky survey (as apposed to
a survey directed at a list of known targets such as nearby stars) is
that \emph{the JVLA was not optimized for sky surveys}: (1) Its
Field-of-View (FoV) is limited by the small primary beam of the large
($D = 25$~m) dishes, and each dish has only one primary beam.  (2) The
JVLA performance improvement over the original VLA is greatest at high
frequencies, where the wider bandwidths and new receivers multiply the
sensitivities of targeted observations by an order of magnitude.
However, FoVs and most source flux densities are lower at high
frequencies, so sky-survey performance is only slightly better.

\section{Flux Density, Peak Flux Density, and Brightness}
\label{fluxbrightnesssec}

The VLASS proposal \citep{ssg15} lists the VLASS rms noise levels
$\sigma_{\rm n}$ and source detection limits ($5\sigma_{\rm n}$) only
in terms of ``peak'' flux densities $S_{\rm p}$ that are defined as
flux density per beam ($\mu{\rm Jy~beam}^{-1}$).  ``Beam'' is
shorthand for the restoring beam solid angle, which is
\begin{equation}\label{eqn:beam}
\Omega_{\rm b} = {\pi \theta^ 2 \over 4 \ln 2}
\end{equation}
for a Gaussian restoring beam with half-power diameter $\theta$.

Peak flux densities ($S_\mathrm{p}$) are not flux densities ($S$).  Flux
densities are properties of {\em astronomical sources} but not images,
while peak flux densities depend on image resolution.  Astrophysically
useful source samples (e.g., samples used to construct luminosity
functions and track galaxy evolution) should be complete above well-defined
flux-density limits and not seriously biased against resolved sources
whose peak flux densities are significantly lower than their flux
densities.  Peak flux densities are numerically equal to flux
densities only for sources having angular diameters $\phi \ll \theta$.
\emph{Incompleteness for extended sources is the biggest weakness of the
high-resolution VLASS.}

Conflating $S_{\rm p}$ and $S$ systematically overestimates what a
high-resolution survey can do.  For example, ``...the star-forming
galaxy population becomes detectable at flux densities around
$S_\mathrm{1.4~GHz} \sim 1 \mathrm{~mJy~beam}^{-1}$...'' on page 8 of
\citet{ssg15} should really be ``...the star-forming galaxy population
becomes detectable at flux densities around $S_\mathrm{1.4~GHz} \sim 1
\mathrm{~mJy}$...''  Although All-Sky can easily detect $ S_\mathrm{p}
\sim 1 \mathrm{~mJy~\allowbreak beam}^{-1}$ sources, most star-forming
galaxies with $S _\mathrm{1.4~GHz} \geq 1 \mathrm{~mJy}$ fall below
its $S_\mathrm{p} = 5 \sigma_\mathrm{n} \approx 0.35
\mathrm{~mJy~beam}^{-1}$ detection limit.  Another example is
Figure~2 in \citet{ssg15}, which shows the predicted Poisson counting
errors in 3~GHz luminosity functions of star-forming galaxies at $z
\sim 1$ and $z \sim 3$ that could be derived from surveys covering 2,
4, and 10~deg$^2$ to a depth of $S = 7.5 \,\mu \mathrm{Jy}$, in order
to justify the Deep survey tier.  However, the Deep detection limit is
not $S = 7.5 \,\mu \mathrm{Jy}$; it is $S_\mathrm{p} = 7.5 \,\mu
\mathrm{Jy~beam}^{-1}$ in a $\theta \approx 0\,\farcs8$ FWHM beam.  If these 
faint sources are resolved enough to 
measure weak-lensing shear (Section~\ref{weaklensingsec}), their peak
flux densities $S_p$ must be significantly lower than their total flux
densities $S$ and many sources with $S \geq 7.5 \,\mu \mathrm{Jy}$
will not be detected.  \emph{A single survey can either detect most sources
  stronger than a certain flux density or it can resolve most sources,
  but it cannot do both.}

Peak flux densities are not really brightnesses either, even though
they have the same {\em dimensions} as brightness.  Brightnesses are
conserved
properties of {\it sources alone} and are independent of image
resolution.  Examples of proper brightness units are the ${\rm
  MJy~sr}^{-1}$ preferred by infrared astronomers and the K (Kelvins)
of Rayleigh-Jeans brightness temperature frequently used by radio
astronomers.  The apparent Rayleigh-Jeans brightness temperature in an
image pixel with peak flux density $S_{\rm p}$ is
\begin{equation}\label{imagetbeqn}
T_{\rm b} = {2 \ln(2) c^2 S_{\rm p} \over \pi k \theta^2 \nu^2}~,
\end{equation}
where $c \approx 3.00 \times 10^8 {\rm ~m~s}^{-1}$ and $k \approx 1.38
\times 10^{-23}{\rm ~J~K}^{-1}$.  In Table~\ref{tab:VLASSspecs} the
rms noises in K (Column 8) were calculated from
Equation~\ref{imagetbeqn}.  See Table~\ref{tab:FNEspecs} to compare
them with the noise parameters of the 1.4~GHz FIRST
\citep{bec95}, NVSS \citep{con98}, and EMU \citep{nor11} surveys.

Two surveys having the same sensitivity in $\mu{\rm Jy~beam}^{-1}$ but
different beam solid angles $\Omega_\mathrm{b}$ will have different
brightness sensitivities in K.  For example, the $5\sigma_{\rm n}$
detection limits of FIRST ($\theta = 5\,\farcs4$) and NVSS ($\theta =
45''$) are $1{\rm~mJy~beam}^{-1}$ and $2.3{\rm~mJy~beam}^{-1}$
respectively, so FIRST is more than twice as sensitive as the NVSS to
sources with $\phi \ll 5\,\farcs4$ (Table 2). In terms of brightness,
their rms image noises are $\sigma_{\rm n} = 3.2$1~K and 0.14~K
respectively, so the NVSS is $\sim 20$ times as sensitive as FIRST to
very extended sources ($\phi \gg 45''$).  Equations~\ref{spoverseqn}
and \ref{spoverslineareqn} in Section~\ref{extsourcesec}
imply that FIRST and NVSS sensitivities are
comparable for circular Gaussian sources of angular diameter $\phi
\sim 6''$ (e.g., radio emission powered by star formation in a face-on
disk galaxy) or narrow linear sources of length $\phi \sim 12''$
(e.g., radio jets powered by an AGN).  Both VLASS tiers have
higher angular resolutions and consequently lower brightness
sensitivities.

If the brightness detection limit of a survey image is greater than
the brightness of a source, that source will not be detected no matter
how close it is, because source brightness is distance-independent.
Most extragalactic sources have spectral indices $\alpha$
close ($\sigma_\alpha \sim 0.13$) to $\langle \alpha
\rangle = -0.7$ \citep{con84}, so their Rayleigh-Jeans
brightness temperatures vary
with frequency as $T_{\rm b} \propto \nu^{-2.7}$.  The brightness
temperature most relevant to sensitive extragalactic radio surveys is
the median brightness temperature of radio sources powered by star
formation in normal face-on spiral galaxies.  It is $\langle T_{\rm b}
\rangle \sim 1$~K at 1.4~GHz \citep{hum81}, and at nearby frequencies
it is
\begin{equation}\label{spiralsurfbrteq}
{\langle T_{\rm b} \rangle \over {\rm K}}
\approx 2.5 \Biggl({\nu \over {\rm GHz}}\Biggr)^{-2.7}~.
\end{equation}
Surveys must have brightness sensitivity limits $5\sigma_{\rm n} <
\langle T_{\rm b}\rangle$ to detect astrophysically complete samples
of normal star-forming galaxies and measure their flux densities
accurately.  Thus the 1.4~GHz NVSS ($5\sigma_{\rm n} \approx 0.7{\rm
  ~K}$) can detect and measure the flux densities of most nearby
spiral galaxies \citep{con02} but FIRST ($5\sigma_{\rm n} \approx
16{\rm ~K}$) cannot.  The median $\nu = 2.682$~GHz brightness
temperature of nearby spiral galaxies is $\langle T_{\rm b} \rangle
\sim 0.17{\rm ~K}$. The $5\sigma_{\rm n} < \langle T_{\rm b} \rangle$
requirement means that 2.682~GHz VLASS images must have $\sigma_{\rm n}
< 0.035{\rm ~K}$ to detect most low-redshift spiral galaxies.  Both
VLASS tiers miss this requirement by an order of magnitude (Table 1),
preventing VLASS images from
tracing the star-formation history of the universe
(Section~\ref{extsourcesec}).

For a survey to detect most radio sources powered by star formation,
its beamwidth must satisfy
\begin{equation}\label{brtdeteq}
\Biggl({\theta \over {\rm arcsec}}\Biggr)^2 \geq
2.44 \Biggl({\sigma_{\rm n} \over \mu{\rm Jy~beam}^{-1}}\Biggr)
\Biggl({\nu \over {\rm GHz}}\Biggr)^{+0.7}~.
\end{equation}
At the point-source sensitivities of the VLASS tiers
(Table~\ref{tab:VLASSspecs}), these minimum beamwidths are $\theta =
18''$ (All-Sky) and $2\,\farcs7$ (Deep).  A JVLA survey that could
detect most star-forming galaxies might use the C configuration at L
band to replace All-Sky, but it would just be an inferior version of
EMU ($\theta = 10''$, $\sigma_{\rm n} = 10\,\mu{\rm Jy~beam}^{-1}$,
$\nu = 1.4{\rm ~GHz}$).  Deep would better probe the cosmic evolution
of star formation if it were done with the B configuration at either S
band or L band.  Neither tier of the proposed VLASS can detect a
complete sample of nearby (e.g., $z < 0.5$) spiral galaxies and
measure the flux densities needed to calculate their luminosity functions.


Columns 6--9 of Table~\ref{tab:FNEspecs} list the FIRST, NVSS, and EMU
survey parameters converted to their 2.682~GHz equivalents for sources
with flux-density spectral index $\alpha = -0.7$ (temperature spectral
index $-2.7$) for direct comparison with the 2.682~GHz VLASS
parameters in Table~\ref{tab:VLASSspecs}.  The equivalent rms
noise of the 1.4~GHz FIRST survey at 2.682~GHz is $\sigma_{\rm n} =
150\,\mu{\rm Jy~beam}^{-1} \times (2.682/1.4)^{-0.7} = 95\,\mu{\rm
  Jy~beam}^{-1}$ so All-Sky ($\sigma_{\rm n} = 65\,\mu{\rm
  Jy~beam}^{-1}$) is only $1.5 \times$ as sensitive for point sources.  EMU
($\sigma_{\rm n} = 5.9\,\mu{\rm Jy~beam}^{-1}$ at 2.682~GHz) is an
order-of-magnitude more sensitive to point sources than any other
``all sky'' survey.  For extended sources with $\phi \gg
5\,\farcs4$, FIRST has 2.682~GHz brightness noise $\sigma_{\rm n} =
3.21{\rm ~K} \times (2.682/1.4)^{-2.7} = 0.56{\rm ~K}$ rms, which is
$2.5 \times$ as sensitive as All-Sky.  Both EMU ($\sigma_{\rm n} =
11$~mK) and NVSS (24~mK) have much lower 2.682~GHz brightness noise
levels.

Such tradeoffs involving point-source sensitivity, angular resolution,
and sensitivity to extended sources are illustrated
by the \citet{hod11} 1.4~GHz Stripe 82 survey. It has rms
noise $\sigma_{\rm n} = 52 \,\mu{\rm Jy~beam}^{-1}$ (equivalent to
$\sigma_\mathrm{n} \approx 33 \,\mu{\rm Jy~beam}^{-1}$ at 2.682~GHz
for $\alpha = -0.7$) and $\theta \gtrsim 1\,\farcs8$, which is $3
\times$ the point-source sensitivity, $3 \times$ the angular
resolution, and $3^{-1} \times$ the brightness sensitivity of FIRST.
The Stripe 82 survey failed to detect 22\% of the FIRST sources with
$S \geq 1000\,\mu{\rm Jy}$, although some of the missing FIRST ``sources''
may only be FIRST sidelobes of stronger sources. 
Comparison with the far more sensitive
COSMOS survey \citep{bon08} indicates the Stripe 82 source-detection
completeness is $\lesssim 0.5$ for $S \leq 500\,\mu{\rm Jy}$.

Figure~1 and Table~2 of \citet{ssg15} use the \citet{wil08} SKADS sky
simulation to predict the sky densities of extragalactic sources that
should be detected by All-Sky and Deep.  That figure shows a total of
$\approx 380 \mathrm{~sources~deg}^{-2}$ stronger than $S = 350 \,\mu
\mathrm{Jy}$ at 3~GHz, but only $290 \mathrm{~sources~deg}^{-2}$
brighter than the All-Sky detection limit $S_\mathrm{p} = 350 \,\mu
\mathrm{Jy~beam}^{-1} = 46 \,\mu\mathrm{Jy~arcsec}^{2}$, for a
\emph{cumulative} completeness of only 75\%.  That is, All-Sky is
predicted to miss about 25\% of \emph{all} sources stronger than $S =
350 \, \mu \mathrm{Jy}$.  Even the predicted All-Sky detection
rate of $290 \mathrm{~sources~deg}^{-1}$ seems optimistic, given that
FIRST detected only $\approx 90 \mathrm{~sources~deg}^{-2}$.  Table~2
of \citet{ssg15} indicates that All-Sky will detect about $10^7$
sources.  EMU \citep{nor11} is expected to detect about $7 \times
10^7$ sources.


\section{Survey Performance Metrics}\label{surveymetricssec}

The ubiquitous ``survey speed'' performance metric is useful only for
comparing the rates at which different telescopes can cover the sky
down to a specified point-source rms noise in
$\mu\mathrm{Jy~beam}^{-1}$ at a given frequency.
Section~\ref{surveymetricssec} defines the meaning(s) of ``frequency''
in targeted and survey observations made with large fractional
bandwidths and introduces speed metrics for comparing targeted and
sky-survey observations with a given array
(Sections~\ref{instrumentalsensitivitysec} and
\ref{instrumentaldirectedspeedsec}), metrics for comparing sky surveys
made at different frequencies
(Section~\ref{pointsourcesurveyspeedsec}), and metrics for comparing
the rates that surveys can detect extended sources as well as point
sources (Section~\ref{extsourcesec}). Section~\ref{alphasec} covers
``in band'' spectral indices and their noise uncertainties.

\subsection{Point Sources}\label{pointsourcesec}

\subsubsection{Instrumental Sensitivity}\label{instrumentalsensitivitysec}
The noise variance in an image made with ``natural'' weighting 
in the $(u,v)$ plane
is given by the simple
radiometer equation for interferometers:
\begin{equation}\label{eqn:noisevariance}
\sigma_{\rm n}^2 = 
{ S_{\rm sys}^2 \over \eta_{\rm c}^2 n_{\rm p} N (N-1) \tau B}~,
\end{equation}
where $S_{\rm sys}$ is the ``system equivalent flux density''
(SEFD) of noise for each antenna, $\eta_{\rm c} \geq 0.8$ is the
correlator efficiency, $n_{\rm p} = 2$ is the number of polarization
channels
contributing to the image, $N$ is the number of working antennas in
the array (the VLA exposure calculator assumes $N = 25$), $\tau$ is
the integration time, and $B$ is the instantaneous total bandwidth
(the VLA exposure calculator should be used with
 $B = 1.5 \mathrm{~GHz}$ at S band
to allow for RFI excision).  Both $\sigma_{\rm n}$ and $S_{\rm sys}$
are really peak flux densities, not flux densities.  The ``robust''
$(u,v)$ weighting used in the VLA exposure calculator at {\tt
  https://obs.vla.nrao.edu/ect/} multiplies the natural-weighting rms
noise by $\approx 1.20$.  

Equation~\ref{eqn:noisevariance} is valid for images with small
fractional bandwidths, but the VLA S band spans the octave frequency
range $\nu_\mathrm{min} = 2 \mathrm{~GHz}$ to $\nu_\mathrm{max} = 4
\mathrm{~GHz}$.  Equation~\ref{eqn:noisevariance} also applies to
the pointing centers of
broadband images in which $S_\mathrm{sys}^2$ and $\tau$ are
independent of frequency, as is often the case for targeted
observations imaging sources much smaller than the primary beamwidth,
and the spectral channels are unweighted (that is, channels with equal
bandwidths have equal weights).  

It appears that the variation of $S_\mathrm{sys}$ with frequency
across the JVLA S band has not been measured accurately enough (Condon
2015, private communication), so all VLASS sensitivity calculations
may have to be revised slightly.  Such a measurement is easy to make,
and it should be made before the VLASS external review.  For now, I
assume $S_\mathrm{sys}$ is nearly independent of frequency.

Source flux density is a narrowband quantity that can vary with
frequency.  What is the \emph{apparent} flux density of a point source in
an image made from data spanning a large fractional bandwidth? 
Most continuum radio sources have nearly power-law spectra
$S(\nu) / S_0 = (\nu / \nu_0)^\alpha$, where $\nu_0$
is any arbitrary reference frequency.  The (spectrally) unweighted
image flux density $S_\mathrm{u}$ of a point source is
\begin{equation}
S_\mathrm{u} = \frac {\int S(\nu) d \nu} {\int d \nu}
\approx S_0 \nu_0^{-\alpha} \Biggl( \frac{ \nu^{\alpha + 1}} {\alpha+1}
\Bigg|_{\nu_\mathrm{min}}^{\nu_\mathrm{max}} \Biggr) \Bigg/ 
(\nu_\mathrm{max} - \nu_\mathrm{min})~, \quad (\alpha \neq -1)
\end{equation}
so
\begin{equation}
\frac {S_\mathrm{u}} {S_0} = 
\Biggl( \frac {\nu_0^{-\alpha}} {\alpha+1} \Biggr)
\Biggl( \frac {\nu_\mathrm{max}^{\alpha+1} - \nu_\mathrm{min}^{\alpha+1}}
{\nu_\mathrm{max} - \nu_\mathrm{min}} \Biggr), \quad (\alpha \neq -1)
\end{equation}
For any spectral index $\alpha$ there is an ``effective'' frequency
$\nu_\mathrm{u}$ at which the unweighted image flux density
$S_\mathrm{u}$ equals the actual flux density of the source
$S(\nu_\mathrm{u})$:
\begin{equation}
\frac {S_\mathrm{u}} {S(\nu_\mathrm{u})} = 
\Biggl( \frac {\nu_\mathrm{u}^{-\alpha}} {\alpha+1} \Biggr)
\Biggl( \frac {\nu_\mathrm{max}^{\alpha+1} - \nu_\mathrm{min}^{\alpha+1}}
{\nu_\mathrm{max} - \nu_\mathrm{min}} \Biggr) = 1~, \quad (\alpha \neq -1)
\end{equation}
The effective frequency of an unweighted image for sources
with spectral index $\alpha$ is
\begin{equation}
\nu_\mathrm{u} = \Biggl[ 
\Biggl( \frac {1} {\alpha+1} \Biggr)
\Biggl( \frac { \nu_\mathrm{max}^{\alpha+1} - \nu_\mathrm{min}^{\alpha+1} } 
{ \nu_\mathrm{max} - \nu_\mathrm{min} } \Biggr) \Biggr]^{1/\alpha} ~, 
\quad (\alpha \neq -1)
\end{equation}
Except for sources with strongly
inverted spectra $\alpha \geq +1$,  $\nu_\mathrm{u}$
is lower than the arithmetic mean frequency $\bar{\nu} =
(\nu_\mathrm{max} + \nu_\mathrm{min})/2$ ~($= 3 \mathrm{~GHz}$ for
the JVLA at S band) that has 
conventionally been called the observing frequency.

\begin{figure}
\includegraphics[angle=0, scale=0.8]{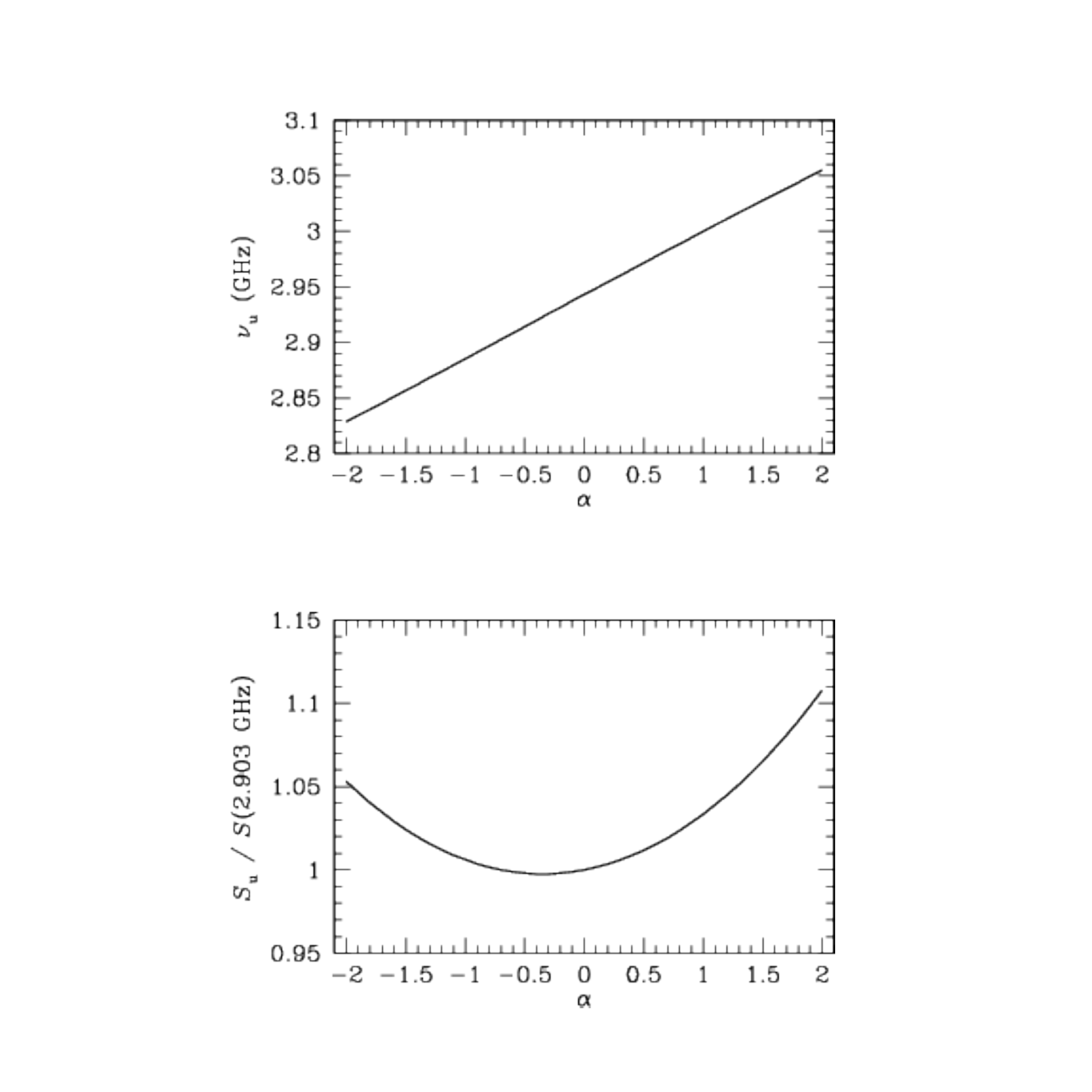}
\caption{The top panel shows the frequency $\nu_\mathrm{u}$ at which
  the spectrally unweighted VLA S-band ($2 \leq \nu \mathrm{\,(GHz)} \leq
  4$) image flux density is correct, as a function of source spectral
  index $\alpha$.  For sources with typical spectral index $\alpha =
  -0.7$, $\nu_\mathrm{u} \approx 2.903 \mathrm{~GHz}$.  If
  $\nu_\mathrm{u} = 2.903 \mathrm{~GHz}$ is taken to be the nominal
  image frequency, the ratio of the source flux density $S_\mathrm{u}$
  near the center of the broadband  image will differ from its true $S(2.903
  \mathrm{~GHz})$ flux density by the flux-density ratio plotted as a
  function of source spectral index in the bottom panel.
  $S_\mathrm{u}$ is within 1\% of $S(2.903 \mathrm{~GHz})$ over the
  spectral range $-1.2 < \alpha < +0.5$ that includes nearly all
  sources in samples complete at frequencies near $\nu \sim 3
  \mathrm{~GHz}$.}
\label{freqeffufig}
\end{figure}

It is useful to choose a single effective image frequency that best
characterizes the unweighted VLA S-band images for most sources.  The
natural choice is $\nu_\mathrm{u}(\langle \alpha \rangle)$, where
$\langle \alpha \rangle = -0.7$; it is $\nu_\mathrm{u} \approx 2.903
\mathrm{~GHz}$.  The ratio $S_\mathrm{u} / S(2.903\mathrm{~GHz})$
of the unweighted image flux density to the true 2.903~GHz flux
density varies slowly with source $\alpha$ as shown in the bottom
panel of Figure~\ref{freqeffufig}.  This ratio is in error by $< 1$\%
for all $\,-1.2 < \alpha < +0.5$, the spectral range encompassing nearly
all sources in a flux-limited sample selected at frequencies near $\nu
\sim 3 \mathrm{~GHz}$.  Thus an unweighted VLA S-band image spanning
$2 \leq \nu \mathrm{\,(GHz)} \leq 4$ should yield accurate and
meaningful 2.903~GHz flux densities for nearly all radio sources 
\emph{close
to the pointing center}. 

However, for broadband mosaiced sky surveys like the VLASS, the integration 
time $\tau$ at each position varies with frequency as $\tau(\nu) \propto
\nu^{-2}$ because the primary-beam solid angle is proportional to
$\nu^{-2}$.  Minimum survey image noise is attained by weighting each
spectral channel by the inverse of its noise variance, so the
weight assigned to the $i$th spectral channel should be $W_\mathrm{i} \propto
\tau(\nu_\mathrm{i}) \propto \nu_\mathrm{i}^{-2}$ if the channels all
have the same bandwidth.  Comparing the noise variance
$\sigma_\mathrm{w}^2$ of a survey image based on weighted spectral channels
spanning the frequency range from $\nu_\mathrm{min} = 2 \mathrm{~GHz}$
to $\nu_\mathrm{max} = 4 \mathrm{~GHz}$ with the VLASS sensitivity originally
calculated from the unweighted Equation~\ref{eqn:noisevariance} at
$\bar{\nu} = 3 \mathrm{~GHz}$ for the VLASS gives
\begin{equation}\label{eqn:surveyweights}
\frac{\sigma_\mathrm{w}^2} {\sigma_\mathrm{n}^2} =
\frac{\int d\nu} {\int (\nu/\bar{\nu})^{-2} d\nu}
= \frac{\nu_\mathrm{min}\, \nu_\mathrm{max}} {\bar{\nu}^2} = \frac{8}{9}
\end{equation}
Properly weighting the spectral channels by $(\nu/\bar{\nu})^{-2}$
lowers the originally calculated VLASS image noise by the factor
$\sigma_\mathrm{w} / \sigma_\mathrm{n} \approx 0.943$.

Spectral weighting $W \propto \nu^{-2}$ 
also lowers the weighted arithmetic mean frequency $\bar{\nu}_\mathrm{w}$ to
\begin{equation}
\bar{\nu}_\mathrm{w} = \frac {\int (\nu/\bar{\nu})^{-2} \nu\, d \nu}
{\int (\nu/\bar{\nu})^{-2}\, d \nu} = 
\frac{\int \nu^{-1} d \nu}{\int \nu^{-2} d\nu}
= \Biggl( \frac{\nu_\mathrm{min}\, \nu_\mathrm{max}}
{\nu_\mathrm{max} - \nu_\mathrm{min}}\Biggr) \ln\Biggl(
\frac{\nu_\mathrm{max}}{\nu_\mathrm{min}} \Biggr)~.
\end{equation}
In the case of the VLASS,
\begin{equation}
\bar{\nu}_\mathrm{w} = \biggl( \frac{2 \cdot 4}{4 - 2}\biggr) \ln \biggl(
\frac{4}{2}\biggr) \mathrm{~GHz} = 4 \ln 2 \mathrm{~GHz}
\approx 2.773 \mathrm{~GHz}
\end{equation}
Note that this weighted frequency is a property of the instrument and survey
frequency range only; it is independent
of radio source spectra. 

\begin{figure}
\includegraphics[angle=0, scale=0.8]{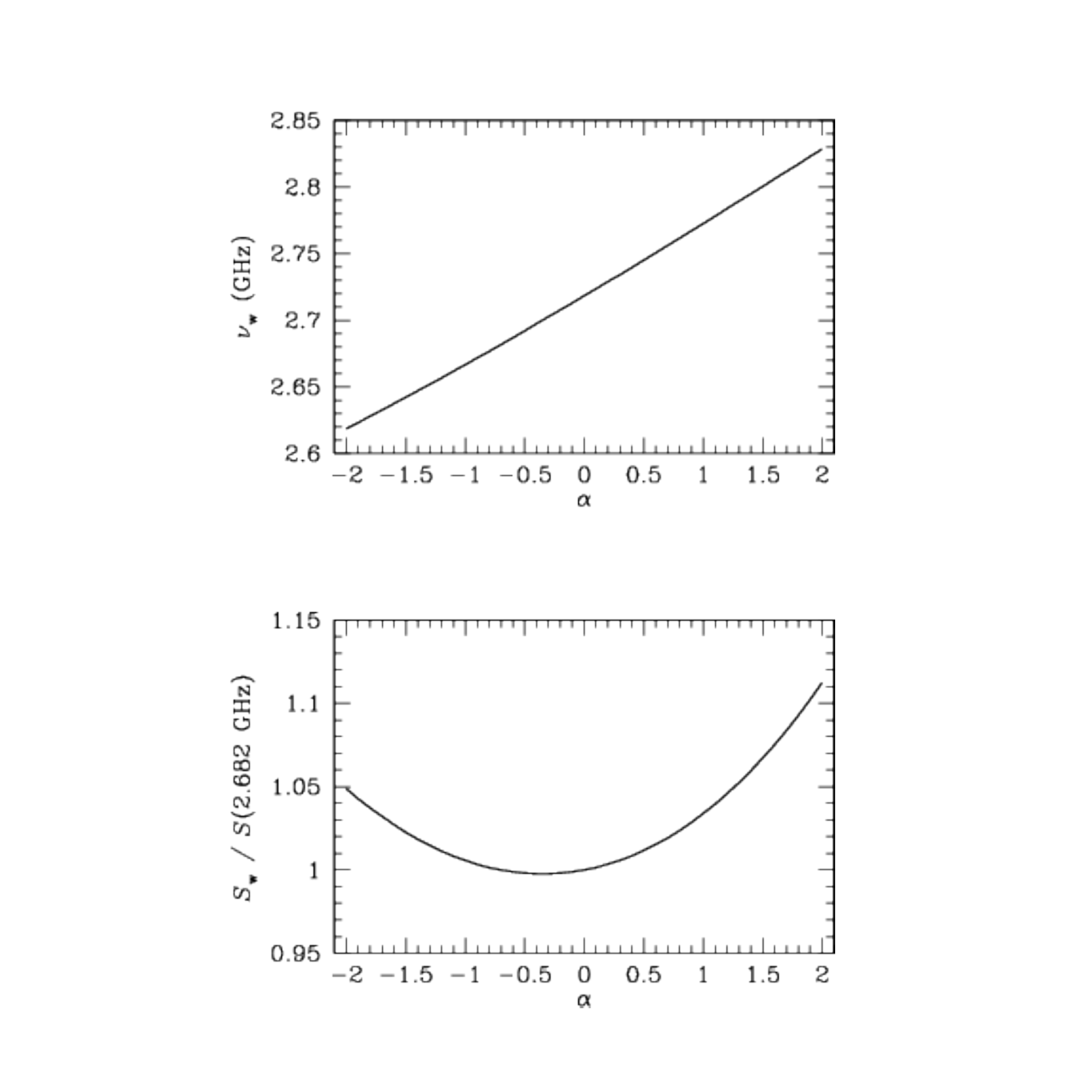}
\caption{The top panel shows the frequency $\nu_\mathrm{w}$ at which
  the spectrally weighted VLASS image flux density is correct, as a
  function of source spectral index $\alpha$.  For sources with
  typical spectral index $\alpha = -0.7$, $\nu_\mathrm{w} \approx
  2.682 \mathrm{~GHz}$.  If $\nu_\mathrm{w} = 2.682 \mathrm{~GHz}$ is
  taken to be the nominal VLASS image frequency, the ratio of the
  source flux density $S_\mathrm{w}$ in the broadband survey image to
  its true $S(2.682 \mathrm{~GHz})$ flux density is plotted as a
  function of source spectral index in the bottom panel.
  $S_\mathrm{w}$ is within 1\% of $S(2.682 \mathrm{~GHz})$ over the
  spectral range $-1.2 < \alpha < +0.5$ that includes nearly all
  sources in samples complete near frequencies $\nu \sim 3
  \mathrm{~GHz}$.}
\label{freqeffwfig}
\end{figure}

If a point source has flux density $S(\nu) / S_0 =  (\nu /
\nu_0)^\alpha$, 
its apparent flux density in a weighted broadband survey image is
\begin{equation}
S_\mathrm{w} = \frac{ \int (\nu/\nu_0)^{-2} S(\nu) d \nu} {\int
  ((\nu/\nu_0)^{-2} d \nu} = S_0 \nu_0^{-\alpha} \Biggl(
\frac{\nu^{\alpha -1}}{\alpha - 1}
\Bigg|_{\nu_\mathrm{min}}^{\nu_\mathrm{max}} \Biggr) \, \Bigg/\,
(-\nu^{-1})
\Bigg|_{\nu_\mathrm{min}}^{\nu_\mathrm{max}} ~,\quad (\alpha \neq -1)
\end{equation}
\begin{equation}
\frac {S_\mathrm{w}} {S_0} = \Biggl(\frac {\nu_0^{-\alpha}}{1 - \alpha} \Biggr)
\Biggl(\frac{\nu_\mathrm{max}^{\alpha - 1} - \nu_\mathrm{min}^{\alpha - 1}}
{\nu_\mathrm{max}^{-1} - \nu_\mathrm{min}^{-1}}\Biggr)
 ~,\quad (\alpha \neq -1)
\end{equation}
The frequency
$\nu_\mathrm{w}$ at which the weighted image flux density is equals
the actual source flux density satisfies
\begin{equation}
\frac{S_\mathrm{w}}{S(\nu_\mathrm{w})} = 
\Biggl( \frac {\nu_\mathrm{w}^{-\alpha}}{1 - \alpha} \Biggr)
\Biggl(\frac{\nu_\mathrm{max}^{\alpha - 1} - \nu_\mathrm{min}^{\alpha - 1}}
{\nu_\mathrm{max}^{-1} - \nu_\mathrm{min}^{-1}}\Biggr) = 1
 ~,\quad (\alpha \neq -1)~.
\end{equation}
It is
\begin{equation}\label{eqn:freqeff}
\nu_\mathrm{w} = \Biggl[\Biggl( \frac {1} {1 - \alpha} \Biggr)
\Biggl(\frac{\nu_\mathrm{max}^{\alpha - 1} - \nu_\mathrm{min}^{\alpha - 1}}
{\nu_\mathrm{max}^{-1} - \nu_\mathrm{min}^{-1}}\Biggr) \Biggr]^{1 / \alpha}
 ~,\quad (\alpha \neq -1)
\end{equation}
The top panel of Figure~\ref{freqeffwfig} shows
$\nu_\mathrm{w}(\alpha)$ for the VLASS.

The single effective frequency that best characterizes spectrally
weighted survey images for most sources is
$\nu_\mathrm{w}(\langle \alpha \rangle)$, where $\langle \alpha
\rangle = -0.7$; it is $\nu_\mathrm{w} \approx 2.682 \mathrm{~GHz}$ 
for the VLASS.
Then $S_\mathrm{w} / S(2.682\mathrm{~GHz})$ varies
slowly with $\alpha$ as shown in the bottom panel of
Figure~\ref{freqeffwfig}.  This ratio is in error by $< 1$\% for all
$-1.2 < \alpha < +0.5$, the spectral range encompassing nearly all
sources in a flux-limited sample at frequencies near $\nu \sim 3
\mathrm{~GHz}$.  Equation~\ref{eqn:freqeff} demonstrates that
the broadband VLASS images should yield accurate and meaningful 2.682~GHz
flux densities for nearly all point sources.

In summary, the original unweighted VLASS with nominal
$\bar{\nu} = 3 \mathrm{~GHz}$ is really a frequency-weighted survey
with effective frequency $\nu_\mathrm{e} \approx 2.682 \mathrm{~GHz}$
and about 6\% lower rms noise.  At $2.682 \mathrm{~GHz}$ the typical
source with $\alpha = \langle\alpha\rangle = -0.7$ is $(2.682 /
3)^{\langle\alpha\rangle} \approx 1.082$ times stronger than it is at
3~GHz, so the frequency-weighted VLASS images should reveal about as
many point sources per steradian as a 3~GHz unweighted survey in about
76\% of the originally proposed observing time.

For the revised All-Sky survey with three scans and robust $(u,v)$
weighting, the VLA exposure calculator predicts an rms noise
$\sigma_\mathrm{n} \approx 69~\mu\mathrm{Jy~beam}^{-1}$
at $\bar{\nu} = 3 \mathrm{~GHz}$ 
appropriate for unweighted targeted observations.  
With optimum survey frequency weighting, 
the rms noise drops to $\sigma_\mathrm{w} \approx 65~\mu\mathrm{Jy~beam}^{-1}$,
the effective frequency becomes $\nu_\mathrm{w} = 2.682~\mathrm{GHz}$, and
the source detection rate should match that of an unweighted 3~GHz survey
with rms noise $\sigma_\mathrm{n} \approx 60~\mu\mathrm{Jy~beam}^{-1}$.

\subsubsection{Instrumental Speed for Directed Observations}
\label{instrumentaldirectedspeedsec}

The integration time $\tau$ required to reach a given noise
$\sigma_{\rm n}$ in a single pointing can be obtained from the
radiometer Equation~\ref{eqn:noisevariance}; it is
\begin{equation}
\tau = 
{1 \over \sigma_{\rm n}^2}
\Biggl[{S_{\rm sys}^2 \over \eta_{\rm c}^2 n_{\rm p}
N (N-1)}\Biggr] {1 \over B}
\end{equation}
so the {\em instrumental} speed $\dot{n}$ for directed observations 
 defined by
\begin{equation}\label{eqn:imagingspeed}
\dot{n} \equiv {1 \over \tau} = \sigma_{\rm n}^2
\Biggl[{\eta_{\rm c}^2 n_{\rm p} N (N-1) 
\over S_{\rm sys}^2}\Biggr]  B
\end{equation}
is a purely instrumental figure-of-merit equal to the rate at which directed
observations of point sources can be made with rms noise $\sigma_{\rm
  n}$.  At frequencies $\nu \lesssim 10{\rm ~GHz}$ the quantity in
brackets is nearly the same for both the old VLA and the new JVLA, but
the JVLA's usable bandwidth (ranging from $B \sim 600{\rm ~MHz}$ at
1.5~GHz to $B \sim 8{\rm ~GHz}$ above 18~GHz) is one or two orders of
magnitude larger than the correlator-limited bandwidth of the old VLA,
which was 50 or 100 MHz depending on the array configuration, field
size, and the tolerable amount of bandwidth smearing.  Thus 
\emph{the JVLA
is faster than the old VLA by one to two orders of magnitude for a
typical user program} making a directed observation or a
directed survey targeting a sample
of $n$ discrete  sources.  

\subsubsection{Instrumental Survey Speed}
\label{instrumentalsurveyspeedsec}

For a  continuum sky survey covering a much larger solid angle than a single
field-of-view $\Omega_{\rm FoV}$, the figure-of-merit corresponding to
$\dot{n}$ in Equation~\ref{eqn:imagingspeed} is the {\em
  instrumental survey speed} $\dot{\Omega}$ defined by
\begin{equation}\label{surveyspeedeqn}
\dot{\Omega} \equiv {\Omega_{\rm FoV} \over \tau} = \dot{n} {\Omega_{\rm FoV}}
\end{equation}
Instrumental survey speed measures the rate of sky coverage for a
given rms noise $\sigma_{\rm n}$, and it is usually expressed in units
of deg$^2$~hour$^{-1}$.  Note that this traditional definition of
survey speed does not include channel weighting
(Section~\ref{pointsourcesec}), so it slightly underestimates the
potential speed of surveys with large fractional bandwidths.

Primary beams are approximately Gaussian,
so the primary beam solid angle is
\begin{equation}
 \Omega_{\rm pb} \approx {\pi \theta_{\rm pb}^2 \over 4 \ln 2}
\approx 1.13 \theta_{\rm pb}^2~,
\end{equation}
where 
\begin{equation}
\theta_{\rm pb} \approx \Biggl({1.09 \lambda \over D}\Biggr) {\rm ~rad} 
\approx 0.75 \Biggl( {\nu \over {\rm GHz}}\Biggr)^{-1}{\rm ~deg}
\end{equation}
is the FWHM primary beamwidth of the VLA's $D = 25$~m antennas.  For sensitivity
calculations, the
effective field-of-view of a single Gaussian primary beam is exactly half the
primary beam solid angle \citep{con98}:
\begin{equation}
\Omega_{\rm FoV} = {\Omega_{\rm pb} \over 2} \approx 
0.32 \Biggl({\nu \over {\rm GHz}}\Biggr)^{-2} \mathrm{~deg}^{-2}~.
\end{equation}
It turns out that the instrumental survey speeds of both the old VLA and 
the new JVLA are well
approximated by the convenient expression
\begin{equation}
\label{eqn:VLAomegadot}
\Biggl({\dot{\Omega} \over {\rm deg}^2{\rm ~hr}^{-1}}\Biggr)
\approx 
\Biggl({\sigma_{\rm n} \over \mu{\rm Jy~beam}^{-1}}\Biggr)^2
\Biggl({S_{\rm sys} \over {\rm Jy~beam}^{-1}}\Biggr)^{-2}
\Biggl({B \over {\rm MHz}}\Biggr)
\Biggl({\nu \over {\rm GHz}}\Biggr)^{-2}
\end{equation}

Because $\Omega_{\rm FoV}$ is proportional to $\nu^{-2}$,
Equations~\ref{surveyspeedeqn} and \ref{eqn:VLAomegadot} imply that
the speed of the JVLA (or any other radio telescope, even one with
multiple beams like ASKAP) for surveys falls as $\nu^{-2}$ relative to
the speed for directed observations.  \emph{This pushes large surveys to
lower frequencies when competing for telescope time against targeted
user proposals.}

\subsubsection{Point-Source Survey Speed}\label{pointsourcesurveyspeedsec}

Note that Equations~\ref{eqn:imagingspeed} and \ref{surveyspeedeqn}
respectively defining the speeds for directed observations and for
blind surveys describe purely {\it instrumental} properties; the
spectra of actual radio sources were not considered.  If the goal of a
sky survey at frequency $\nu$ is to detect the largest number of point
sources per steradian in a given observing time, source spectra must be
taken into account.
Nearly all discrete radio sources are
extragalactic at mJy and $\mu$Jy levels, with effective spectral index
$\langle \alpha \rangle
\approx -0.7$ \citep{con84}. 
Thus the rms image noise  of a survey at frequency $\nu_1$
should be multiplied by $(\nu_2 /
\nu_1)^{\langle\alpha\rangle}$ to compare with another survey at 
 frequency $\nu_2$. 

This leads
to the definition of the {\em point-source survey speed} $\dot{\Omega}_{\rm
  s}$ needed to compare the point-source detection
rates of sky surveys made at two different
frequencies $\nu_1$ and $\nu_2$:
\begin{equation}\label{sourcesurveyspeedeqn}
\frac {\dot{\Omega}_{\rm s}(\nu_1)} {\dot{\Omega}_{\rm s}(\nu_2)}  
\equiv \frac {\dot{\Omega}(\nu_1)} {\dot{\Omega}(\nu_2)} 
\Biggl({\nu_1 \over \nu_2}\Biggr)^{2\langle\alpha\rangle}~.
\end{equation}
Because $\dot{\Omega} \propto \nu^{-2}$ and $\langle \alpha \rangle =
-0.7$, $\dot\Omega_{\rm s} \propto \nu^{-3.4}$ for fixed bandwidth $B$
and $S_\mathrm{sys}$.  If $B \propto \nu$, then $\dot{\Omega_{\rm s}}
\propto \nu^{-2.4}$, which is still a steep frequency dependence
strongly favoring surveys at lower frequencies.  For the VLA and JVLA
at frequencies $1 \mathrm{~GHz} < \nu < 10 \mathrm{~GHz}$,
\begin{equation}\label{jvlassseq}
\Biggl({\dot{\Omega}_{\rm s} \over {\rm deg}^2{\rm ~hr}^{-1}}\Biggr)
\approx 
\Biggl({\sigma_{\rm n} \over \mu{\rm Jy~beam}^{-1}}\Biggr)^{-2}
\Biggl({S_{\rm sys} \over {\rm Jy}}\Biggr)^{-2}
\Biggl({B \over {\rm MHz}}\Biggr)
\Biggl({\nu \over {\rm GHz}}\Biggr)^{-2}
\Biggl({\nu \over \nu_0}\Biggr)^{-1.4}
\end{equation}

\begin{deluxetable}{rcrrccr}
\tablecolumns{7}
\tablewidth{0pc}
\tablecaption {JVLA Instrumental and Point-Source Survey Speeds}
\tablehead{
\colhead{$\nu$} & 
\colhead{$S_{\rm sys}$} & \colhead{$B$} & 
\colhead{$\Omega_{\rm FoV}$} & \colhead{$\dot{\Omega}$} & 
\colhead{\hphantom{x}$\dot{\Omega}_{\rm s}$} & VLASS{\rlap /} \\
\colhead{(GHz{\rlap )}} & 
\colhead{(Jy~beam)$^{-1}$} & \colhead{(MHz{\rlap )}} & 
\colhead{(deg$^2$)} & \colhead{(deg$^2$/hr)} & 
\colhead{(deg$^2$/hr{\rlap )}} & FIRST}
\startdata
1.5  & 420 &  600 & 0.142 & {\llap 1}5.1 & \llap{1}5\rlap{.1}  & 10 \\
3.0  & 370 & 1500 & 0.035 & {\llap 1}6.5 &  6\rlap{.3} & 4 \\
6.0  & 310 & 3400 & 0.0089 & 7.2 & 1\rlap{.0} & 0.7 \\
10.0 & 250 & 3400 & 0.0032 & 3.0 & 0\rlap{.21} & 0.14 \\
\enddata
\label{tab:JVLAspeeds}
\end{deluxetable}

Column 5 of Table~\ref{tab:JVLAspeeds} shows the instrumental survey
speeds $\dot{\Omega}$ at 1.5~GHz (L band), 3.0~GHz (S band), 6.0~GHz
(C band), and 10.0~GHz (X band) for $\sigma_{\rm n} = 100 \,\mu{\rm
  Jy~beam}^{-1}$.  The L and S band instrumental survey speeds are
nearly equal and much faster than the shorter-wavelength bands, but
that doesn't mean that the L and S bands are nearly equal for
detecting sources in sky surveys.  Column 6 of
Table~\ref{tab:JVLAspeeds} was calculated from
Equation~\ref{sourcesurveyspeedeqn} using $\nu_0 = 1.5{\rm ~GHz}$ for
the fiducial frequency at which $\dot{\Omega} = \dot{\Omega_{\rm s}}$.
The source survey speed $\dot{\Omega}_{\rm s}$ is highest at 1.5~GHz
and falls rapidly at higher frequencies because the field-of-view is
getting smaller and the sources are getting weaker.

The choice of S band instead of L band for the VLASS costs about a
factor of 2.4 in observing time (e.g., 9000 hours versus 3800 hours
needed to detect the same number of sources), so it needs to be
justified in terms of dynamic-range limitations (a potential problem for
``snapshot'' surveys like FIRST, NVSS, and All-Sky), angular resolution,
spectral-index information, Faraday rotation, cadence for transients,
etc.  It is not sufficient to dismiss L band on the grounds that the
instrumental survey speed $\dot{\Omega}$ is about the same as at S
band.

The original ($\sigma = 100 \,\,\mu \mathrm{Jy~beam}^{-1}$) version
of All-Sky can also be compared with 1.4~GHz
VLA surveys such as FIRST \citep{bec95}.  The VLA observational status
summary for July 1996 gives $\sigma_{\rm n} = 60\,\mu{\rm Jy ~beam}^{-1}$
after $\tau = 10 {\rm ~min}$ for a naturally weighted 1.4~GHz image
made with the nominal 100 MHz bandwidth.  The nominal bandwidth of
FIRST was $B \approx 50 {\rm ~MHz}$ so $\sigma_{\rm n} \approx 35\,
\mu{\rm Jy~beam}^{-1}$ after $\tau = 1{\rm ~hr}$.  With $\Omega_{\rm
  FoV} \approx 0.163 {\rm ~deg}^2$ at 1.4~GHz, the FIRST point-source
survey speed relative to $\nu_0 = 1.5 {\rm ~GHz}$ is
$\dot{\Omega}_{\rm s} \approx 1.5 {\rm ~deg}^2{\rm ~hr}^{-1}$.

The (original) VLASS source survey speed is only a $4
\times$ improvement over FIRST, while regular users competing for the
same JVLA time are getting a $10 \times$ to $100 \times$ speed
improvement over the VLA on their directed observations, making it
$2.5 \times$ to $25 \times$ harder to justify taking time from
directed observations to make VLASS than it was to justify FIRST and
NVSS.  \emph{This factor should be considered in comparisons that use
  FIRST and NVSS to justify taking time from directed observations to
  make the VLASS (Section~\ref{sec:justification}).}

Column~7 of Table~\ref{tab:JVLAspeeds} lists the JVLA/FIRST source
survey speed ratios for point sources with $\langle \alpha \rangle =
-0.7$.  A JVLA point-source survey should be faster than the original
FIRST by a factor of 10 at L band and by a factor of 4 at S band, but
slower than FIRST at C band and X band.  For surveys of $\langle
\alpha \rangle = -0.7$ point-source populations, the JVLA at 3 GHz is
only $\sim 4\times$ as fast or $\sim 2 \times$ as sensitive as the VLA
was at 1.4 GHz.  That's why the originally proposed 3 GHz VLASS
All-Sky survey exclusive of WIDE covered just over twice the FIRST
survey area with slightly worse sensitivity to point sources, so it
would have needed about half of the 4000 hours spent on FIRST.  The
originally proposed 3 GHz VLASS WIDE survey covered about the same
solid angle as FIRST with $\sim 1.8 \times$ better point-source
sensitivity, so it would have needed about 3000 hours.

The revised VLASS survey speeds with survey-weighted frequency channels
and natural $(u,v)$ weighting would be about 32\% higher, but the
increased noise from robust $(u,v)$ weighting actually slows down
the revised VLASS survey speed by about 9\%.

\subsubsection{Extended-Source Sensitivity}\label{extsourcesec}

The preceding parts of Section~\ref{pointsourcesec} apply only to
sources much smaller than the survey point-spread function (PSF).  The
apparent brightness distribution of a source on an image is the
convolution of the PSF and the actual source brightness distribution.  Under
convolution, the total flux density is conserved and the source area on
the image is the sum of the PSF and source areas, so the peak flux
density = flux/area 
on the image falls accordingly.  For simplicity, consider a circular
Gaussian source (e.g., the disk of a face-on star-forming galaxy) of
angular diameter $\phi$ and total flux density $S$.  Its peak flux
density $S_{\rm p}$ in an image with a circular Gaussian PSF of FWHM
diameter $\theta$ will be
\begin{equation}\label{spoverseqn}
\Biggl({S_{\rm p} \over S}\Biggr) = \Biggl({\theta^2 \over
\phi^2 + \theta^2}\Biggr)~.
\end{equation}
The analog of Equation~\ref{spoverseqn} for a narrow linear source
(e.g., a radio core plus twins jets and/or lobes powered by an AGN, or
an edge-on thin disk) with an approximately Gaussian brightness profile is
\begin{equation}\label{spoverslineareqn}
\Biggl({S_{\rm p} \over S}\Biggr) = 
\Biggl({\theta^2 \over \phi^2 + \theta^2}\Biggr)^{1/2}
\end{equation}

For example, the Deep survey has $\theta = 0\,\farcs9$ resolution 
at 2.682~GHz and
a detection limit $S_{\rm p} = 7.5\,\mu{\rm Jy~beam}^{-1}$.
If a face-on star-forming galaxy at redshift $z = 0.2$ has a FWHM
diameter of 5 kpc, its angular size will be $\phi = 3\,\farcs1$.
Deep can detect such a galaxy only if its flux density is greater than
\begin{equation}
S = S_{\rm p} \Biggl({\phi^2 + \theta^2 \over \theta^2}\Biggr) = 
7.5\,\mu{\rm Jy~beam}^{-1} \cdot
\Biggl({3.1^2 + 0.9^2 \over 0.9^2}\Biggr)\,{\rm beams} =  96\, \mu{\rm Jy}
\end{equation}
If the disk were edge-on, Deep could detect it
only if $S \geq 27 \,\mu{\rm Jy}$.


In the limit of sources much
  larger than the beam ($\phi \gg \theta$), $S_{\rm p} / S \rightarrow
  (\theta / \phi)^2$ and it makes sense to describe sources in terms of
  their brightness temperatures
\begin{equation}
T_{\rm b}({\rm K}) = {2 \ln(2) c^2 S \over \pi k \phi^2 \nu^2}
\approx 1.22 \Biggl({S \over \mu{\rm Jy}}\Biggr)
\Biggl({\phi \over {\rm arcsec}}\Biggr)^{-2}
\Biggl({\nu \over {\rm GHz}}\Biggr)^{-2}~.
\end{equation}
The rms sensitivities in Kelvins of brightness temperature at 2.682~GHz are
listed for both VLASS tiers and for comparable surveys in the
rightmost columns of Tables~\ref{tab:VLASSspecs} and
\ref{tab:FNEspecs}, respectively.

\subsection{Point-Source Spectral Indices}\label{alphasec}

The JVLA's octave bandwidth is large enough to yield useful VLASS
in-band spectral indices, at least for fairly strong point sources.
Most radio sources have nearly power-law spectra, so their spectral
indices $\alpha \equiv d \ln S / d \ln \nu$
can be derived from linear fits of many narrow-band channel
flux densities to $\ln (S/S_0) = \alpha \ln (\nu/\nu_0)$, where $S_0$
is the flux density at some fiducial frequency $\nu_0$.  The accuracy
with which the slope $\alpha$ of the linear fit can be
determined from the noisy channel flux densities is proportional the
length
\begin{equation}\label{eqn:armlength}
l = \ln (\nu_\mathrm{max}) - \ln
(\nu_\mathrm{min}) = \ln (\nu_\mathrm{max} / \nu_\mathrm{min})
\end{equation}
 of the bandwidth
``lever arm'' in ln(frequency) space.  Ideally $l = \ln (4 \mathrm{~GHz} / 2
\mathrm{~GHz}) \approx 0.693$ for the VLASS; in practice $l$ may shrink
if RFI preferentially eliminates channels near the ends of the frequency range.

Traditional two-point spectral indices $\alpha$
determined from two narrow-band flux densities
$S_1$ and $S_2$ measured at frequencies $\nu_1$ and $\nu_2$
with rms noises $\sigma_1$ and $\sigma_2$
have rms noise errors
\begin{equation}\label{eqn:2pointalpha}
\sigma_\alpha = 
\frac {\bigl[(\sigma_1 / S_1)^2 + (\sigma_2 / S_2)^2 \bigr]^{1/2}}
{ \vert \ln(\nu_1 / \nu_2) \vert }
\end{equation}
For example, the $\nu_1 = 0.327 \mathrm{~GHz}$ WENSS survey has
$\sigma_1 \approx 3.6 \mathrm{~mJy}$ and the $\nu_2 = 1.4
\mathrm{~GHz}$ NVSS survey has $\sigma_2 \approx 0.45 \mathrm{~mJy}$
for point sources.  The rms width of the spectral-index distribution
of ``normal spectrum'' sources in flux-limited samples is
$\sigma_\alpha \approx 0.13$ \citep{con84}, so the measured spectral
indices of sources with $\alpha \approx -0.7$ should have errors
smaller than this to be useful.  If we require $\sigma_\alpha \leq
0.1$ for a source with $\alpha = -0.7$, Equation~\ref{eqn:2pointalpha}
implies $S_2 > 9.5 \mathrm{~mJy}$ at $\nu_2 = 1.4 \mathrm{~GHz}$
or $S(2682 \mathrm{~GHz}) > 6.0 \mathrm{~mJy}$.  Can
All-Sky ``in-band'' or ``instantaneous'' spectra
lower this sensitivity limit significantly?

The ideal (ignoring frequencies lost to RFI) VLASS All-Sky would span
2 to 4~GHz uniformly with a large number $N \gg 1$ of narrow frequency
channels.  
For comparison with standard equations for linear least-squares
fits \citep{bev69}, the power-law approximation
\begin{equation}\label{eqn:powerlaw}
\ln (S) = \ln (S_0) + \alpha \ln (\nu / \nu_0)~,
\end{equation}
can be written as
\begin{equation}
y = a + bx ~,
\end{equation}
where
\begin{equation}
y = \ln(S), \quad a = \ln (S_0), \quad b = \alpha, \quad \mathrm{and} \quad
x = \ln (\nu/\nu_0)~.
\end{equation}
In the $i$th frequency channel, $x_\mathrm{i} = \ln (\nu_\mathrm{i} /
\nu_0)$, $y_\mathrm{i} = \ln(S_\mathrm{i})$ is the measured flux
density, and $\sigma_\mathrm{i}$ is the rms uncertainty in
$y_\mathrm{i} = \ln(S_\mathrm{i})$.  
Because  $d \ln (S_\mathrm{i}) =
d (S_\mathrm{i}) / S_\mathrm{i}$, the uncertainty $\sigma_\mathrm{i} =
\sigma(S_\mathrm{i}) / S_\mathrm{i} = 1 / \mathrm{SNR}_\mathrm{i}$
 is the channel noise divided by the source flux density,
or the reciprocal of the channel SNR.  Thus $\sigma_\mathrm{i}$ is dimensionless
and is
not  the same as  the usual channel noise in $\mu \mathrm{Jy~beam}^{-1}$.  
The spectral channel weights $W_\mathrm{i} =
1 / \sigma_\mathrm{i}^2$ that optimize the least-squares fit 
depend on the individual source
spectrum in addition to instrumental parameters---the least-squares
fit automatically 
determines the optimum channel weights for each
source spectral index. 

The coefficients $a$ and $b$ of the least-squares fit are
\citep{bev69}
\begin{equation}\label{eqn:generala}
a =  \frac{1}{\Delta} \Biggl( 
\sum \frac {x_\mathrm{i}^2} {\sigma_\mathrm{i}^2}
\sum \frac {y_\mathrm{i}} {\sigma_\mathrm{i}^2} -
\sum \frac {x_\mathrm{i}} {\sigma_\mathrm{i}^2} 
\sum \frac {x_\mathrm{i} y_\mathrm{i}} {\sigma_\mathrm{i}^2}
\Biggr) 
\end{equation}
and
\begin{equation}\label{eqn:generalb}
b =  \frac {1} {\Delta} \Biggl( 
\sum \frac {1} {\sigma_\mathrm{i}^2}
\sum \frac {x_\mathrm{i} y_\mathrm{i}} {\sigma_\mathrm{i}^2} -
\sum \frac {x_\mathrm{i}} {\sigma_\mathrm{i}^2}
\sum \frac {y_\mathrm{i}} {\sigma_\mathrm{i}^2} 
\Biggr) \rlap{~,} 
\end{equation}
where
\begin{equation}\label{eqn:generalDelta}
\Delta = 
\sum \frac {1} {\sigma_\mathrm{i}^2}
\sum \frac {x_\mathrm{i}^2} {\sigma_\mathrm{i}^2} -
\Biggl( \sum \frac {x_\mathrm{i}} { \sigma_\mathrm{i}^2} \Biggr)^2
\end{equation}
and all summation indices go from $i = 1$ through $i = N$.
The variances of the fitted $a$ and $b$ values are
\begin{equation}\label{eqn:generalsigmaa}
\sigma_a^2 =  \frac {1} {\Delta} 
\sum \frac {x_\mathrm{i}^2} {\sigma_\mathrm{i}^2} 
\end{equation}
and 
\begin{equation}\label{eqn:generalsigmab}
\sigma_b^2 =  \frac {1} {\Delta}
\sum \frac {1} {\sigma_\mathrm{i}^2}~.
\end{equation}

It is advantageous to use the weighted harmonic mean
frequency $\bar{\nu}_\mathrm{h}$ defined by
\begin{equation}
\ln (\bar{\nu}_\mathrm{h}) = 
\sum W_\mathrm{i} \ln (\nu_\mathrm{i}) \bigg/ \sum W_\mathrm{i}
\end{equation}
as the fiducial frequency $\nu_0$
because
\begin{equation}
\sum W_\mathrm{i} \ln (\bar{\nu}_\mathrm{h}) =
\sum W_\mathrm{i} \ln (\nu_\mathrm{i})
\end{equation}
\begin{equation}
\sum W_\mathrm{i} [\ln (\nu_\mathrm{i}) - \ln (\bar{\nu}_\mathrm{h})]
= \sum W_\mathrm{i} \ln (\nu_\mathrm{i} / \bar{\nu}_\mathrm{h}) = 
\sum W_\mathrm{i} x_\mathrm{i}
= 0~,
\end{equation}
leads to
\begin{equation}\label{eqn:harmonicmean}
\sum \frac {x_\mathrm{i}} {\sigma_\mathrm{i}^2} = 0~.
\end{equation}
Equation~\ref{eqn:harmonicmean} simplifies
Equations~\ref{eqn:generala}, \ref{eqn:generalb}, and
\ref{eqn:generalDelta} to
\begin{equation}\label{eqn:a}
a =  \frac{1}{\Delta} \Biggl( 
\sum \frac {x_\mathrm{i}^2} {\sigma_\mathrm{i}^2}
\sum \frac {y_\mathrm{i}} {\sigma_\mathrm{i}^2} 
\Biggr) 
\end{equation}
and
\begin{equation}\label{eqn:b}
b =  \frac {1} {\Delta} \Biggl( 
\sum \frac {1} {\sigma_\mathrm{i}^2}
\sum \frac {x_\mathrm{i} y_\mathrm{i}} {\sigma_\mathrm{i}^2}
\Biggr) \rlap{~,} 
\end{equation}
where
\begin{equation}\label{eqn:Delta}
\Delta = 
\sum \frac {1} {\sigma_\mathrm{i}^2}
\sum \frac {x_\mathrm{i}^2} {\sigma_\mathrm{i}^2}~.
\end{equation}
Equations~\ref{eqn:generalsigmaa} 
and \ref{eqn:generalsigmab} 
simplify to
\begin{equation}\label{eqn:sigmaa}
\sigma_a^2 =  \biggl( \sum \frac {1} {\sigma_\mathrm{i}^2} \biggr)^{-1} 
\end{equation}
and 
\begin{equation}\label{eqn:sigmab}
\sigma_b^2 = \biggl( \sum \frac {x_\mathrm{i}^2} {\sigma_\mathrm{i}^2}
\biggr)^{-1} \rlap{~.}
\end{equation}

\begin{figure}
\includegraphics[angle=0, scale=0.8]{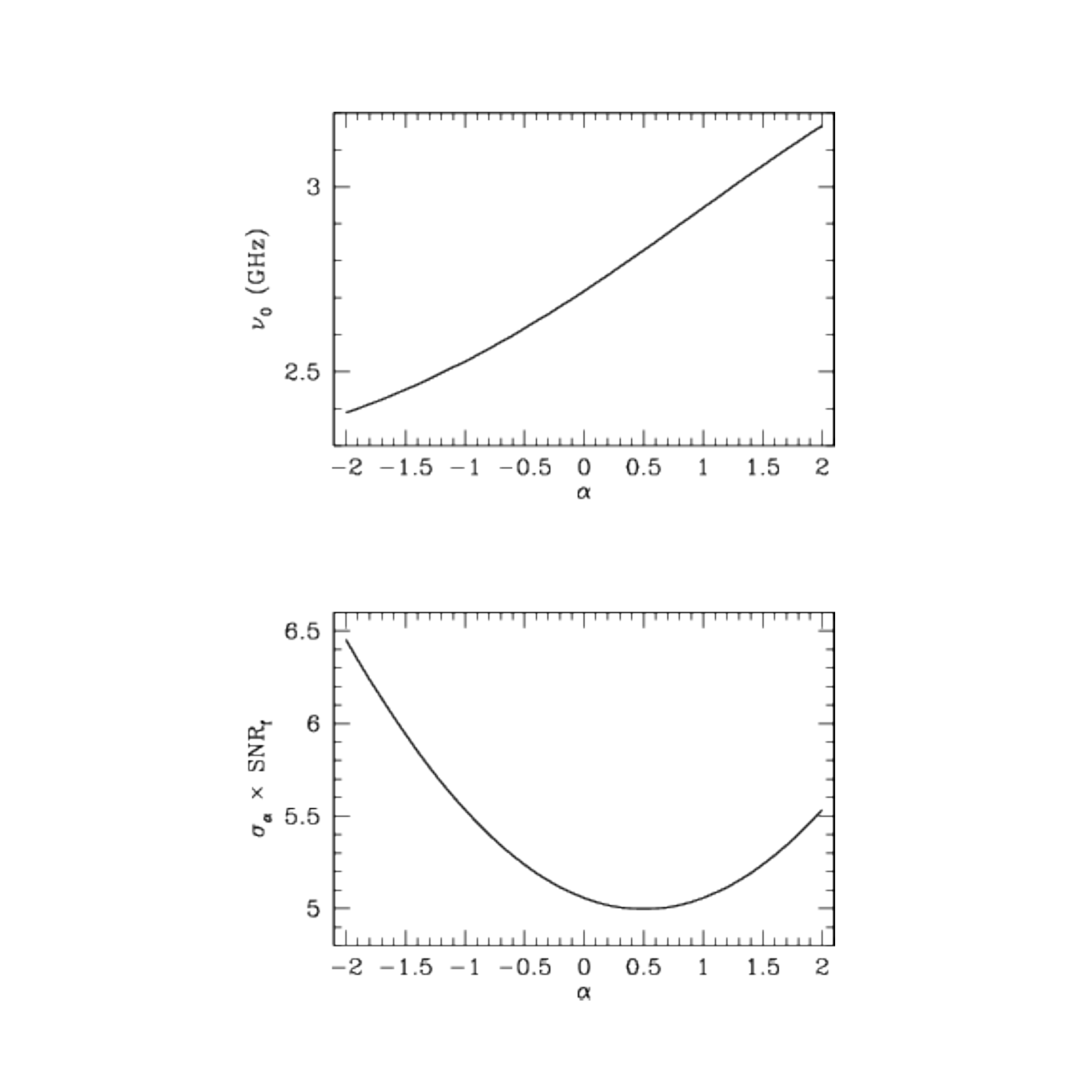}
\caption{The top panel shows the fitting ``pivot'' frequency
  $\nu_0 = \bar{\nu}_\mathrm{h}$
 (Equation~\ref{eqn:harmonicmean}) as a function of source
  spectral index $\alpha$ for the VLASS survey in which the
  integration time $\tau \propto \nu^{-2}$.  The bottom panel shows
  the product of the rms uncertainty $\sigma_\alpha$
in the spectral index $\alpha$
  and the signal-to-noise ratio SNR$_\mathrm{f}$ of the fitted flux density
  at frequency $\nu_0 = \bar{\nu}_\mathrm{h}$ as a function of
  source spectral index $\alpha$ for the VLASS.  The same curves apply
  to S-band  pointed observations of sources much smaller than the primary
  beam if the values of $\alpha$ on the abscissae are reduced by
  exactly 1.0; e.g., a survey $\alpha = 0$ corresponds to a pointed
  $\alpha = -1.0$.  }
\label{alpharmsfig}
\end{figure}

Even more importantly, Equation~\ref{eqn:harmonicmean} makes $a$
proportional to the weighted $y_\mathrm{i}$ terms and eliminates the
weighted cross-correlation terms $x_\mathrm{i} y_\mathrm{i}$ (compare
Equation~\ref{eqn:generala} with \ref{eqn:a}).  Similarly, comparing
Equation~\ref{eqn:generalb} with \ref{eqn:b} shows that $b$ is
proportional to the weighted $x_\mathrm{i} y_\mathrm{i}$ terms and free
of the the weighted $y_\mathrm{i}$ terms.  That is, the weighted
harmonic mean frequency $\bar{\nu}_\mathrm{w}$ of the data is the
unique ``pivot'' frequency at which the intercept (the fitted
$\ln[S(\nu_0)]$) and slope (the fitted $\alpha$) are
uncorrelated.  It is the frequency at which the fitted source flux density
has the highest SNR. The pivot frequency $\nu_0 = 
\bar{\nu}_\mathrm{h}$ is plotted
as a function of $\alpha$ for the VLASS ($2 \mathrm{~GHz} < \nu < 4
\mathrm{~GHz}$) in the top panel of Figure~\ref{alpharmsfig}.

For convenience, and with no loss of generality, assume that the
spectral channels are uniformly spaced in $x_\mathrm{i} =
\ln(\nu_\mathrm{i} / \nu_0)$; that is, channel
bandwidth $B_\mathrm{i} \propto \nu_\mathrm{i}$.  For a pointed
observation, the integration time $\tau$ is independent of frequency,
so the channel rms noise is proportional to $B_\mathrm{i}^{-1/2}
\propto \nu_\mathrm{i}^{-1/2}$.  For sources with flux densities
proportional to $\nu^{-1/2}$ ($\alpha = -0.5$), $\sigma_\mathrm{i}
\propto \mathrm{SNR}_\mathrm{i}^{-1}$
will be the same for all channels.  For survey observations with
$\tau_\mathrm{i} \propto \nu_\mathrm{i}^{-2}$, the channel rms noise
is proportional to $(B_\mathrm{i} \tau_\mathrm{i})^{-1/2} \propto
\nu_\mathrm{i}^{-1/2} \cdot \nu_\mathrm{i}^{+1} \propto
\nu_\mathrm{i}^{+1/2}$, and sources with $\alpha = +0.5$ will have
$\sigma_\mathrm{i}$ the same for all channels.  For these special
cases, the ratio of Equations~\ref{eqn:sigmaa} and \ref{eqn:sigmab}
is
\begin{equation}\label{eqn:sumeq}
\biggl( \frac {\sigma_\mathrm{a}} {\sigma_\mathrm{b}} \biggr)^2
=  \sum x_\mathrm{i}^2 \biggl/ \, \sum 1^2~.
\end{equation}
In the limit of large $N$, sums can be replaced by integrals over
$x$ and Equation~\ref{eqn:sumeq} becomes
\begin{equation}
\biggl( \frac {\sigma_\mathrm{a}} {\sigma_\mathrm{b}} \biggr)^2
\rightarrow \frac {\int x^2 dx} {\int dx} = 
\frac {x^3}{3}  \Bigg|_{-l/2}^{+l/2} \,
\Biggl/ \,x\, \Bigg|_{-l/2}^{+l/2} = \frac {l^2}{12}~,
\end{equation}
where $l = \ln(\nu_\mathrm{max} / \nu_\mathrm{min})$ is the
logarithmic width of the full frequency band (Equation~\ref{eqn:armlength})
and $1 / 12$ is the variance of a unit rectangle.  Let
$S_\mathrm{f} = S(\bar{\nu}_\mathrm{h})$ be the fitted flux density and
$\mathrm{SNR}_\mathrm{f}$ be the signal-to-noise ratio of the 
linear fit.  Then
$\sigma_\mathrm{a} = \sigma(\bar{S}_\mathrm{w}) /\bar{S}_\mathrm{w} =
 1 / \mathrm{SNR}_\mathrm{f}$, $\sigma_\mathrm{b} =
\sigma_\alpha$ is the rms error in the fitted spectral index, and
\begin{equation}
\sigma_\alpha \times \mathrm{SNR}_\mathrm{f} 
= \frac {\sqrt{12}}{\ln(\nu_\mathrm{max}/\nu_\mathrm{min})}
\end{equation}
In the case of the VLASS with $\nu_\mathrm{max} / \nu_\mathrm{min} = 2$,
\begin{equation}\label{eqn:alphasnr}
\sigma_\alpha \times \mathrm{SNR}_\mathrm{f} = 
\frac {\sqrt{12}} {\ln(2)} \approx 5.0
\end{equation}
for the special case of a source with $\alpha = +0.5$. Likewise for a
pointed S-band observation of a source with $\alpha = -0.5$.  For
other spectral indices, the channel weights are not equal and the
product $\sigma_\alpha \times \mathrm{SNR}_\mathrm{f}$ is slightly
larger, as shown in the bottom panel of Figure~\ref{alpharmsfig}.
Roughly speaking, Equation~\ref{eqn:alphasnr} indicates that a source fit
$\mathrm{SNR}_\mathrm{f} \approx 50$ is needed for the VLASS to measure
an in-band spectral index with rms uncertainty $\sigma_\alpha \approx
0.1$. For the EMU frequency limits $\nu_\mathrm{min} = 1.1 \mathrm{~GHz}$
and $\nu_\mathrm{max} = 1.4 \mathrm{~GHz}$, $\sigma_\alpha \times 
\mathrm{~SNR}_\mathrm{f} \approx 14.4$.  While EMU's required
$\mathrm{SNR}_\mathrm{f}$ is $2.9 \times$ higher than All-Sky's,
EMU is about $10 \times$ as sensitive as All-Sky, so EMU
should be  about $3.5 \times$ more sensitive than All-Sky for
measuring spectral indices of faint sources.

\begin{figure}
\includegraphics[angle=0, scale=0.8]{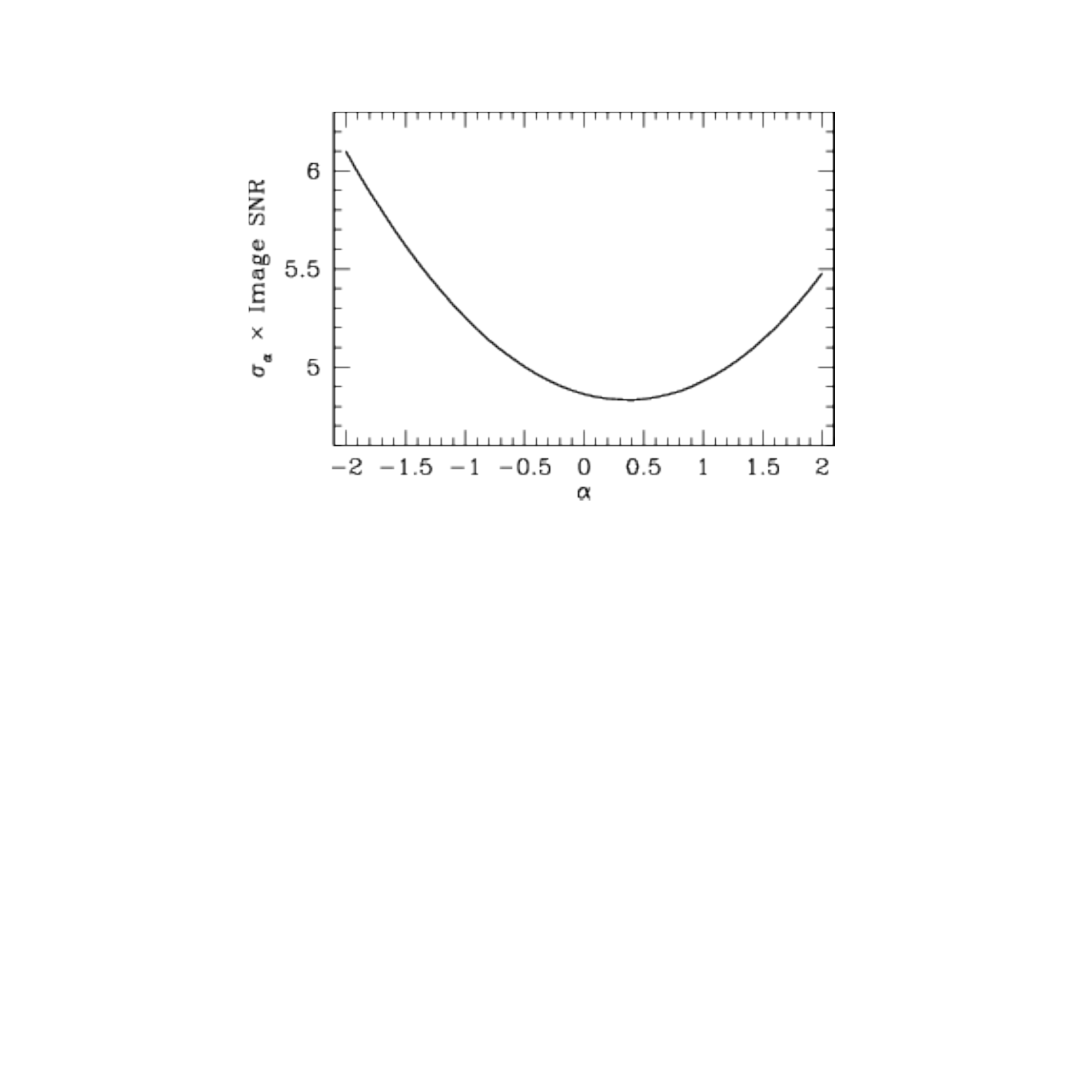}
\caption{The product of the rms uncertainty $\sigma_\alpha$
in the spectral index $\alpha$
  and the signal-to-noise ratio  on a VLASS image (Image SNR)
is plotted  as a function of
  source spectral index $\alpha$.}
\label{sigaalphaimagesnrfig}
\end{figure}

To predict the accuracy of \emph{all} in-band VLASS spectral indices, it is 
necessary to relate SNR$_\mathrm{f}$ to the SNR of the source on 
the VLASS image.  Again it is easier to work with logarithmic channel widths.
In the $i$th channel, 
\begin{equation}
(d S_\mathrm{i})^2 = \biggl( \frac{\nu_\mathrm{i}} {\bar{\nu}}
\biggr)^2 \biggl( \frac{\nu_\mathrm{i}} {\bar{\nu}}
\biggr)^{-1} = \biggl( \frac{\nu_\mathrm{i}} {\bar{\nu}}
\biggr)^1
\end{equation}
so
\begin{equation}
\Biggl( \frac {d S_\mathrm{i}} {S_\mathrm{i}} \Biggr)^2= 
\biggl( \frac {\nu_\mathrm{i}} {\bar{\nu}} \biggr) [S(\bar{\nu})]^{-2}
\biggl( \frac {\nu_\mathrm{i}} {\bar{\nu}} \biggr)^{-2 \alpha}
= \mathrm{SNR}_\mathrm{i}^{-2}
\end{equation} 
\begin{equation}
\mathrm{SNR}_\mathrm{i}^2 = [S(\bar{\nu})]^2 
\biggl( \frac {\nu_\mathrm{i}} {\bar{\nu}} \biggr)^{2 \alpha - 1}
\end{equation}
\begin{equation}
\mathrm{SNR}_\mathrm{f}^2 = \sum_\mathrm{i=1}^\mathrm{N} \mathrm{SNR}_\mathrm{i}^2
=  [S(\bar{\nu})]^2 
\sum_\mathrm{i=1}^\mathrm{N} 
\biggl( \frac {\nu_\mathrm{i}} {\bar{\nu}} \biggr)^{2 \alpha - 1}
\end{equation}
Replacing the sum over channels by an integral over $\ln(\nu)$ gives
\begin{equation}
\mathrm{SNR}_\mathrm{f}^2 =  [S(\bar{\nu})]^2 
\int  \biggl( \frac {\nu} {\bar{\nu}} \biggr)^{2 \alpha - 1}
d \ln \nu \Bigg/ \int d \ln \nu
\end{equation}
\begin{equation}
\mathrm{SNR}_\mathrm{f}^2 =  \frac {[S(\bar{\nu})]^2} {\bar{\nu}^{2 \alpha - 1}} 
\int \nu^{2 \alpha - 2} d \nu \bigg/ 
\ln(\nu_\mathrm{min}/\nu_\mathrm{max})
\end{equation}
\begin{equation}
\mathrm{SNR}_\mathrm{f} = [S(\bar{\nu})]^2\,
\Biggl( \frac {\nu_\mathrm{max}^{2 \alpha - 1} - \nu_\mathrm{min}^{2 \alpha - 1}}
{2 \alpha - 1} \Biggr)
\Biggl[ \frac {\bar{\nu}^{(1 - 2 \alpha)}} 
{\ln(\nu_\mathrm{min}/\nu_\mathrm{max})}
\Biggr]~, \quad (\alpha \neq +1/2)
\end{equation}
The source SNR on the weighted image multiplied by $\sigma_\alpha$ is
shown in Figure~\ref{sigaalphaimagesnrfig}.  To yield $\sigma_\alpha =
0.1$, a point source with $\alpha = -0.7$ must have an
$\mathrm{image~SNR} \approx 51$.  For All-Sky, with $\sigma_\mathrm{w}
\approx 65 \, \mu\mathrm{Jy~beam}^{-1}$, the minimum point-source flux
density is $S(2.682 \mathrm{~GHz}) \approx 3.3 \mathrm{~mJy}$.
\emph{Thus All-Sky alone is about twice as sensitive as the
  combination of WENSS and NVSS} [$S(2.682 \mathrm{~GHz}) \approx 6.0
  \mathrm{~mJy}$] \emph{for measuring spectral indices of point
  sources with $\alpha \sim -0.7$. EMU alone is about} $3.5 \times$
\emph{as sensitive as All-Sky.} These conclusions are actually true
for any value of the required accuracy so long as $\sigma_\alpha \ll
1$, not just for the example of $\sigma_\alpha = 0.1$.

\section{VLASS Performance and Science Applications}
\label{sciencesec}

The final VLASS proposal \citep{ssg15} lists six VLASS science themes
(Imaging Galaxies Through Time and Space, Radio Sources as
Cosmological Probes, Hidden Explosions, Faraday Tomography of the
Magnetic Sky, Peering Through Our Dusty Galaxy, and Missing Physics)
and their goals in Section 3 and ``additional science enabled by
VLASS'' in Appendix C. However, the proposal contains little
quantitative evidence that VLASS can deliver the proposed science or that
the VLASS is the best survey to do so.  This section addresses the
questions ``Can the VLASS do the proposed science?'' and ``Will it do
better than other surveys, EMU in particular?'' 

\subsection{Imaging Galaxies Through Time and Space}

The headline topic in this section is ``The star-formation history of
the Universe,'' but the surface-brightness sensitivity of All-Sky is
not good enough to detect radio emission from most star-forming
galaxies (see Equation~\ref{spiralsurfbrteq} and
Table~\ref{tab:VLASSspecs}).  All-Sky has a lower surface-brightness
sensitivity than FIRST.  If All-Sky could trace the evolution of star
formation at redshifts $z \leq 0.5$ as claimed, why hasn't FIRST already
done better?  Actually, even FIRST doesn't have the necessary
surface-brightness sensitivity, as demonstrated by the fact that the
low-redshift 2dFGRS/FIRST sample yields a local luminosity function
for star-forming galaxies that is a factor of ten too low
\citep{mag02}.  The low-resolution NVSS can detect nearby star-forming
galaxies \citep{con02}, and the far more sensitive EMU survey was designed to
extend those results to higher redshifts.

Section~3.1 of \citet{ssg15} says ``The optimal
combination of sensitivity and spatial resolution of VLASS allows the
study of the entire AGN population from classical radio-loud sources
down to the realm of radio-quiet AGNs ($P \sim 10^{22-23} \mathrm
{~W~Hz}^{-1}$) \dots from $z \sim 0 - 6$.'' However, a source
with $P = 10^{23} \mathrm {~W~Hz}^{-1}$ and $\alpha = -0.7$ at $z = 6$
has a flux density $S = 0.4 \,\mu \mathrm{Jy}$, well below the Deep
point-source detection limit $S_\mathrm{p} = 7.5
\,\mu\mathrm{Jy~beam}^{-1}$.  Deep could not detect such a point
source beyond $z \approx 1.6$, and the redshift limit
for an extended source is even lower.


To answer its question ``Why are some AGN strong radio emitters and
others not?'', \citet{ssg15} points out that ``better demographics \dots
are key.''
To highlight the impact the high-resolution All-Sky survey will have on quasar
science, Section 4.1.3 of \citet{ssg15} cites the \citet{hod11} survey
of Stripe 82 to suggest that All-Sky will yield far better
demographics than FIRST, noting ``The FIRST survey detects barely
10\% of SDSS quasars...'' while ``Hodge et al.~(2011) report on
A-array observations (i.e., with better resolution than FIRST) to $3
\times$ the depth of FIRST in SDSS Stripe-82 and found that 97\% of
known SDSS quasars are recovered in the higher resolution data.''

\citet{hod11} didn't actually
detect 97\% of the SDSS quasars in Stripe 82, which would have
been a huge jump over ``barely
10\%.'' The qualifier ``known SDSS quasars'' apparently
means ``only the subset of
 SDSS radio quasars previously detected by FIRST.''  Section 9
of \citet{hod11} reports ``\dots the FIRST catalog matches to 229
out of 3885 quasars from the SDSS DR7 Quasar Catalog (Schneider et
al. 2010) within a matching radius of $5''$.  Of those, the Stripe 82
catalog recovers 223.  In addition to those 223, the new catalog also
has radio sources matching 76 quasars not previously detected by
FIRST.  The total fraction of spectroscopic quasars that are radio
sources to the depth probed here is therefore 7.7\% ($\pm 0.4$\%).''
Thus  in Stripe 82 \citet{hod11} recovered $223/229 = 97$\% of the SDSS radio
quasars previously detected by FIRST, but the $3 \times$ higher
resolution and $3 \times$ higher
point-source sensitivity of the \citet{hod11} A-array
images increased the radio detection rate of all SDSS Strip 82 quasars from
5.9\% to  7.7\%.  This $30$\% increase is consistent with independent
observations \citep{con13} showing that optically selected quasars
have  extremely flat source counts.  All-Sky is less sensitive
than \citet{hod11}, so it probably won't improve the quasar
detection rate by even $30$\%.  

The ``Quasar Science'' section in \citet{ssg15} notes that surface
density of optically selected quasars is low and then claims that
``Efficient matching of radio sources to surveys at other wavelengths
(particularly in the optical) requires $\sim$ arcsec resolution.''
Not so.  Matching radio sources having low surface density (e.g.,
$\sim 290 \mathrm{~deg}^2$ for All-Sky) with quasars not having
significantly higher surface densities ($\sim 40 \mathrm{~deg}^2$ was
quoted) requires only that the the synthesized beam solid angle be
much smaller than $(290 \mathrm{~deg}^{-2})^{-1}$, or $\theta \ll
200''$ resolution.  Even the low-resolution NVSS ($\theta =45''$) has
efficiently matched radio sources to optically selected quasar samples
\citep{con13}.  EMU's ``poor'' resolution ($\theta = 10''$) is more
than adequate for efficient, reliable, and complete matching with any
foreseeable sample of optically selected quasars (see
Section~\ref{IDCRsec}), and EMU's $10 \times$ higher point-source
sensitivity than All-Sky's ensures that EMU will be the preferred
survey for detecting radio emission from more optically selected
quasars.

``To investigate quasars as a function of orientation, we need robust
spectral indices.'' \citep{ssg15}.  All-Sky ``internal'' S-band
spectral indices are robust only for fairly strong sources [$S(2.682
\mathrm{~GHz}) > 3.3 \mathrm{~mJy}$], many of which are so extended
that All-Sky will not be able to measure their flux densities accurately
\citep{con13}.  EMU will be able to measure spectral indices of quasars
about a factor of 3.5 fainter (Section~\ref{alphasec}). 

The key to statistical studies of galaxies through time and space is
indeed demographics.  Evolving populations of extragalactic radio
sources span a wide range of flux densities, angular sizes, surface
brightnesses, and redshifts.  Unbiased demographics requires
flux-limited samples not biased against sources with large angular
sizes or low surface brightnesses, a problem for the VLASS.  Radio
demographers have known for decades that any single survey is
dominated by sources near the survey sensitivity limit, so ``wedding
cake'' surveys with several layers of different sensitivity and sky
coverage are needed to obtain statistically useful samples spanning
the range of source flux densities.  The VLASS ``wedding cake'' shown
in Figure~\ref{cakefig} is clearly not optimized for studying galaxies
through time and space.  If that is a primary science goal of the VLASS,
the VLASS should be redesigned.

\subsection{Radio Sources as Cosmological Probes}\label{weaklensingsec}

The high angular resolutions of Deep ($\theta \approx 0\,\farcs9$ in
the two northern fields and $\approx 2\,\farcs2 \times 0\,\farcs9$ in
the southern ECDFS field) are comparable with the typical angular size
$\phi$ of faint ($\geq 10 \,\mu \mathrm{Jy}$) galaxies.  The
angular-size distribution of such faint galaxies is not well known,
and the median angular size of $\mu \mathrm{Jy}$ radio sources may
range from $\langle \phi \rangle \approx 0\,\farcs5$ to $1\,\farcs2$.
Detecting most faint galaxies to measure their two-point correlation
function or luminosity function requires that the beam be larger than
the galaxies, while resolving most galaxies to probe dark energy by
measuring weak gravitational lensing shear requires that the beam be
smaller than the galaxies.

The resolution, point-source sensitivity,
and sky coverage of Deep are only marginally sufficient to exploit the
``unique and powerful added value [offered by the radio band] to the
field of weak lensing.''  Section 3.2 of \citet{ssg15} forthrightly
 portrays Deep
as a pilot survey ``\dots aimed at delivering a $5\sigma$ detection of
cosmic shear in the radio.'' \citet{cha04} made a $3.6\sigma$ detection
using FIRST, while ``Current optical weak lensing
surveys \dots are expected to deliver detections with a significance
of 15, 30, and 43\,$\sigma$, respectively.''  The main value of
Deep will be to study radio systematics in preparation for
possible SKA surveys with much higher sensitivity and wide sky
coverage.

Figure~2 of \citet{ssg15} shows the accuracy with which luminosity
functions of star-forming galaxies near $z = 1$ and $z = 3$ can be
determined by 3~GHz continuum surveys covering 2, 4, and 10~deg$^2$
with a common detection limit $S = 7.5\,\mu\mathrm{Jy}$.  The
detection limit of Deep is $S_\mathrm{p} =
7.5\,\mu\mathrm{Jy~beam}^{-1}$, not $S = 7.5\,\mu\mathrm{Jy}$.  The
3~GHz resolution of Deep is about $0\,\farcs8$, so Deep will miss most
star-forming galaxies with angular diameters $\phi > 0\,\farcs8$, as
shown in Section~\ref{extsourcesec}.  The ability of Deep to match the
luminosity functions shown Figure~2 of \citet{ssg15} and the effects
of resolution bias is unknowable until the angular-size
distribution of $\mu\mathrm{Jy}$ sources is specified and taken into
account.

\subsection{Hidden Explosions}

Radio transients often signal explosive events, and Section 3.3 of
\citep{ssg15} points out that ``the most numerous radio transients
with the greatest potential impact are actually those populations that
have been hidden from view at these other wavebands, being largely
detectable only at radio wavelengths.''  Multi-epoch blind radio surveys
are needed to find them. All-Sky will cover $\Omega
\approx 3.4 \times 10^4 \mathrm{~deg}^2$ of the sky $N = 3$ times over
a span of seven years (once every 32+ months), and each epoch will
have an rms noise $\sigma_\mathrm{e} \approx \sigma_\mathrm{n} N^{1/2}
\approx 69 \, \mu \mathrm{Jy~beam}^{-1} \times 3^{1/2} \approx 120
\,\mu \mathrm{Jy~beam}^{-1}$, which is $ \approx 120 \,\mu \mathrm{Jy}$ for
unresolved transient sources.  The transient detection limit must be a
fairly high $S \approx 10 \sigma_\mathrm{e} \approx 1.2 \mathrm{~mJy}$
because there are $\approx 10^{11}$ synthesized beams in $N \Omega \approx 10^5
\mathrm{~deg}^2$.  Figure 5 of \citep{ssg15} shows the extragalactic
transient ``phase space'' (estimated instantaneous areal density
$\rho$ of transients stronger than flux density $S$ as a function of
$S$).  On it, a survey covering solid angle $\Omega$ with detection
limit $S$ in each of $N$ epochs is characterized by the point at $S =
10 \sigma_\mathrm{e} = 10 \sigma_\mathrm{n} N^{1/2}$, $\rho = (N \Omega)^{-1}$.

The product $S^2 \rho = 100 \, \sigma_\mathrm{n}^2 / \Omega$ of any given
survey (fixed telescope, frequency, and total observing time) is
independent of the number of epochs 
$N$, so changing $N$ moves the survey parallel to the
diagonal going from the upper left to lower right on Figure~5 of
\citet{ssg15}.  A simple figure-of-merit for comparing transient
surveys is $M \equiv (\Omega/ \sigma_\mathrm{n}^2)$.  By this criterion,
All-Sky is three or four orders-of-magnitude better than the other
transient surveys plotted on Figure 5.  
EMU covers $\Omega \approx 3.1 \times 10^4
\mathrm{~deg}^2$
with rms noise $\sigma_\mathrm{n} \approx 10 \,\mu\mathrm
{Jy~beam}^{-1}$,
so its transient
figure-of-merit is about $40 \times$ higher than All-Sky's.

The dashed line on \citep{ssg15} Figure~5 shows the areal density of
the usual variable but persistent extragalactic radio sources that
must be distinguished ``genuine'' transients.  The areal density of
variable sources is $\sim 10^5$ above the All-Sky point and even
farther above the EMU point.  To exploit the full potential of these
surveys, it will be necessary to find the one transient needle in a
haystack of $\sim 10^5$ variable straws plus the large expected number
of transient ``sources'' stronger than $10\sigma_\mathrm{e}$ that are
actually sidelobes in the All-Sky snapshot images.  This will be
difficult, especially for surveys with low $N$, so I suspect that
neither survey will perform as well as indicated by Figure~5, and that
the apparent advantage of EMU over All-Sky will be much less than $40
\times$ unless EMU, with its much larger field-of-view, can be
scheduled to yield $N \gg 3$ epochs.

\subsection{Faraday Tomography of The Magnetic Sky}

The drivers of All-Sky polarization science are ``to characterize
properties of the magneto-ionic medium in AGNs and in galaxies across
a wide range of redshifts'' and ``the use of the VLASS [All-Sky] for
studies of Faraday foregrounds'' with ``a 6-fold increase in the
background polarized source densities'' available today from the NVSS.
The proposal doesn't compare this All-Sky prediction
with the more sensitive EMU.
These two surveys are complementary in the sense that the 3~GHz
All-Sky can penetrate much higher Faraday depths than the 1.4~GHz EMU,
which has higher Faraday resolution.

The polarization science section of \citet{ssg15}
conflates polarized flux density and
polarized brightness.  The statements ``At the angular resolution of
the VLASS [All-Sky], most objects in the mJy regime will be resolved
allowing tomographic exploration of the structure of AGN and radio
galaxies, \dots'' and ``Based on the NVSS we expect over $10^5$ sources
with polarized fluxes $> 0.75 \mathrm{~mJy}$ \dots'' are actually in
conflict. Detection rates are limited by image brightness in
$\mathrm{mJy~beam}^{-1}$, not by flux density in mJy.  If most objects
are resolved by All-Sky, the All-Sky detection rate will be much lower
than calculated from polarized flux densities measured with the
low-resolution NVSS.  The statement ``Average fractional polarization of
unresolved Milky-Way type galaxies is a factor of 3--4 higher at 2 GHz
than at 1.4 GHz (Stil et al., 2009; Braun et al., 2010; Sun \& Reich,
2012), a 2--4 GHz survey with sufficient sensitivity thus opens
enormous potential for characterizing the development of galactic
magnetic fields'' suggests that the VLASS can detect polarization from
Milky-Way type galaxies despite the fact that it doesn't even have
the surface-brightness sensitivity to detect Milky-Way type galaxies
in total intensity.  Figure~8 of \citet{ssg15} shows the redshift and
luminosity distributions of galaxies ``expected to be detected in one
square degree with polarized flux density greater than $10
\,\mu\mathrm{Jy}$.''  The detection limit of Deep is not $<
10\,\mu\mathrm{Jy}$, it is $< 10\,\mu\mathrm{Jy~beam}^{-1}$.  Deep
will heavily resolve and hence not detect polarization in most of the
low-redshift or low-luminosity galaxies plotted.

\subsection{Peering Through Our Dusty Galaxy}
\label{sec:Galaxy}

The headline Galactic science in \citet{ssg15} Section 3.5 is discovering
exotic radio pulsars, where ``The value of the VLASS is that it will
serve as a finding survey in which \emph{candidate radio pulsars} can
be used to winnow the large number of radio sources detected in the
VLASS to a feasible number on which to conduct a periodicity search.''
Of the $\rho \sim 380 \mathrm{~sources~deg}^{-2}$ with flux densities
above the All-Sky \emph{peak} flux-density limit, between 25\% and
50\% will be resolved by All-Sky, leaving $\rho \sim 240 \pm 50
\mathrm{~deg}^{-2}$ point sources as pulsar candidates.

Winnowing is efficient to the extent that it can easily reduce the sky
density of candidates well below one per primary beam solid angle of
the telescope used for the periodicity search.  In the case of the
GBT, $\Omega_\mathrm{pb} \approx [20 / \nu\mathrm{(GHz)}]^{2}$, or $\rho \ll 13
\mathrm{~deg}^{-2}$ ($\rho \ll 2.5 \mathrm{~deg}^{2}$) at 820~MHz
(350~MHz), the pulsar search frequencies most often used at the GBT.
However, \citet{ssg15} claims only ``With a combination of
multi-wavelength counterpart comparisons, polarization, future high
angular resolution observations, and other criteria, we expect it to
be feasible to reduce the source density to of order $30
\mathrm{~deg}^{−2}$.''

Even efficient winnowing doesn't help if the radio
candidate catalog isn't deep enough to match the periodicity search
sensitivity.  In the \citet{ssg15} example of
identifying new $\gamma$-ray pulsars in the FERMI catalog
\citep{ray12}, it turns out that none of the new pulsars would have
been seen by All-Sky because all are a factor of 5 or more below the
All-Sky detection limit at 3~GHz.  Part of the problem for the 3~GHz
All-Sky is that pulsars have much steeper radio spectra ($\alpha \sim
-1.6$) than most radio sources ($\alpha \sim -0.7$), so high-frequency
surveys have low needle/haystack ratios.

Section 3.5 of \citet{ssg15} covers All-Sky detections of radio stars.
\emph{Detecting} radio stars in our Galaxy is fairly easy, but reliably
\emph{identifying} them (Section~\ref{IDCRsec}) requires extremely accurate
positions because the fraction $f$ of all radio sources that are powered
by stars is
very small ($f < 10^{-4}$).  All-Sky really should greatly improve on
existing flux-limit samples of radio stars, even though it is not much
more sensitive than FIRST or NVSS, because its high angular resolution
allows it measure accurate ($\sigma \sim 0\,\farcs3$ in each
coordinate) positions of even the faintest detectable point sources.
It has little competition from FIRST, which does not cover the
Galactic Plane, from the NVSS, whose position errors are too large, or
possibly even from EMU, which may be confusion limited near the
Galactic Center.  More weight should have been given to this part
of the All-Sky proposal.

Section 3.5 notes that ``\dots the total number of known [planetary]
nebulae is far lower than even the most conservative
expectations. Consequently, large samples of these objects are
required to trace evolutionary sequences.''  Although the proposal
explicitly recognizes the need for adequate brightness temperature
sensitivity, it uses the \citet{aaq90} $\lambda = 6 \mathrm{~cm}$ VLA
survey of PNe to estimate that PNe tend to be a few to several
arcseconds in size.  The \citet{aaq90} survey is strongly biased
against detecting extended, low-brightness PNe because it targeted an
optical sample of PNe smaller than $\phi \sim 4''$ and was made with
$\theta = 0\,\farcs4$ resolution.  By comparing the IRAS colors of
their radio-detected PNe with those of evolved PNe, \citet{aaq90}
concluded that they had ``\dots identified a group of nebulae with
high radio surface brightness temperatures which are excellent
candidates for young planetary nebulae.''  The typical PN is about
0.3~pc in diameter, or about $\phi \approx 8''$ at the distance of the
Galactic center.  To trace evolutionary sequences of PNe, a radio
survey should have resolution $\theta > \phi$.  With $\theta \approx
2\,\farcs8$, All-Sky will detect a $\phi \approx 8''$ PN only if it is
stronger than $S \approx 3 \mathrm{~mJy}$, which makes it no more
sensitive than the NVSS survey of PNe
\citep{con98pne} and far less sensitive than EMU.

Figure~9 in \citet{ssg15} is a logarithmic plot showing relative
sensitivities in the L, S, C, and Ku bands to sources with different
spectral indices, \emph{normalized to 2.8 GHz}, the S-band logarithmic
mean frequency.  In general, nonthermal sources are stronger at lower
frequencies and thermal sources are stronger at higher frequencies.
The caption concludes ``The Galactic radio source population contains
both thermal and non-thermal emitters, and the 2–-4~GHz observing
frequency range for the VLASS provides a balance in the sensitivity
for these two classes of sources.''  Not so.  The ``balance''
apparently provided by observing in the 2--4~GHz frequency range is an
artifact generated the choice of 2.8~GHz as the normalizing frequency,
and has nothing to do with actual populations of Galactic sources.
Likewise, the claim in Section~3.5 ``Thus, observation frequency of
2--4 GHz, as planned for the VLASS, balances sensitivity to thermal
and non-thermal sources, both of which are found in Galactic radio
source populations.'' is without foundation.  Both Figure~9 and the
claim in Section~3.5 should be deleted from the proposal.

\subsection{Angular Resolution and Optical Identifications}

The main drawback of All-Sky is its poor brightness sensitivity, which
is the inevitable consequence of its unusually high angular resolution, $\theta
\approx 2\,\farcs 8$.  The point-source detection limit of All-Sky is
$S \approx 350 \,\mu \mathrm{Jy}$, but so many sources have angular
diameters $\phi > \theta \approx 2\,\farcs8$ that All-Sky will
generate incomplete and biased samples of most source populations.
All-Sky will detect nuclear starburst galaxies like Arp 220 but
completely miss normal star-forming galaxies.  All-Sky will detect
young PNe but miss most older PNe.  All-Sky will detect compact sources
in radio galaxies and quasars, but it will miss most of the flux from jets and
lobes larger than their host galaxies because there is \emph{no}
redshift at which sources larger than 24~kpc have angular diameters
smaller than $2\,\farcs8$.

All-Sky was forced to such high angular resolution by the claim in
Section~4.1.1 and Appendix D of \citet{ssg15} that $>95$\% reliable
position-coincidence identifications with optically faint galaxies can
be made if and only if $\theta$ is very small, regardless of source
signal-to-noise ratio S/N.  That claim is based on a bad
identification method that tries to match the individual components of
a resolved double source, instead of the source centroid, with its
host optical galaxy.  For example, the radio source Cyg A has two
strong lobes, each about 1 arcmin from the host optical galaxy.
Matching each lobe to a deep optical catalog will yield two 
bad ``matches'' with
radio/optical separations $\sim 1$~arcmin.  The problem of
mis-identifying individual lobes of double sources has been addressed
by ``collapsing'' close pairs of radio components into a single
component whose centroid position is much closer to the optical galaxy.
\citet{lin14} tried this and found ``when matching collapsed FIRST
sources or NVSS sources directly to the optical catalogues, we find no
evidence of a significant difference between the results.''  That is,
the high resolution ($\theta = 5\,\farcs4$) of FIRST and the low
resolution ($\theta = 45''$) of NVSS give identification results 
on extended sources that are not
significantly different.

Only in the rare cases that the high-resolution survey can detect and resolve
a weak radio core does a high-resolution survey do better.  Surveys
such as EMU with lower resolution but higher sensitivity will yield
identifications of most sources that are at least as reliable as
All-Sky's.  Appendix~\ref{sec:ids} below presents a detailed review of
what went wrong in \citet{ssg15} Appemdix D.

The ideal radio survey would yield flux-limited source samples with
optical identifications that are \emph{complete} as well as reliable.
All-Sky optical identifications are necessarily $>25$\% incomplete
because All-Sky radio catalogs are $>25$\% incomplete.  This problem
is not addressed by \citet{ssg15}.  Incompleteness is worse than 
unreliability because it cannot be corrected by follow-up observations.
I expect that EMU will yield more useful optical identifications
of flux-limited samples than All-Sky will.




\section{Justifying the VLASS}\label{sec:justification}

The VLASS must have a strong justification for its request for about
9,000 hours of valuable time on the new JVLA.  It needs to be justified as (1)
being technically better (e.g., more sensitive, higher position
accuracy, \dots) than other surveys and (2) having a
higher ``reach'' or impact than the regular proposals
it would displace on the VLA schedule. 

 Read the ``Report of the NRAO
Large Proposals Committee'' \citep{bri97} of the committee
created specifically to address the
competition between large surveys and smaller PI observations,
especially their Recommendation 7 ``The NRAO should not make
Announcements of Opportunity for the submission of large
proposals. Large proposals should be submitted at the normal proposal
deadlines, without special solicitation by the observatory.'' The reasoning
behind that recommendation \citep{bri97} is (emphases added):

``The committee considered whether the NRAO should explicitly solicit
proposals for large projects via Announcements of Opportunity,
targeted either to specific disciplines or to special deadlines (other
than those of the regular proposal process.)

``It was our unanimous opinion that this would be undesirable.  It
would separate ``opportunities'' for proposing large projects from the
regular proposal process, whereas we see merit in keeping the
processes for large and small proposals well-coupled. \emph{It is also hard
to see what benefit would come by encouraging the whole user community
to think about large proposals simultaneously.}

``The NRAO-operated telescopes are ground-based and flexible in their
capabilities, so operational and planning considerations differ
greatly from those needed to establish the scientific program of
space-borne instruments, for example. \emph{The AO approach would however
place some obligation on the NRAO to schedule some large projects
after a period in which it had encouraged the whole user community to
make proposals for them.}

``It is particularly undesirable to create an artificial imbalance
between the pressures for large and regular proposals when our
ultimate goal is to find an appropriate balance. \emph{We believe that
balance is more likely to be achieved through a proposal process that
is driven mainly by the scientific interests of individual
investigators, rather than through one driven by ad hoc deadlines}.''

\subsection{The Better Mousetrap}\label{sec:mousetrap}

FIRST and NVSS are prototypes for the 
All-Sky survey.  They were justified primarily on the basis of their
orders-of-magnitude technical superiority over previous large-scale
sky surveys, and not on any specific ``transformational science''
(the 20th century term for today's ``killer app'')  they
would accomplish.  Rather, the bulk of the science would be done
by community users who would come up with new ideas to exploit the 
greatly improved survey
data---build a better mousetrap, and the world will beat a path to
your door.  The original FIRST paper \citep{bec95} could legitimately
claim that ``FIRST represents a factor of $\sim 50$ improvement in
limiting sensitivity over the best available sky survey at any radio
wavelength'' and ``More importantly, however, FIRST also represents a
factor of 50 improvement in angular resolution and concomitant
positional accuracy.''  Of course, to be valuable to the general
scientific community, it also has to cover a large part of the sky;
FIRST would be ``in short, the radio equivalent of the Palomar Sky
Survey for 25\% of the celestial sphere.''

Similarly, the primary selling point for the EMU (Extragalactic Map of
the Universe) survey covering the sky south of $\delta = +30^\circ$
(75\% of the celestial sphere) with $\sigma \approx 10\,\mu{\rm
  Jy~beam}^{-1}$ rms noise and $\theta = 10''$ resolution was that
``EMU has the potential to have the enormous impact that the NVSS had
a decade ago, but at a factor of 40 better in sensitivity and 5 in
angular resolution.''

Unfortunately, All-Sky is only $\sim 1.5 \times$ as sensitive to point
sources as FIRST, and it is less sensitive to extended sources.
Neither All-Sky nor Deep have the surface-brightness sensitivity to
detect the extended radio emission from most spiral galaxies.  EMU
combines the best features of both FIRST (high point-source
sensitivity and the angular resolution to identify weak sources with
faint galaxies) and NVSS (high surface-brightness sensitivity and
nearly ``all sky'' coverage).  The instantaneous sky coverage of EMU
($30 {\rm ~deg}^2$) beats All-Sky by more than two
orders-of-magnitude, so EMU will be a superior finder of bright
transients.  Finally, EMU is likely to be completed at about the same
time scale as the VLASS (2023) because ASKAP will become a dedicated
survey telescope after the second-generation array feeds are
installed.  All of the second-generation feeds have now been funded
for construction during the next two years, and the EMU survey itself
only needs two years for completion, compared with seven years for the
VLASS.  It doesn't matter which survey wins the race to completion by
a year or two; the impacts of large surveys like FIRST and NVSS, VLASS
and EMU are felt on time scales of decades.

The Deep survey \emph{is} a better mousetrap for point sources,
with $\sigma \approx 1.5 \, \mu{\rm Jy~beam}^{-1}$ and $\theta \approx
0\,\farcs8$ resolution, specifications that EMU
and WODAN will never reach.  Deep covers only a tiny fraction of the
sky ($\Omega = 10{\rm ~deg}^2 \approx 
0.024$\% of the sky), so it is 
analogous to the various ``deep'' surveys by Hubble, Chandra, and
Spitzer.  If Deep is done at 3~GHz with the A configuration, it will partially
resolve most faint star-forming galaxies.  This is good for studying
the shear produced by weak gravitational lensing but bad for obtaining
the complete samples of low-brightness sources needed for studying the
evolution of star formation.  Equation~\ref{brtdeteq} indicates that
Deep could detect most star-forming galaxies if its beamwidth were at
least $\theta \approx 2\,\farcs8$, which is possible with the B
configuration at either 3 or 1.4~GHz.

The high scientific impact of the very sensitive multi-band optical/IR
Hubble Deep Fields comes from the fact that they really are ``deep''
in the sense of detecting and distingiushing galaxies at very high
redshifts.  In contrast, the median redshift of radio sources is
nearly independent of flux density, so more sensitive radio surveys
like Deep mainly detect $z \sim 1$ sources with lower radio luminosities
\citep{con89}.  Also, there is no radio spectral signature that
clearly distinguishes high-redshift sources.  The most distant quasars
may already be in the FIRST and NVSS catalogs, but we have to way to
distinguish them from the millions of ``ordinary'' radio
sources. High-redshift starburst galaxies stand out only at sub-mm
wavelengths, where they appear frequently as background sources in
ALMA images.  A Deep proposal for 3,000+ hours of JVLA time to
constrain the evolution of star formation out to $z \sim 1$ or 2 needs
strong and \emph{quantitative} scientific and technical
justifications.  In particular, the angular resolution needs to be
lowered and the ``Prussian hat'' VLASS wedding cake
(Figure~\ref{cakefig}) needs more layers to optimize coverage in that
part of the the redshift-luminosity plane containing the galaxies
responsible for the bulk of star formation.

\subsection{``Reach'' or Citation Impact}\label{sec:reach}

\begin{figure}
\vskip .0in
\includegraphics[angle=0, scale=0.43]{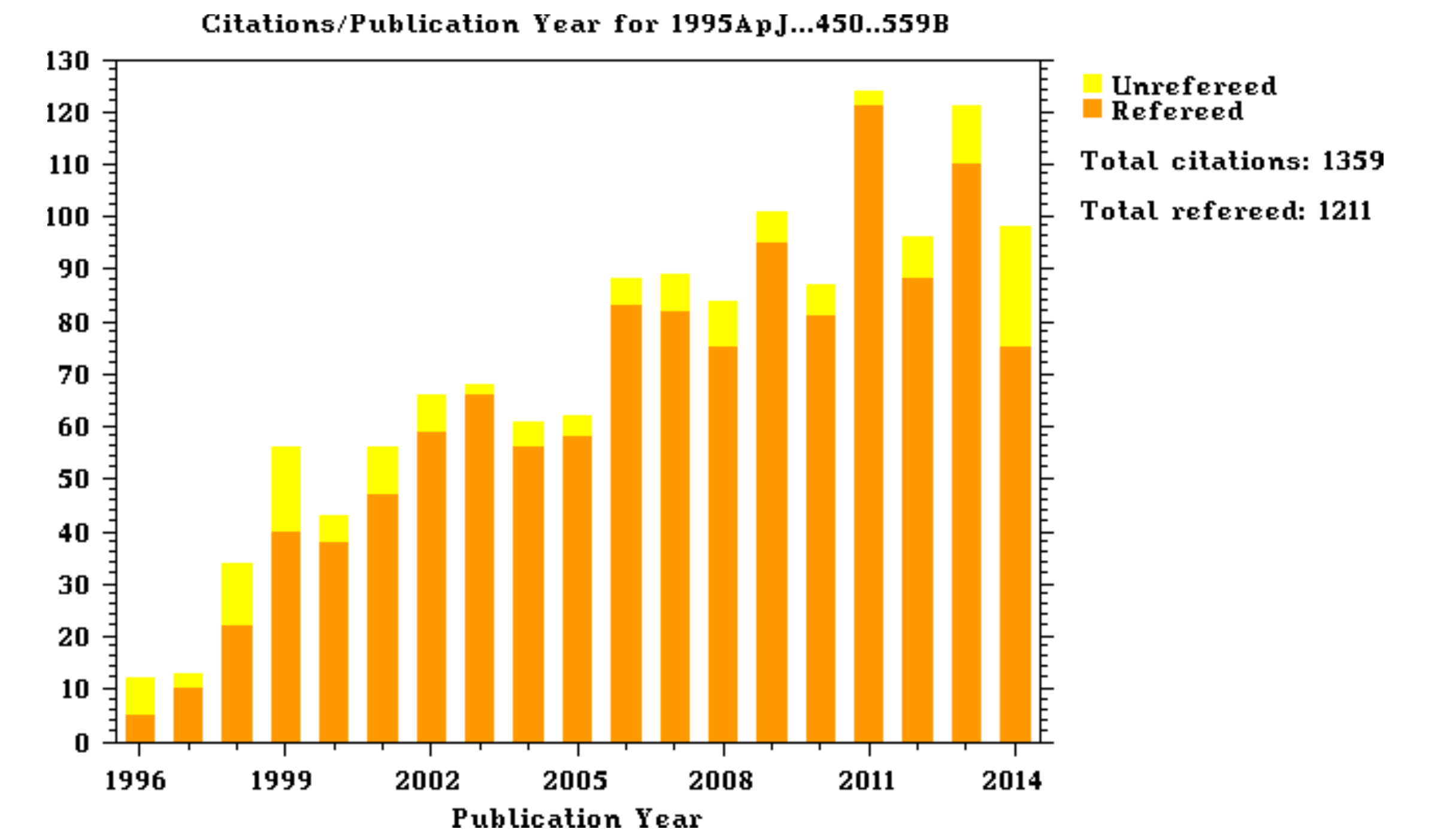}
\vskip .14in
\includegraphics[angle=0, scale=0.43]{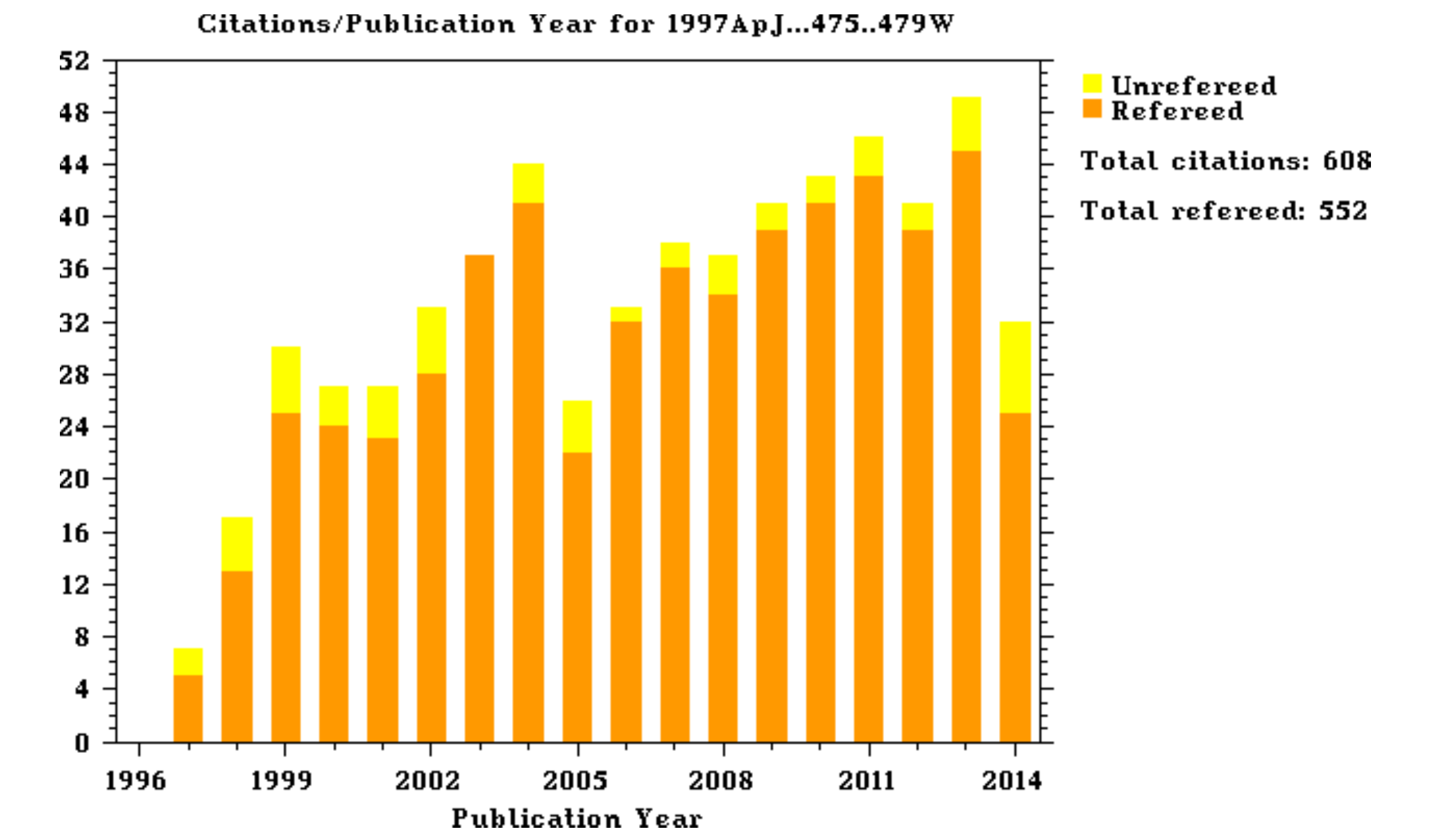}
\vskip .14in
\includegraphics[angle=0, scale=0.43]{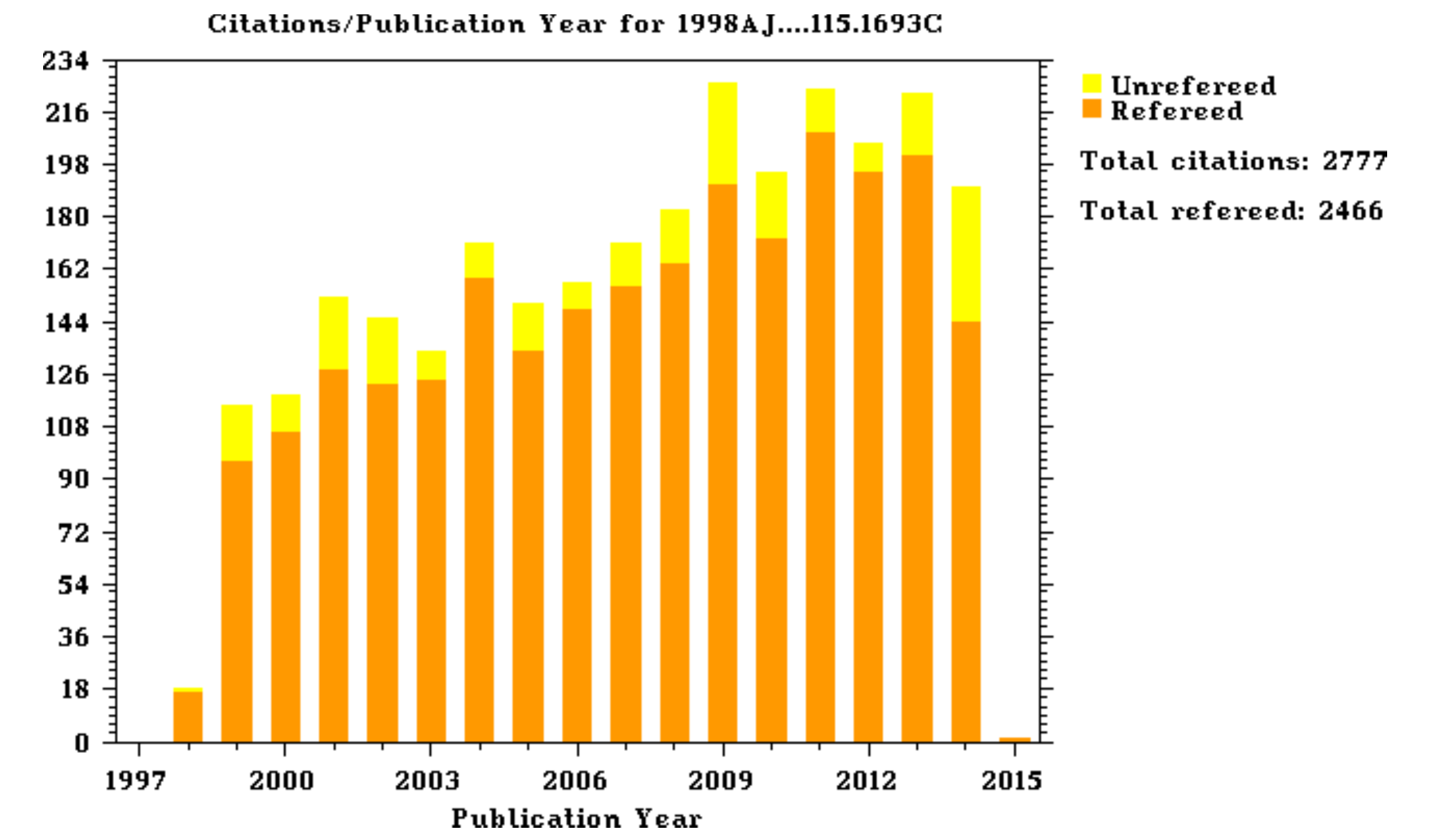}
\caption{Citation histories for the FIRST survey images \citep{bec95},
the FIRST survey catalog \citep{whi97}, and 
the NVSS \citep{con98}
as of 2014 September 23.}
\label{citesfig}
\end{figure}

Appendix A.2 of \citet{mur140808} and Appendix B.2 
of \citet{ssg15} use the citation rates of the FIRST
survey paper \citep{bec95}, the FIRST catalog paper \citep{whi97}, and
the NVSS paper \citep{con98} to estimate
the scientific impact of future large surveys like All-Sky.  The ADS citation
histories of these papers as of 2014 September 23 are shown in
Figure~\ref{citesfig}.  Although these publications are now 16 to
19 years old, their citation rates remain steady or slowly growing,
and their combined refereed citation rate is about $\sim 320$ per
year.  No competitive surveys have been published, so their lifetime
refereed citation numbers have grown to 1211, 552, and 2466,
respectively, for a total of 4229 when all three papers 
are added up.  

The take-home lesson from
Figure~\ref{citesfig} is that large surveys should be designed for the
long run, and surveys that are surpassed after only a few years will
have much less impact.  It makes no sense to rush the VLASS in order to
``scoop'' EMU, only to have it surpassed shortly after it is
completed.

How does the citation impact of NVSS+FIRST compare with the
regular VLA projects that these surveys displaced?  Together these
two surveys used up about 6,000 hours, or one year of VLA observing
time, so their citation numbers should be compared with the citation
numbers of papers produced by one year of ``regular'' VLA projects.

Appendix A.2 of \citet{mur140808}
compared the number of refereed VLA papers published per year (192)
with the number of refereed citations per year of FIRST and/or NVSS (262) and
concluded ``Thus, there is very strong evidence against the argument
that the science out of the VLA is negatively impacted when surveys
displace regular proposals.''  I think this is comparing apples
(number of refereed VLA papers \emph{published} per year) and oranges
(number of refereed papers \emph{citing} FIRST and/or NVSS data per
year).

The relevant comparison for the number of refereed papers citing FIRST
and/or NVSS data per year is the number of refereed papers citing the
192 regular VLA papers per year.  For example, if the average number
of citations per regular VLA paper per year is 4, the 192 regular VLA
papers from one year will result in 768 citations per year. Even more
relevant, and more favorable to FIRST and NVSS, is comparing the
lifetime number of citations attributable to FIRST/NVSS (4229) with
the lifetime number of citations attributable to the 192 regular VLA
papers resulting from one year of VLA observing time.  I checked with Marsha
Bishop, the NRAO librarian, and the lifetime number of citations per VLA
paper published in the years 2007-9 is 34, so the 192 regular VLA
papers should yield a lifetime number of citations about 6500, which
is still somewhat higher than the FIRST/NVSS 4229.  

I conclude that citation rates, either per year or integrated over
paper lifetime, don't clearly favor FIRST/NVSS over small VLA programs.  
The JVLA today is in much heavier demand than the VLA was 20
years ago, and the upgrade has helped targeted observations far more than
it has helped surveys (Section~\ref{surveymetricssec}), so the VLASS
will need a stronger justification than the citation counts presented
in Appendix A.2 of \citet{mur140808} or Appendix B of \citet{ssg15}.

\appendix


\section{Appendix A: Positional Accuracy and Angular Resolution}\label{sec:ids}

There is a still-unresolved debate within the SSG (or more specifically, between
the SSG and me)
 about the radio resolution needed to make
optical identifications of radio sources with faint galaxies and quasars.
See Appendix D of \citep{ssg15} for the latest version of the claim
that 95\% identification 
\emph{reliability} requires a large search radius
$\sim 0.3\,\theta$.  This claim is based on a straw-man
identification procedure that does not distinguish between radio
\emph{components} (peaks on a radio image) and radio \emph{sources}
(the totality of radio emission from a galaxy).  See \citet{con14}, 
Appendix D of \citet{ssg15}, and this Appendix
 for the current state of this debate.  
The identification question needs to be sorted out before the VLASS
design is frozen.  Also, the VLASS proposal still needs to address the
question of identification \emph{completeness}, which decreases at
high angular resolution, as the radio survey catalog completeness
decreases.

------------------------------------------------------------------------

Section 5.1.1 of \citet{mur140808} states that resolution but not
sensitivity 
is a key survey parameter: \newline\hphantom{000}
``While sky area is one of the two key parameters that define the
value of a wide-area radio survey, the other key parameter is
resolution, as this allows counterparts to be identified at other
wavelengths among the dense populations of faint galaxies.''

The reason seems to be the claim in \citet{mur140808} Appendix B that the proper
identification search-circle radius is
$\sim 0.4 \theta$, regardless of (S/N).
The more sensitive  EMU and WODAN surveys are dismissed as
not having sufficient resolution ($10''$ and $15''$, respectively) 
to make deep optical identifications:\newline\hphantom{000}
``For Pan-STARRS, a FWHM resolution better than $7''$ is required for
95\% reliability in radio-PS1 cross-matches, as demonstrated in Figure
14. That criterion is easily met by VLASS, but with resolutions of
$10''$ and $15''$ respectively (FWHM), both ASKAP-EMU and WODAN fall short
(Figure 14). Thus, despite their excellent flux sensitivity, the
forthcoming SKA-precursor surveys will not have adequate spatial
resolution for confident identifications of counterparts in
Pan-STARRS.\newline \hphantom{000}
``A recurring claim is that the excellent $\sim 10\,\mu$Jy rms flux
sensitivity planned for the SKA-precursor surveys (WODAN, ASKAP-EMU)
will lead to good positional accuracy for radio sources despite the
relatively low resolution of those surveys. The evidence, however,
suggests that this is incorrect --- the positions of radio sources
observed at low resolution do \emph{not} actually converge to the optical
counterpart position as (S/N) increases. The wrong
conclusion is reached due to simplistic assumptions about the
structure of radio sources. In Appendix B we provide a detailed
discussion and analysis (i.e., the ``S/N model of positional accuracy'')
demonstrating that half of the optical counterparts to SDSS depth will
be false matches using the matching radius that will be required for
WODAN (e.g., see Figure 14). The false counterparts will obviously be
an even problem for deeper optical surveys, such as the ongoing DES
and HSC surveys and eventually for LSST. Thus the SKA-precursor
surveys can not be substituted for the VLASS all-sky survey. This tier
will have a long, useful, and heavily used lifetime even into the era
of the SKA-precursor surveys.''  (Section 5.1.1)

Likewise for WIDE:
\newline
``At the limits of the DES, HSC, and LSST surveys there are 5, 8, and
14 galaxies in an ASKAP beam---never mind the number of stars. The
higher S/N observations of EMU/Wodan will not mitigate this source
density when it comes to associating optical and radio sources (see
the Appendix), resulting in a high probability of
misidentifications. B-array observations in the S-band provide
sufficient resolution to bring these numbers down to $\lesssim 1$
galaxy in the VLA beam without unnecessarily over-resolving extended
radio sources.''  (Section 5.2)

This Appendix B analysis of angular resolution and optical
identifications has serious problems.  It\newline (1) sets up a
straw-man ``S/N model'' that ignores calibration errors when
calculating position errors, \newline (2) knocks  down the ``S/N model'' by
introducing a faulty identification technique that seems to imply much
larger ``empirical'' position errors, \newline (3) incorrectly
ascribes these large errors to source asymmetry, \newline (4) claims
that the identification search circle radius must be $\sim 0.4
\theta$, regardless of S/N, \newline (5) dismisses more sensitive but
lower resolution surveys as being unable to identify even moderately
bright optical objects (e.g., $r = 22.1$ for EMU, $r = 20.7$ for
WODAN) with 95\% reliability,
\newline
(6) emphasizes identification reliability $R$ but omits the
equally important identification completeness $C$ that is degraded
by high angular resolution, and \newline
(7) ignores the powerful and widely used likelihood-ratio method
\citep{sut92} for improving on position-coincidence identifications.

These problems are addressed in the remainder of Section~\ref{sec:ids}.

\subsection{Positional Accuracy}

\subsubsection{The ``S/N Model'' of Positional Accuracy}

Appendix B in \citet{mur140801} 
presented the straw-man ``S/N model'' for positions
errors and defines it by the rule:
\newline \hphantom{000} ``\dots as the flux density increases, the
positional error will decrease as 1/(S/N), allowing the optical
counterpart to be matched. Specifically, the NVSS description (Condon
et al.~1998) gives this formula for the noise in RA or Dec for point
sources:
\begin{equation}\label{eqn:snmodelsigma1D}
\sigma_{\rm 1D} = {\theta \over (S/N)\sqrt{(2 \ln 2)}}~ .
\end{equation} 
Here $\theta$ is the resolution FWHM ($45''$ for NVSS) and
S/N is the signal-to-noise ratio. The median NVSS rms noise for these
matched sources is 0.47 mJy/beam. Note that this noise equation
already has been increased by an empirical factor of $\sqrt{2}$ compared with
the theoretical equation ``to adjust the errors into agreement with the
more accurate FIRST positions'' (Condon et al.~1998). This
predicts $\sigma_{\rm 1D} \sim 7\,\farcs6$
at the catalog detection limit (S/N = 5) and $\sigma_{\rm 1D} \sim
1''$ at a flux density of 18 mJy/beam.''


Equation (3) in \citet{mur140801} Appendix B
(Equation~\ref{eqn:snmodelsigma1D} here) differs from Equation~25 in
\citet{con98} because it confuses $S/N$ with $\rho$ in Equation~25.
The value of $\rho$ for point sources given in Equation~26 of
\citet{con98} is actually $\rho = \sqrt{2} (S/N)$, so Equation (3)
overestimates $\sigma_{\rm 1D}$ by a factor of $\sqrt{2}$.  Equation
(3) for the noise component of error should read
\begin{equation}\label{eqn:sigma1D}
\sigma_{\rm 1D} = \Biggl({1 \over \sqrt{2}}\Biggr)
 {\theta \over (S/N) \sqrt{(2 \ln 2)}} \approx {0.6\, \theta \over (S/N)} .
\end{equation}

Although presented as ``a long-standing notion'' needing to be
debunked, the straw man ``S/N model'' was first defined in Appendix B.
It is not a long-standing notion because it implies that the
\emph{noise component} of position error is the \emph{total} position
error used for identifying point sources.  In the long-standing
``two component''
calculation of position errors, the noise component of position error
does decrease as 1/(S/N) \citep{con97}, but it is always added
quadratically to the calibration component of error, which is
independent of (S/N). For example, the actual NVSS description
\citep{con98} is:\newline \hphantom{000} 
``The rms uncertainties ($\sigma_\alpha$,
$\sigma_\delta$) in the centroid coordinates ($\alpha$, $\delta$) of
any source with uncorrected peak amplitude $A_P$ can be approximated
by quadratic sums of intensity-independent calibration uncertainties
($\epsilon_\alpha$, $\epsilon_\delta$) and noiselike uncertainties
that are inversely proportional to $A_P$.''

Comparing the ``S/N model'' 90\% confidence errors (shown by the
straight blue line in Figure 21 of Appendix B and reproduced here as
Figure~\ref{plotnvsspos2fig}) with the 50\% confidence curves in
Figure~30 of \citet{con98} for the total NVSS pointing errors
including the calibration errors $\epsilon_\alpha = 0\,\farcs45$ and
$\epsilon_\delta = 0\,\farcs 56$ makes this distinction obvious.  The
failure of the ``S/N model'' to match the data points above $S \sim
0.10$~Jy in Figure~\ref{plotnvsspos2fig} is caused by its omission of
intensity-independent calibration errors.

\subsubsection{The ``Two Component Model'' of Positional 
Accuracy}\label{sec:twocomponent}

To visually fit the dense black band of data points in
\citet{mur140801}
Figure~\ref{plotnvsspos2fig}, 
it is most appropriate to calculate $r_{50}$, the 50\%
confidence radial separation 
between the measured and true positions.  
For Gaussian-distributed 1D position errors
with rms $\sigma \equiv \sigma_{\rm 1D}$ 
in each coordinate, the distribution of 2D position errors is
\begin{equation}
P(r) = {r \over \sigma^2} \exp\Biggl(- {r^2 \over 2 \sigma^2}\Biggr)~,
\end{equation}
and the cumulative probability that $r$ is less than some cutoff
$r_{\rm c}$ is
\begin{equation}\label{eqn:cumprob}
P(<r_{\rm c}) = \int_0^{r_{\rm c}} P(r) dr = 
1 - \exp\Biggl(-{r_{\rm c}^2 \over 2 \sigma^2}\Biggr)~.
\end{equation}
The median cutoff $r_{50}$  defined by $P(<r_{50}) = 50$\% should divide
the black band of data points into equal numbers above and below.
Solving Equation~\ref{eqn:cumprob} with $P(< r_{\rm c}) = 0.50$ gives
\begin{equation}\label{eqn:rconfidence}
r_{50} = \sqrt{2 \ln 2}\, \sigma \approx 1.18\, \sigma~.
\end{equation}
Similarly, the radius of the 90\% confidence error
circle is $r_{90} = \sqrt{2 \ln 10}\, \sigma \approx 2.15\,\sigma$, 
$r_{95} = \sqrt{2 \ln 20}\, \sigma \approx 2.54\, \sigma$, 
$r_{99} = \sqrt{2 \ln 100}\,\sigma \approx 3.03 \,\sigma$, etc.
($r_{90}$ here is the counterpart of $\sigma_{90}$ in Appendix B.)

The NVSS has $\theta = 45''$ resolution, $\sigma_{\rm
  n} = 0\,\farcs45{\rm ~mJy~beam}^{-1}$ rms noise, and $\epsilon \approx
0\,\farcs5$ rms calibration errors in each coordinate.  Most NVSS sources
are unresolved, so the ``two component'' model for position errors
predicts
\begin{equation}\label{eqn:sigmaeq}
\sigma \approx \Biggl[\Biggl({0.6 \theta \over (S/N) }\Biggr)^2
+ (0\,\farcs 5)^2 \Biggr]^{1/2}
\approx \Biggl[\Biggl({12''\over S({\rm mJy})}\Biggr)^2
+ (0\,\farcs 5)^2 \Biggr]^{1/2}~.
\end{equation}
For all $S \gg 24{\rm ~mJy}$, calibration errors dominate noise
errors, so $\sigma \approx 0\,\farcs5$ and $r_{50} \approx 0\,\farcs
6$, in good agreement with the dense horizontal band of data points in
the right half of Figure~\ref{plotnvsspos2fig}.  For $S \ll 24{\rm
  ~mJy}$, $r_{50} \approx 1.18\, \sigma \approx 14'' / S({\rm mJy})$
is noise dominated and inversely proportional to $S$.  Consequently
the $r_{50}$ line has the same slope as the blue line in
Figure~\ref{plotnvsspos2fig}, but its intercept at the left edge of
the plot (where $S \approx 2{\rm ~mJy}$) is $r_{50} \approx 7''$,
which is indeed in the middle of the black band of data points. This
demonstrates that the long-standing ``two component'' method of
calculating position errors fits real data much better than the new
``S/N model''.  The ``two component'' results for the radius of the
NVSS 90\% confidence error circle are $r_{90} \approx 1\,\farcs08$ at
high flux densities and $r_{90} \approx 26'' / S({\rm mJy})$ at low
flux densities.

Sources partially resolved by the NVSS ($\phi \lesssim \theta = 45''$)
will appear broader than $45''$ FWHM on NVSS images.  In that case,
$\theta$ in Equation~\ref{eqn:sigmaeq} should be replaced by
$(\phi^2 + \theta^2)^{1/2}$ and the position error $\sigma$ will be
slightly larger.

\subsubsection{The ``Empirical'' Position Error}\label{sec:empirical}

The red binned curve in Figure~\ref{plotnvsspos2fig} plots the
``empirical'' 90\% confidence FIRST-NVSS separation $r_{90}$ or
$\sigma_{90}$ as a function of flux density.  It ranges from $r_{90}
\approx 5''$ at the highest flux densities to $r_{90} \approx 20''$
below $S \sim 250{\rm ~mJy}$, much larger than $r_{90}$
predicted for point sources by both the ``S/N
model'' and the ``two component'' models. What went wrong?

Large, asymmetrical radio sources are blamed in \citet{mur140801}
Appendix B:\newline \hphantom{000}
``\textbf{Why are the low-resolution positions so inaccurate?} --- Why are the
inaccuracies in the positions so much greater than the S/N model
predictions? Real radio sources are not symmetrical objects. They have
lobes, jets, cores; star-forming galaxies have spiral arms; and there
can be confusion where multiple radio sources get mixed together in
the low resolution beam. A low resolution survey does indeed provide a
measurement, with high accuracy, of the mean flux-weighted position
as the S/N increases. However, the flux-weighted centroid is often not
where the optical counterpart lies. In many cases, the counterpart is
associated with some sharp structure within the radio source, and that
structure may be far from the flux-weighted center.''

My explanation is that the method used to generate the data points in
Figure~\ref{plotnvsspos2fig} is a really bad way to match  NVSS sources  with 
FIRST sources and leads to
bad matches 
with separations approaching $100''$ that contaminate the FIRST/NVSS
separation data.
Consequently the red 90\% confidence ``empirical'' curve is not a
useful indicator of radio-optical offsets and cannot be used to
calculate identification search radii for the NVSS or, more
importantly for the VLASS proposal, for surveys that compete with it
(EMU, WODAN, and even FIRST).

How were the matches in Figure~\ref{plotnvsspos2fig} generated?  
Appendix B explains:\newline \hphantom{000}
``As a large-scale test, we selected a sample of all the FIRST sources
that have an SDSS match within $0\,\farcs7$ and that have an NVSS
match within $100''$. For all these $\sim 135,000$ sources, we
computed the distance to the nearest NVSS source. The important thing
about this sample is that the FIRST source matches the optical source
position. That means that if NVSS is to identify the same counterpart,
it needs to have a position close to the FIRST source position. There
may be several FIRST source components associated with a single NVSS
source, but only the FIRST sources that match optical counterparts are
included.''

The sky density of FIRST sources is $\approx 90 {\rm ~deg}^{-2}$
\citep{whi97} and the sky density of NVSS sources is $\approx 53 {\rm
  ~deg}^{-2}$ \citep{con98}, so $\approx 60$\% of FIRST sources have
true NVSS counterparts and $\approx 40$\% of FIRST sources do not.
Among these unmatched FIRST sources, those having no unrelated NVSS
neighbors within $100''$ were dropped from the sample.  All unmatched
FIRST sources having unrelated NVSS neighbors within $100''$ remain in
the sample and appear as false identifications with NVSS-FIRST
separations up to $100''$.  The probability that an unrelated NVSS
source lies within $100''$ of a random position on the sky is $\approx
0.13$, so we expect about 124,000 good matches and 11,000 bad matches
among the 135,000 plotted data points.  With only 92\% of the matches
being good, the ``empirical'' 90\% confidence limit is bound to be
too high. 

How can we decide between the ``bad sources'' and the ``bad
matches'' explanations for the large 90\% confidence ``empirical''
errors?  

The ``bad sources'' problem was studied observationally by
\citet{fom69}.  He found that, even among strong extragalactic sources
($S \geq 2{\rm ~Jy}$ at 1.4~GHz), only about 30\% would be significantly
resolved ($ \phi > 45''$) by the NVSS, and most extended sources are
fairly symmetric. Among all extended sources, symmetric or not, the
offset of the source's \emph{flux-weighted centroid} from the host
galaxy was usually within 10\% of the overall source extent and never
more than 20\% (see his Figure 8).  Thus the optical identification of a
$50''$ source is likely to lie within $5''$ and almost certainly within
$10''$ of the source centroid.  Any ``bad sources'' should contribute
points centered on and concentrated close to the black band in
Figure~\ref{plotnvsspos2fig}.  Very large FIRST-NVSS separations (e.g.,
$>70''$) should be \emph{extremely} rare, and they would not normally
be accepted as identifications in any case.

In contrast, NVSS ``bad matches'' should be randomly distributed over
$r = 100''$ circles centered on the FIRST sources, and they should
contribute about 8\% of the data points in
Figure~\ref{plotnvsspos2fig}.  The probability that the FIRST/NVSS
separation is between $r$ and $r + dr$ should be nearly proportional
to $r$, so the ``bad matches'' data points should be concentrated near
the $r = 100''$ limit at the top of Figure~\ref{plotnvsspos2fig}, and
nearly half of them should be above $r = 70''$.

The broad distribution of high data points in
Figure~\ref{plotnvsspos2fig}, especially the black band at the very
top, shows that most of the outliers cannot be explained by ``bad
sources'' but are consistent with being ``bad matches''.

The percentage of bad matches is so high because the search circle
radius was set to be $100''$ for all sources, regardless of flux
density or calculated position error.  In actual practice, nobody
blindly identifies an object $100''$ away from a source whose radio
position was measured with a beam of FWHM radius $45''/2 =
22\,\farcs5$.  The radius of the circle in which pure
position-coincidence identifications are accepted is reasonably chosen
to be somewhere between $r_{90} \approx 2.15\,\sigma$ for 90\% and
$r_{99} \approx 3.03\,\sigma$, and identification candidates at larger
separations are rejected.  The Appendix of \citet{con75} derives the
equations actually used to choose search-circle radii for pure
position-coincidence identifications and calculates the resulting
identification completeness and reliability.  For unresolved NVSS
sources, and most NVSS sources are unresolved, these limits are between
$\sim 1\,\farcs1$ and  $1\,\farcs5$ at high flux densities 
and between $\sim 11''$ and $16''$
at the NVSS sensitivity limit.  All candidate matches with larger offsets
would be rejected. Figure~\ref{plotnvsspos2fig} shows that
there are many such points, and counting them as real identifications
is responsible for the very high red ``empirical'' $r_{90}$ line.

\subsubsection{The ``Resolution'' Model for the Matching Radius}
\label{sec:resolution}

The ``empirical'' model attributing large radio position errors to
``bad sources'' was used to
support the ``resolution model'' for identifications in Appendix B.
The ``resolution model'' states that the radio-optical matching radius
should be fixed at  40\% of the FHWM resolution $\theta$ for 
\emph{every}
S/N and $\theta$: \newline \hphantom{0001} ``This analysis shows that
matching at the $45''$ resolution of NVSS requires a matching radius
of $20'' = 40$\% of the NVSS FWHM resolution. Our experience with the
FIRST survey is similar: to get a reasonably complete list of optical
identifications we had to use a matching radius of $2'' \sim 40$\% of
the FIRST FWHM resolution. We argue that is a universal requirement
for radio sources, at least for sources down to the sub-mJy regime:
the matching radius that is required for realistic radio source
morphologies is 40\% of the FWHM resolution.''

This ``resolution model'' drives All-Sky  to higher angular
resolution ($\theta \approx 2\,\farcs5$) than FIRST ($\theta =
5\,\farcs4$), EMU ($\theta = 10''$), and WODAN ($\theta \approx
16''$): \newline \hphantom{0001} ``WODAN will therefore require an
optical matching radius of $6 \times 7''$ and ASKAP-EMU will require
$4''$. A cross-match between SDSS and FIRST shows that 34\% of FIRST
sources have a false (chance) SDSS counterpart within
$6\,\farcs5$. For comparison, 33\% of FIRST sources have a true match
within $2''$.  The conclusion is that half the optical counterparts at
SDSS depth will be false matches when using a $6\,\farcs5$ matching
radius.\newline\hphantom{000} ``The number of false matches can be
reduced somewhat by doing a careful analysis of the likelihood of
association as a function of separation, but when the starting point
is contaminated by 50\% of false matches, the final list of
identifications will not complete or reliable. The false matching
problem will only get worse for deeper optical/IR data.''

Appendix B of \citet{mur140808}
concludes:\newline\hphantom{000} ``The bottom line is that
we need high resolution to get the accurate positions required for
optical identifications. Deeper radio imaging is not a substitute for
the necessary resolution. VLASS will be the survey of choice for
multi-wavelength science, and an all-sky VLASS will have a long and
useful life even after the SKA-precursor surveys are complete.''

This conclusion is an essential part of the VLASS justification, so it
should be examined very carefully.  What if it is wrong, and more
sensitive ($\sigma_{\rm n} = 10\,\mu {\rm Jy~beam}^{-1}$)
lower-resolution ($\theta = 10''$) surveys like EMU are actually
\emph{better} than All-Sky for making complete and reliable
optical identifications?

\subsection{Position-Coincidence Optical Identifications}
\label{IDCRsec}

There is no doubt that accurate radio positions are needed to make
complete and reliable identifications with faint optical objects by
position-coincidence alone.  A pure position-coincidence
identification program chooses the optical object that is both closest
to the radio position and lies within a pre-defined search area; if
there is no such optical object, there is no identification.  The
optimum size for the search area is a compromise determined by the
combined radio and optical position uncertainties and by the surface
density of optical objects. If the chosen search area is too small,
the identifications will be very reliable (the reliability $R$ of a
set of claimed identifications is the fraction of them that are
correct) but incomplete.  If it is too large, completeness $C$
is high bu misidentifications are
more likely to occur and $R$ is low.

The completeness $C$ of an identification program is the fraction of
the radio sources in some complete sample that have identifications
brighter than the magnitude limit and that are correctly identified.
Higher radio resolution reduces position errors and tends to increase
both $R$ and $C$ of sources \emph{in the radio-selected sample}, but
the radio sample itself becomes less complete.

The position-coincidence identification completeness $C$ and reliability $R$
of unresolved sources that do make it into
the radio  catalog are \citep{con75}
\begin{equation}\label{eqn:completeness}
 C = k^{-1} \Biggl[ 1 - \exp\Biggl(-{k m^2 \over 2}\Biggr)\Biggr]
\end{equation}
and 
\begin{equation}\label{eqn:reliability}
R = C \Biggl\{  f^{-1} + (1 - f^{-1})
\exp\Biggl[{m^2 (1 - k) \over 2} \Biggr] 
- \exp\Biggl(-{m^2 k \over 2} \Biggr) \Biggr\}^{-1}~,
\end{equation}
where $k \equiv 1 + 2 \pi \rho \sigma^2$, $\rho$ is the sky density of
identification candidates, $m = r_{\rm c} / \sigma$ is the cutoff
radius of the search circle in units of the rms 1D position error
$\sigma$, $f$ is the fraction of radio sources having counterparts
in the optical catalog, and $(1 - f)$ is the fraction of sources in
``empty fields.''  The free parameter $m$ defining the search area
should be small enough for high reliability and large enough for high
completeness. A good value for $m$ makes $R \sim C$, and it usually
lies in the range $2 \lesssim m \lesssim 3$.
If $\rho$ is large, $\sigma$ must be small
to get good completeness and reliability for pure
position-coincidence identifications; e.g., $R \sim C \gtrsim 90$\%.

\begin{figure}
\vskip -.1in
\includegraphics[angle=0, scale= 0.8]{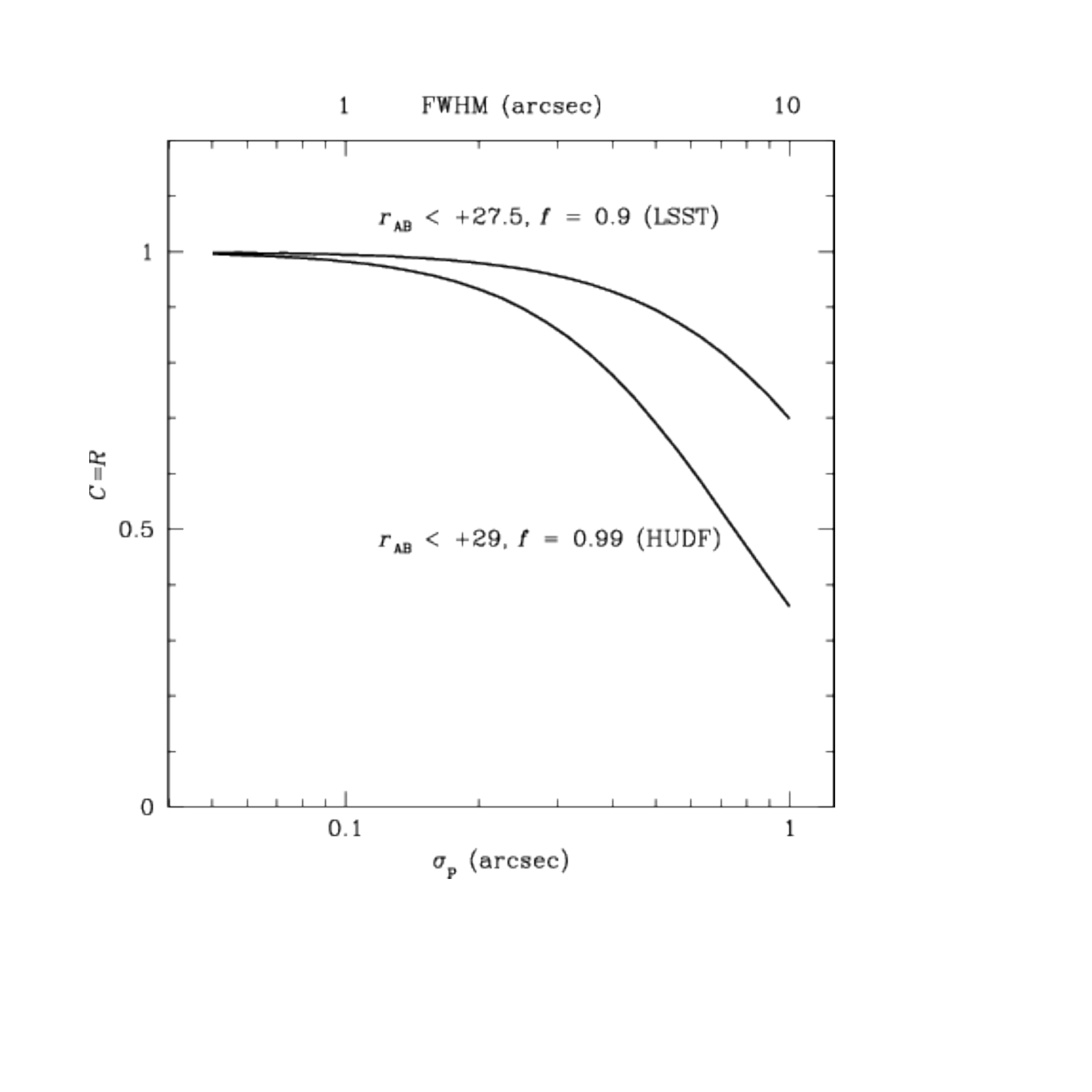}
\caption{If the search-circle radius is chosen such that
identification completeness $C$ and reliability $R$ are about equal,
the rms position error $\sigma_{\rm p}$ is determined by the sky
density of optical identification candidates and the fraction $f$ of
radio sources having optical counterparts. The two curves show
the values of $C = R$ for ``blind'' position-coincidence
identifications to the limits of the planned LSST survey
and the HUDF. Lower abscissa: rms position error in
each coordinate (arcsec).  Upper abscissa: maximum PSF FWHM (arcsec)
for which the noise error on an $SNR=5$ source equals $\sigma_{\rm p}$.
Ordinate: Completeness and reliability.}
\label{idcrfig}
\end{figure}

Figure~\ref{idcrfig} shows $C = R$ for position-coincidence
identifications of sources with the faintest 
optical objects ultimately detectable
by the LSST southern sky survey ($r_{\rm AB} < 27.5$, $f \sim 0.9$)
and in the HUDF \citep{bec06} ($r_{\rm AB} < 29.5$, $f \sim 0.99$) as
a function of $\sigma$ (lower abscissa) and of $\theta$ when $S/N = 5$
(upper abscissa).  The required radio resolutions for 95\% (90\%)
complete and reliable optical identifications of sources at the survey
detection limit $5\sigma_{\rm n}$ are $\theta \le 4''$ ($6''$) and
$\theta \le 2''$ ($3''$) for the LSST and HUDF, respectively.  If the
calibration component of EMU position error is $\epsilon <\theta/25
\approx 0\,\farcs4$ ($\epsilon \sim \theta/90$ for the NVSS), EMU
sources stronger than $\sim 13\sigma_{\rm n} \sim 130 \, \mu{\rm
  Jy~beam}^{-1}$ can be identified with the faintest LSST objects with
90\% completeness and reliability.  It appears that the sensitive but
low-resolution EMU will be significantly better than All-Sky
for completely and reliably identifying radio sources in faint
flux-density limited samples with the faintest optical objects in
LSST.

\subsection{Identifications Using Likelihood Ratios}

\begin{figure}
\includegraphics[angle=0, scale=0.7]{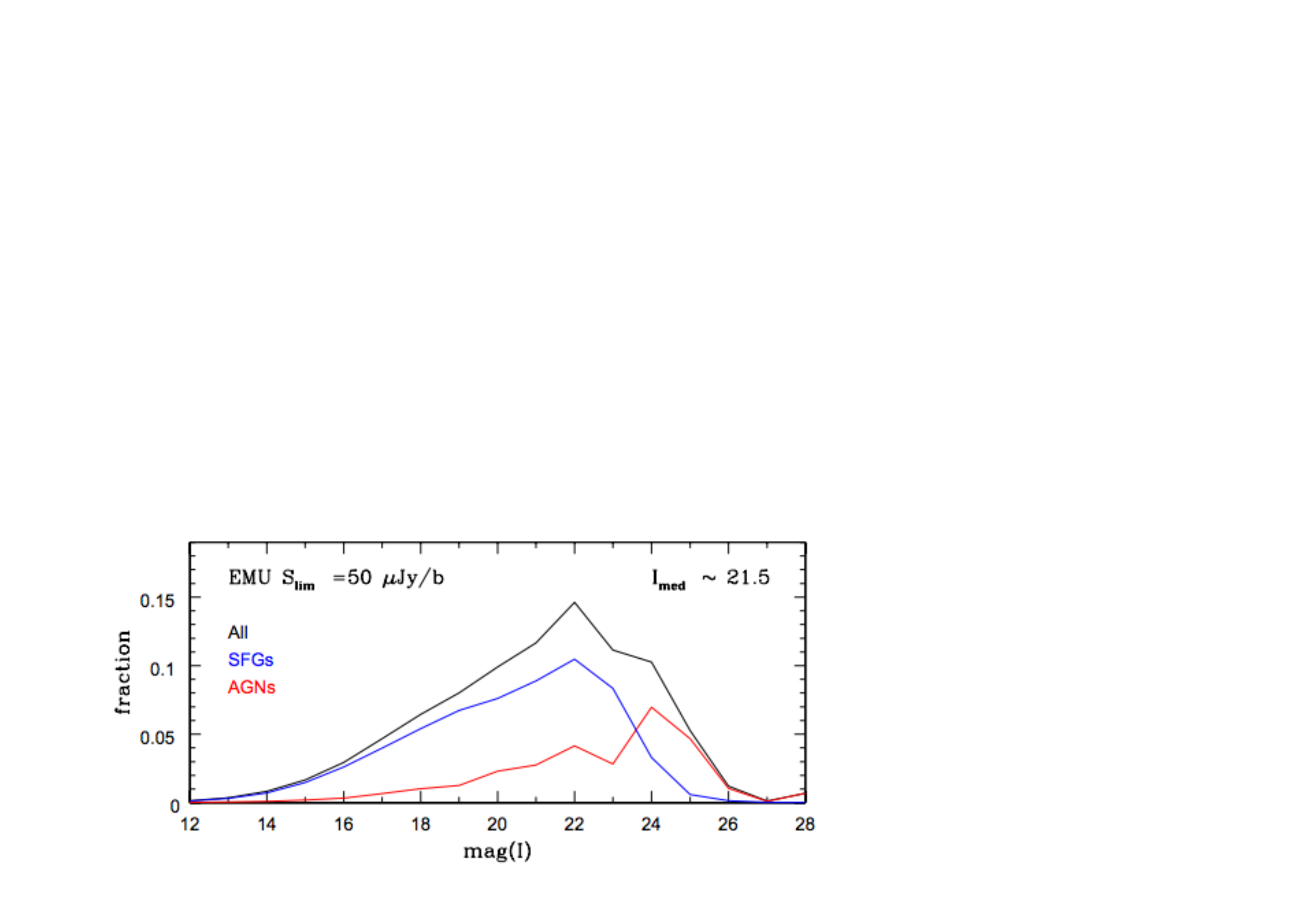}
\caption{The I-band magnitude distributions from the SKADS simulation for  all
radio sources stronger than $50\,\mu{\rm Jy~beam}^{-1}$ (black),
star-forming galaxies (blue),  and AGNs (red). Nearly all of the optical
counterparts are brighter than $I \sim 25$, well above the LSST limit
shown in Figure~\ref{idcrfig}.
}
\label{magdistfig}
\end{figure}

Optical identifications using only position
coincidence ignores prior knowledge about the magnitude distribution
of optical identifications.  All pure position-coincidence identifications fail
when the sky density of optical candidates becomes too high.  For
example, suppose the entire sky were covered by optical images as
sensitive as the HUDF, which contains $\sim 10^3$ objects per square
arcmin brighter than $r_{\rm AB} \sim 29.5$ \citep{bec06}.  Would that
be a disaster for identifying radio sources in All-Sky or EMU?  No,
because we already know that most sub-mJy radio sources have
counterparts brighter than $r_{\rm AB} \sim 26$ \citep{cil03,bon12},
so most of their identifications can be recovered by ignoring the countless
significantly fainter optical objects.  This is effectively what is
done when identifications are made with shallower optical images from
SDSS, Pan-STARRS, LSST, etc. that can't see the faintest optical
objects.  The SKADS simulation also predicts that
most radio sources stronger than the EMU detection limit will have
optical counterparts brighter than $I \sim 25$
(Figure~\ref{magdistfig}).  The sky density of such bright galaxies is
$50{\rm ~arcmin}^2$, so even the very conservative EMU $0.15 \,\theta =
1\,\farcs5$ search radius is good enough for making $>90$\% reliable
identifications by position-coincidence alone.

Likelihood ratios \citep{sut92,cil03} formally optimize this Bayesian
approach, using prior information about the magnitude distributions,
types, colors, etc. of the optical identifications.  They often allow
identifications to be made even when the position errors are so large
that Equations~\ref{eqn:completeness} and \ref{eqn:reliability}
predict mediocre completeness and reliability.  For example,
\citet{sut92} developed the mathematics of likelihood ratios to
identify galaxies with \emph{IRAS} sources, which have extremely large
position errors.  Likelihood ratios will certainly be used to
supplement position coincidence when identifying weak radio sources in
the new surveys with faint optical objects.  They will relax the
position accuracy requirements and favor lower-resolution surveys that
can generate more nearly complete flux-limited samples over
high-resolution surveys that cannot.  No proposal for high-resolution
but incomplete surveys like All-Sky can credibly claim to yield better
optical identifications than more complete lower-resolution surveys
like EMU without a thorough discussion of likelihood ratios.

\subsection{An Empirical Test of the ``Empirical'' Position Errors}
\label{sec:test}

\begin{figure}
\vskip -.1in
\includegraphics[angle=0, scale=0.8]{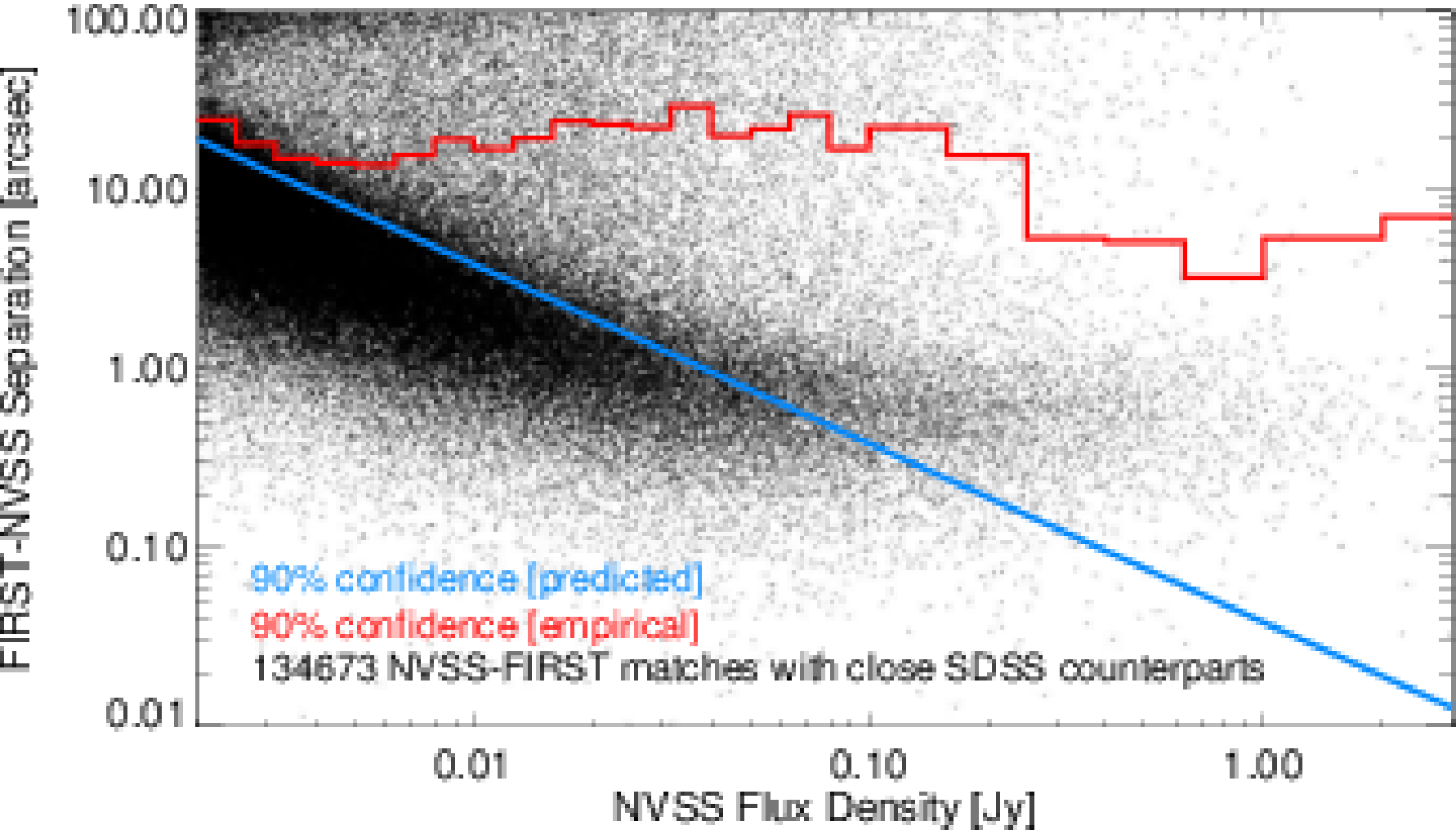}
\caption{This is the original version of the FIRST-NVSS separation plot
in Appendix B of \citet{mur140808}.  The red line shows the
90\% confidence ``empirical'' separation as a function of NVSS
flux density. It suggests that the identification search radius
needs to  be about 0.4 times the NVSS FWHM resolution $\theta = 45''$ even at
fairly high flux densities $S \sim 100{\rm ~mJy}$.}
\label{plotnvsspos2fig}
\end{figure}

The ``empirical'' 90\%-confidence position errors discussed in
Appendix~\ref{sec:empirical} are plotted as a red line  in
Figure~\ref{plotnvsspos2fig}. They are much larger than
predicted by the ``two-component'' model
(Appendix~\ref{sec:twocomponent}).  Figure~\ref{plotnvsspos2fig} shows
FIRST-NVSS radio position separations for only those FIRST source components
that lie within $0\,\farcs7$ of SDSS optical identifications, so these
FIRST positions are accurate proxies for the positions of the optical
identifications.  

Appendix~\ref{sec:empirical} suggested two
explanations for this discrepancy: ``bad sources'' and ``bad
matches.'' In the ``empirical'' model, there are many ``bad sources''
whose flux-weighted radio centroids are quite far from their optical
identifications (host galaxy or quasar).  For bad sources, the
``resolution model'' (Appendix~\ref{sec:resolution}) states that the
identification search-circle radius needs to be a large fraction
of the FWHM resolution 
($0.4\, \theta$), so only high-resolution surveys
can make reliable identifications with faint optical objects, regardless
of (S/N).  The
``bad matches'' model argues that most of the large FIRST-NVSS
separations shown in Figure~\ref{plotnvsspos2fig} are caused by FIRST
misidentifications of NVSS sources.  One reason for bad matches
is that there are more
FIRST sources per square degree ($\sim 90$) than NVSS sources ($\sim
53$) because FIRST is more sensitive than the NVSS to point sources,
so not all FIRST sources actually have real NVSS matches.  Appendix B
originally included all FIRST-NVSS coincidences having separations up
to 100~arcsec, but some of these are just random NVSS sources near
faint FIRST sources.

\begin{figure}
\vskip -.1in
\includegraphics[angle=0, scale=0.5]{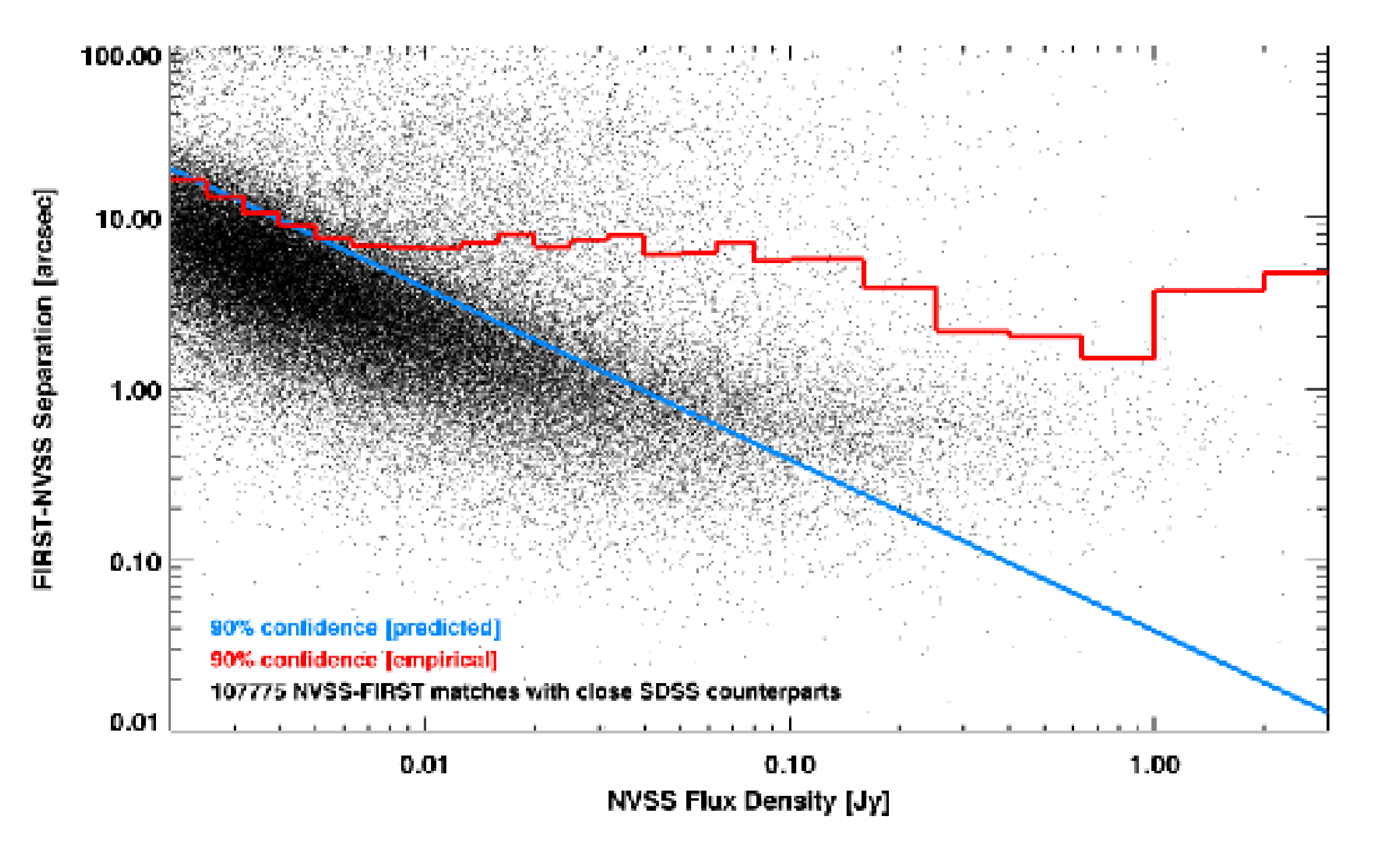}
\caption{Rick White's revised version of the FIRST-NVSS separation
  plot.  The red line is lower and corresponds to about 0.15 times the
  NVSS FWHM resolution at high flux densities, but it is still higher
  than expected from the ``two component'' model for identifying compact or
  symmetric sources.}
\label{plotnvsspos5bfig}
\end{figure}

Rick White recently redid the FIRST identifications of NVSS sources by
excluding FIRST sources fainter than the NVSS sensitivity limit.
Figure~\ref{plotnvsspos5bfig} is his revised version of
Figure~\ref{plotnvsspos2fig}.  A number of ``bad matches'' have been
eliminated, so the new 90\% confidence ``empirical'' limit shown by the
red line is now $\approx 0.15\,\theta \approx 7''$ rather than
$\approx 0.4\,\theta$. However, the red line is still much higher than
the ``two-component'' model prediction for high (S/N).

To best way to distinguish between ``bad sources''
and ``bad matches'' is to inspect the individual source images.
Rick produced a list of 200 FIRST matches with NVSS sources
in the narrow NVSS integrated flux-density range $100 \leq S {\rm
  ~(mJy)} \approx 104$. 
 The
table below shows the first ten and the last 30 sources ordered by
increasing FIRST-NVSS separation.  
\vfill\eject

\noindent {\tt 
Fri Aug 29 14:02:18 2014 \\
NVSS-FIRST matches with FIRST-SDSS sep < 0.7" sorted by increasing separation  \\
200 sources with NVSS flux density $\sim$ 100 \\ 
\hphantom 
{0000000000000000000000000000000}-----FIRST--~~ ~~~ ---NVSS----  --SDSS- \\
No.
 \hphantom{0000}   RA  \hphantom{00000000}         Dec     
\hphantom{0000}     Fpeak    ~ Fint   ~~Sep   
\hphantom{0} Fint  ~Sep   ~~~i  Cl \\
001 16 56 05.204 +15 15 14.00   ~99.52   103.38  ~0.07   101.11  0.21 20.31 s \\
002 10 50 03.740 +51 08 24.13   100.56   104.23  ~0.08   101.83  0.13 21.01 g \\
003 16 20 00.478 +40 43 19.04   103.11   103.93  ~0.11   101.24  0.25 22.39 s \\
004 11 48 40.790 +54 06 37.22   ~86.25   ~88.24  ~0.14   101.28  0.26 21.77 s \\
005 09 41 52.435 +27 22 17.84   ~84.39   ~85.99  ~0.14   101.79  0.21 20.07 s \\
006 09 04 44.337 +23 33 54.05   105.35   107.31  ~0.17   103.94  0.04 17.00 s \\
007 22 03 00.068 +02 16 46.04   ~95.10   102.61  ~0.17   101.14  0.20 21.81 s \\
008 14 54 41.019 +20 40 03.17   104.54   109.30  ~0.19   101.87  0.03 17.58 g \\
009 16 20 33.436 +17 39 55.46   104.23   111.44  ~0.19   103.87  0.08 16.41 g \\
010 12 45 58.839 +54 35 15.85   101.58   104.42  ~0.19   103.78  0.30 20.28 g \\
\\
\dots\dots \\
\\
171 16 20 56.294 +27 34 02.67   ~~2.56   ~10.75  ~2.66   100.74  0.11 20.37 s \\
172 11 03 21.859 +20 48 35.76   ~60.56   ~81.34  ~3.08   100.63  0.49 20.87 g \\
173 22 42 06.671 +07 31 48.31   ~~7.46   ~~4.19  ~3.12   103.87  0.10 17.56 g \\
174 10 03 59.946 +22 52 45.36   ~55.41   ~57.74  ~3.14   101.12  0.22 21.22 s \\
175 16 35 28.177 +49 08 15.95   ~24.92   ~26.76  ~3.15   100.94  0.24 16.28 g \\
176 21 19 05.336 -08 11 43.15   ~96.03   ~98.30  ~3.37   103.69  0.25 18.12 s \\
177 13 41 15.291 +28 16 05.18   ~87.12   ~88.26  ~3.43   100.74  0.17 20.02 s \\
178 02 50 48.692 +00 02 07.93   ~11.54   ~19.17  ~3.76   100.24  0.60 18.88 s \\
179 11 52 32.881 +49 39 38.83   ~41.11   ~47.08  ~4.24   101.02  0.18 16.97 s \\
180 02 53 56.006 -01 13 45.19   ~26.33   ~27.00  ~7.52   100.69  0.40 21.65 s \\
181 14 05 28.404 +20 50 16.59   ~16.83   ~18.57  ~8.64   103.49  0.16 19.25 g \\
182 23 11 36.939 -02 09 07.36   ~29.34   ~33.31  ~9.60   103.82  0.05 21.40 g \\
183 08 32 00.174 +19 53 12.33   ~59.80   ~64.31  10.14   101.16  0.29 18.24 s \\
184 17 09 06.105 +42 01 57.49   ~~5.61   ~~5.67  13.20   103.17  0.19 18.26 g \\
185 12 01 25.725 +55 32 20.27   ~~6.37   ~~6.07  13.28   101.73  0.29 22.01 g \\
186 16 49 52.366 +31 08 07.20   ~21.19   ~25.84  13.50   100.16  0.41 21.50 g \\
187 11 05 18.032 +30 09 15.61   ~~2.83   ~15.05  16.10   101.51  0.26 21.48 g \\
188 15 53 11.942 +27 33 20.28   ~10.92   ~10.30  17.01   103.76  0.40 15.27 g \\
189 09 02 35.299 -00 21 12.77   ~~6.79   ~~7.48  17.84   102.79  0.30 20.52 g \\
190 10 36 41.440 +12 33 32.82   ~~2.54   ~~2.20  18.68   103.76  0.02 20.23 g \\
191 07 38 27.603 +24 15 23.36   ~17.46   ~58.34  19.48   100.88  0.51 16.49 g \\
192 13 44 36.420 +16 07 27.86   ~~8.86   ~10.64  19.96   101.93  0.12 20.61 g \\
193 13 11 04.669 +27 28 07.46   ~~2.25   ~~3.13  20.57   103.43  0.22 16.99 g \\
194 14 14 08.431 +48 41 56.23   ~~8.16   ~~7.06  22.18   103.61  0.22 17.84 g \\
195 14 05 00.812 +29 25 14.03   ~~7.12   ~~7.24  24.68   102.55  0.03 19.77 s \\
196 16 44 52.560 +37 30 09.26   ~~3.56   ~~3.08  25.05   101.68  0.18 18.07 s \\
197 08 17 16.145 +07 08 46.49   ~~9.53   ~10.87  27.63   101.65  0.23 18.15 g \\
198 10 23 13.604 +63 57 09.23   ~21.25   ~21.39  29.39   103.74  0.06 16.95 s \\
199 12 03 23.678 +13 20 06.77   ~~2.75   ~~3.70  56.56   102.49  0.65 19.15 g \\
200 11 51 27.937 +36 12 32.20   ~~6.05   ~~5.49  58.39   102.31  0.16 16.73 g \\
}

 The position errors of $S \gg 24{\rm ~mJy}$ NVSS sources are
 completely dominated by calibration errors so the ``two-component''
 model (Equation~\ref{eqn:sigmaeq}) predicts $\sigma = \sigma_{\rm 1D}
 \approx 0\,\farcs 5$ for unresolved NVSS sources with $S \sim
 100{\rm~mJy}$.  The FIRST position errors of such strong sources are
 much smaller and can be ignored to first order.  If the NVSS position
 error distribution is roughly Gaussian and the sources are ``good''
 (the radio centroid is close to the optical identification), the FIRST-NVSS
 separations should have a Rayleigh distribution. About 100 of the 200
 matches sources should have FIRST-NVSS separations less than $r_{50}
 \approx 1.18 \sigma \approx 0\,\farcs6$
 (Equation~\ref{eqn:rconfidence}), and 94 actually do.  Likewise, the
 180 smallest separations among the the 200 matches should have
 FIRST-NVSS separations just a little over $r_{90} \approx 2.15 \sigma
 \approx 1\,\farcs1$, but source number 180 actually has FIRST-NVSS separation
 $7\,\farcs 52$, putting it close to the red line in
 Figure~\ref{plotnvsspos5bfig}, and the 200th source has a FIRST-NVSS
 separation of $58''$.  Thus the discrepancy between the ``empirical''
 separations and the separations predicted by the ``two-component''
 model is confined to the large-separation tail of the separation
 distribution.  Are these ``bad sources'' or ``bad identifications''?
 One clue is their FIRST flux densities, which should be fairly close
 to the NVSS flux densities, $S \approx 100 {\rm ~mJy}$.  Most of the
 200 FIRST flux densities are indeed close to 100~mJy, but sources 181
 through 200 (the 10\% with the largest separations) have much lower
 integrated FIRST flux densities, and ten of them are are below
 10~mJy.  The 20 sources with the largest
separations are worth investigating individually.
 
\begin{figure}
\vskip -.1in
\plottwo{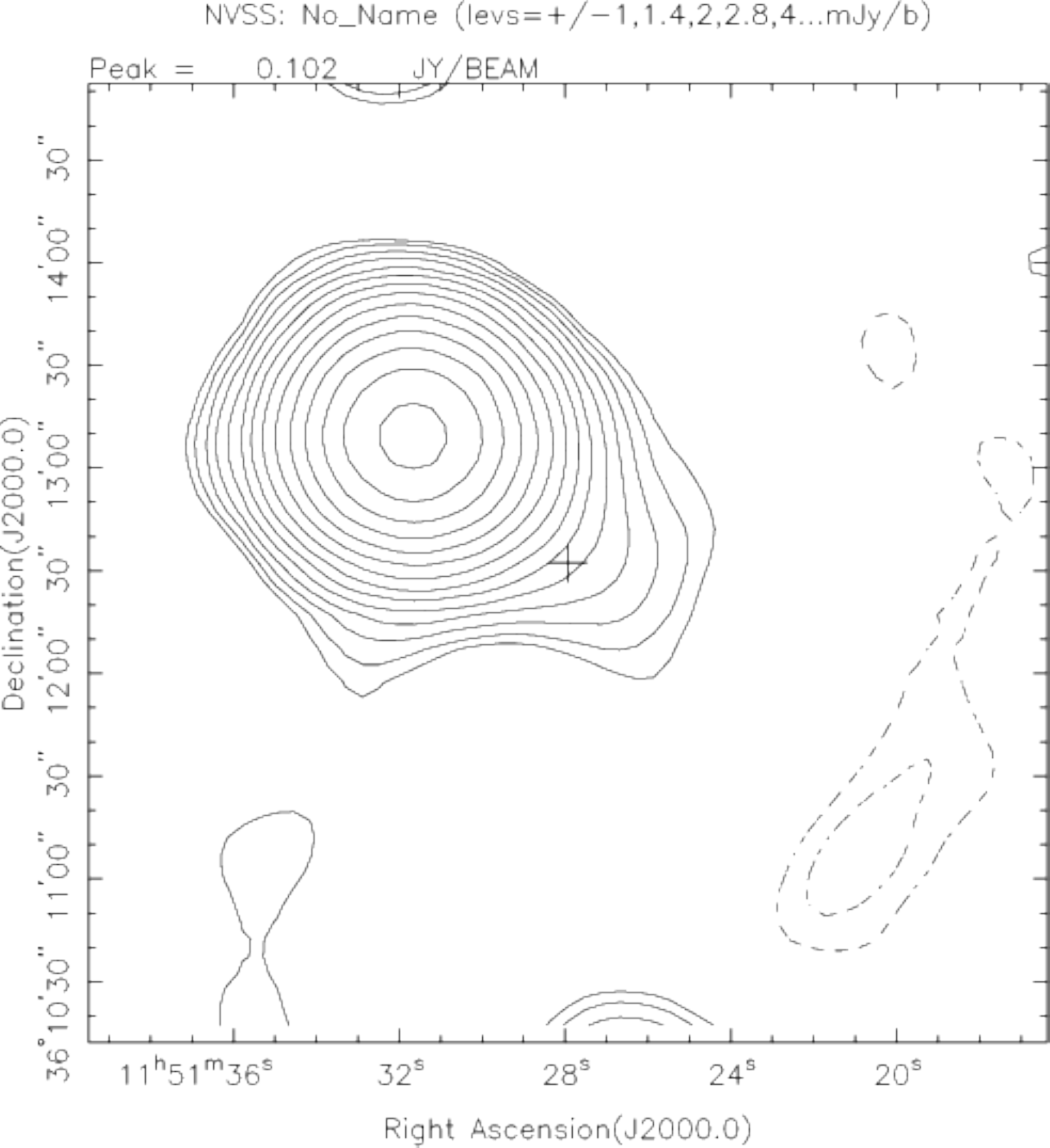}{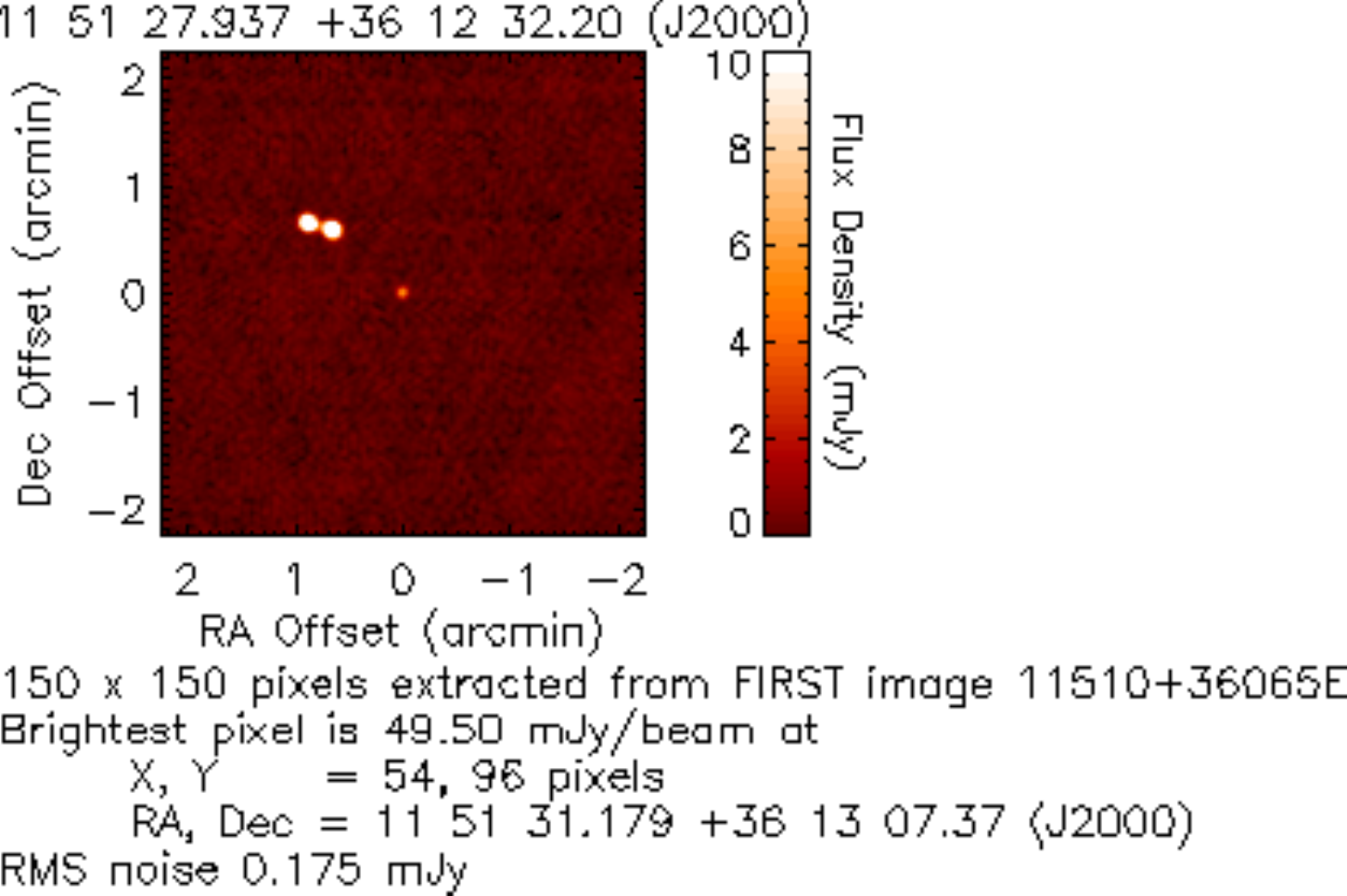}
\caption{This NVSS contour plot and FIRST cutout are both 4.5 arcmin
  on a side and centered on FIRST 115127.94+361232.2, a 6~mJy FIRST
  point source identified with an SDSS galaxy. This faint FIRST source
at the position of the cross
  is visible only as a tail extending from the main NVSS source.
  The correct NVSS source match
is the 113~mJy FIRST double source to the northeast, which is
a background source unrelated
to the FIRST point source.
}
\label{115127fig}
\end{figure}

Source number 200 has the biggest FIRST-NVSS separation ($58''$).  Its
NVSS contour plot and FIRST image cutout centered on the optically
identified FIRST source are shown in Figure~\ref{115127fig}.  This
source is instructive at two levels: \\ (1) It is a ``bad match'' in
that the optically identified 6 mJy FIRST source is not related to the
$S \approx 100{\rm ~mJy}$ NVSS source, which is a completely unrelated
radio source that happens to lie about 58 arcsec away on the sky.  The
113~mJy NVSS source at 11 51 31.66 +36 13 09.5 is closely matched to
the strong (also 113 mJy total flux density) FIRST double whose
centroid is at 11 51 31.71 +36 13 09.0, for a true FIRST-NVSS
separation of only $0\,\farcs 8$.  This source should be moved from
the ``tail'' of large separations, and moving just this one source to
$0\,\farcs8$ separation means that the 90\% confidence ``empirical''
red line should be lowered from the $7\,\farcs52$ separation of source
180 to the FIRST-NVSS separation of source number 179, $4\,\farcs24$.
Clearly the red lines in Figures~\ref{plotnvsspos2fig} and
\ref{plotnvsspos5bfig} can be sensitive to
small numbers of bad matches. \\ (2) The NVSS and FIRST catalogs do
not actually list ``radio sources,'' defined as the total radio
emission from a single galaxy or quasar.  They only list ``radio
components,'' defined as radio image peaks defined by Gaussian fits.
The two lobes of the double radio source were not resolved by the NVSS
so it is cataloged as a single NVSS component, but the lobes were
resolved by FIRST and are cataloged as two FIRST components.  The
double radio source might also have a ``core'' component centered on
its host galaxy or quasar, but the core is too weak to appear in the
FIRST catalog.  To make an optical identification of this source, the
FIRST components would have be be associated with each other as being the
lobes of a single radio source and their centroid position measured.
This is an example of using prior knowledge and likelihood ratios
instead of just pure position coincidence to make optical
identifications.  We know that a double source is more likely than two
strong, nearly equal sources being so close together, so the
identification is likely to lie between the two radio components. The
separation of the two lobes is about $14''$.  If the centroid is
typically displaced from the identification by $\sim10$\% of the lobe
separation \citep{fom69}, the $1\,\farcs4$ radio-optical offset is
larger than the centroid position error of the NVSS, and having the higher
resolution FIRST image would not significantly lower the search-circle
radius.

\begin{figure}
\vskip -.1in
\plottwo{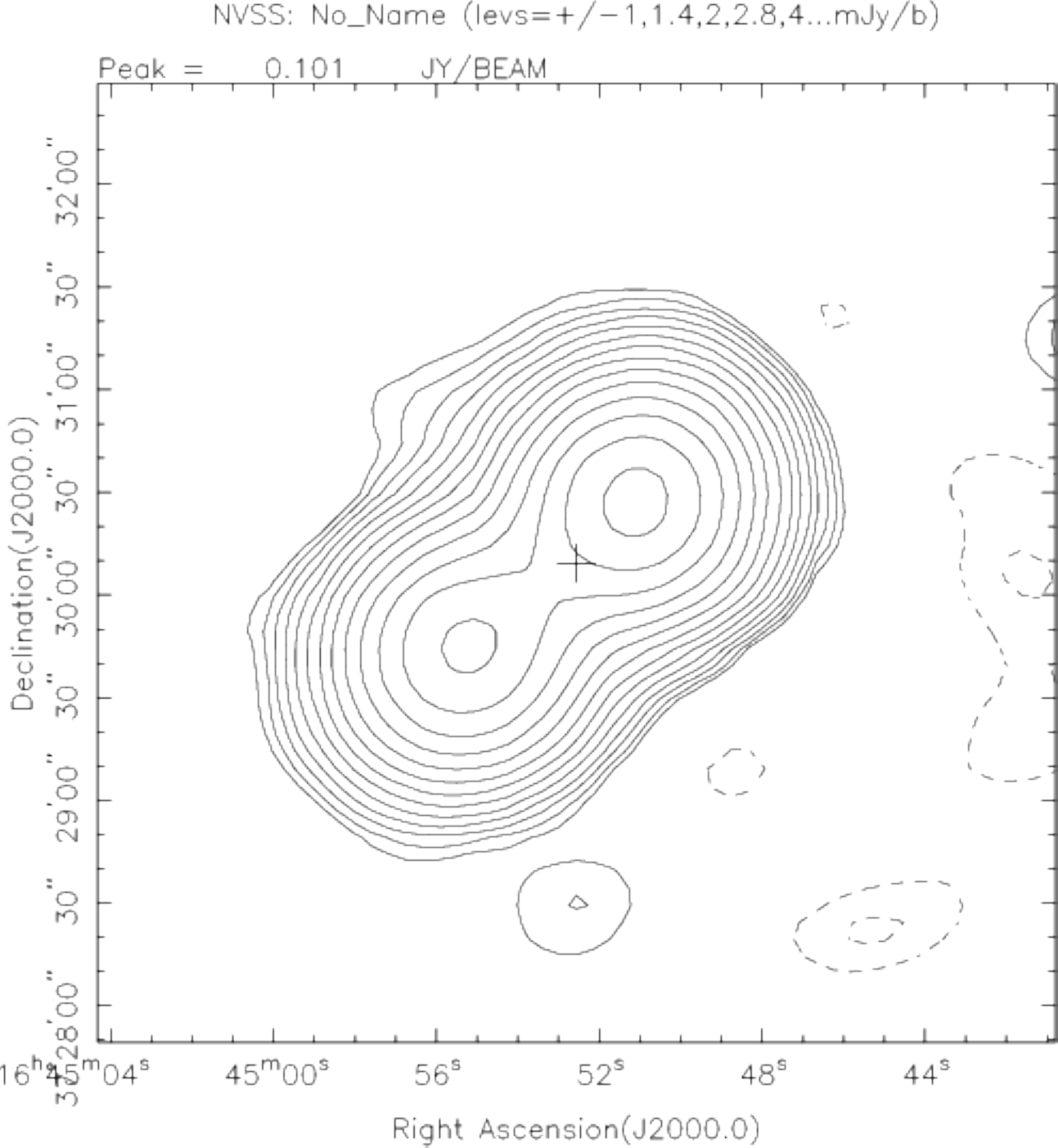}{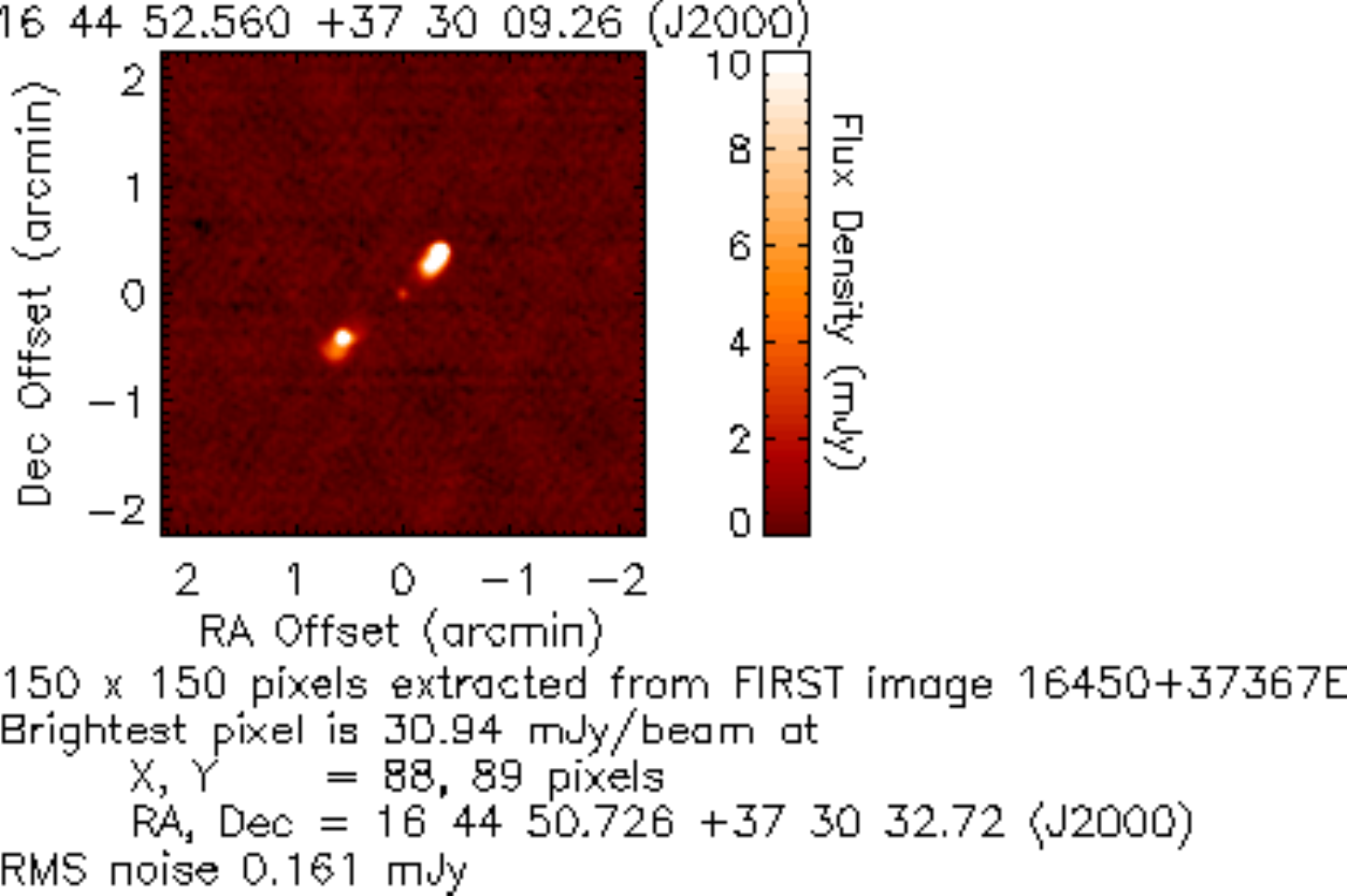}
\caption{This NVSS contour plot and FIRST cutout are both 4.5 arcmin
  on a side and centered on FIRST 164452.56+373009.3, a 3~mJy FIRST
  point source correctly identified with an SDSS galaxy. It is the
  core of a triple source whose total FIRST flux density is 137~mJy.
  It was matched with the northeast component only of the 188~mJy NVSS
  double, so it is another ``bad match''.  }
\label{164452fig}
\end{figure}

A different kind of ``bad match'' is illustrated by the 3 mJy FIRST
component at 16 44 52.560 +37 30 09.26 (number 196 in the table).  It
is the core of a triple radio source (Figure~\ref{164452fig}), and its
optical identification with a $g = 18.1$ SDSS quasar at 16 44 52.57
+37 30 09.2 is secure.  The NVSS does not detect the core, only the
two lobes of this 188~mJy radio source.  The bad FIRST-NVSS match
is with the northwest radio lobe 25 arcsec away in position angle
$-44^\circ$, not with the flux-weighted centroid.  
The centroid of the NVSS double source is at 16 44 52.83
+37 30 09.5,  $3\,\farcs2$ from the SDSS quasar.  This offset is
only 5\% of the 66 arcsec lobe separation and quite sufficient to make
the identification.  Thus this is not a ``bad source,'' but the
ability of FIRST to detect the core separately is still an advantage,
leading to a subarcsec radio-optical offset, while the NVSS
radio-optical offset is 3.2 arcsec.  On the other hand, the core flux
density is only $3 \times$ the FIRST detection limit, so if this
source were $4X$ less luminous or twice as far away, FIRST would not
detect the core.  Then FIRST would be at a disadvantage because it
misses about 50 mJy of the lobe flux, so the FIRST centroid
position at 16 44 52.44 +37 30 14.5 is less accurate, being
$5\,\farcs2$ from the optical position.

Which of the final 20 sources in the list are bad matches with
completely unrelated stronger sources?  The 3.7~mJy FIRST source
number 199 at 12 02 23.678 +13 20 06.77 has the second-largest
FIRST-NVSS separation ($56''$).  The correct match for the unresolved
103~mJy NVSS source at 12 03 20.68 +13 19 30.9 is the single-component
105~mJy FIRST source at 12 03 20.649 +13 19 31.45, and their
FIRST-NVSS separation is $0\,\farcs6$.  This association is confirmed
by the optical identification of the NVSS source with the galaxy 2MASX
J12032061+1319316, also $0\,\farcs6$ from the NVSS position.  The
3~mJy FIRST source number 192 at 13 44 36.420 +16 07 27.86 is
unrelated to the much stronger NVSS/FIRST source at 13 44 36.06 +16 07
08.7.  The strong source has a flat radio spectrum, so it unlikely to
be a lobe of source 192.  The 2.2 mJy FIRST source number 190 at 10 36
41.440 +12 33 32.82 is unrelated to the $\sim 100{\rm ~mJy}$
NVSS/FIRST source which is identified with a 2MASS $z = 2.145$ quasar
at 10 36 40.36 +12 33 38.9.  The 7~mJy FIRST source number 189 at 09
02 35.299 $-$00 21 12.77 appears to be unrelated to the stronger
NVSS/FIRST double source to the east. The weak, extended FIRST source
number 187 is in the lobe of the unrelated triple source identified
with a bright galaxy whose core is at 11 05 22.861 +30 09 41.46.
Removing these bad matches from the list lowers the 90\% 
FIRST-NVSS separation to $\approx
3\,\farcs2$, or $\approx 0.07\,\theta$.

Most of the remaining sources in the ``top 20'' are bad matches
with individual NVSS lobes of double sources larger than $\sim 60''$
rather than with the NVSS centroids. Many of these NVSS centroids
may be farther than $3\,\farcs2$ to the radio cores, so they won't
lower the estimated 90\% separation by much.  Replacing the
component offsets with centroid offsets will, however, greatly shorten
the long ``tail'' of large offsets and eliminate the conspicuously high
points in Figures~\ref{plotnvsspos2fig} and \ref{plotnvsspos5bfig}.

Further refining the Appendix B approach to calculating the resolution
$\theta$
required to make deep optical identifications is probably not
worthwhile because: \\ (1) The sample is biased by the high resolution
of FIRST, so it excludes many extended sources, spiral galaxies
in particular. \\ (2) The optical identifications were made with
cataloged FIRST source components, not with complete radio sources.  Thus
triple sources with sufficiently strong cores have been identified, but
most double sources have not.

I conclude that 90\% of the identifications conservatively lie within
$0.1\,\theta$; a $0.4\,\theta$ search radius is unnecessarily large.
The ``resolution'' model in Appendix B of \citet{mur140808} indicated
that a VLASS with $\theta = 3''$ resolution and a $0.4\,\theta =
1\,\farcs2$ identification search radius
can make  reliable deep
position-coincidence identifications while EMU with $\theta = 10''$
and a $4''$ search radius
cannot. However, a $1\,\farcs2$ search radius is $0.12\,\theta$ for
EMU, which easily satisfies the conservative new $0.1\,\theta$
requirement, so the  high resolution of
All-Sky  is not needed for identification reliability
and will hurt identification completeness.  EMU is far more sensitive
than All-Sky to both unresolved cores and  extended
emission from lobes and spiral galaxies, making it better
 than All-Sky for complete as well as
reliable optical identifications.

\clearpage



\clearpage

\clearpage

\clearpage

\clearpage

\clearpage

\clearpage

\clearpage

\clearpage

\clearpage

\clearpage

\clearpage


\begin{thebibliography}{}

\bibitem[Aaquist \& Kwok(1990)]{aaq90}
  Aaquist, O.~B., \& Kwok, S.~1990, A\&AS, 84, 229

\bibitem[Becker et al.(1995)]{bec95} Becker, R.~H., White, R.~L., \&
  Helfand, D.~J.~1995, ApJ, 450, 559

\bibitem[Beckwith et al.(2006)]{bec06}
    Beckwith, S.~V.~W.~et al. 2006, AJ, 132, 1729 



\bibitem[Bevington(1969)]{bev69}
    Bevington, P.~R.~1969, ``Data Reduction and Error Analysis for the
    Physical Sciences,'' (McGraw-Hill, New York)

\bibitem[Bondi et al.(2008)]{bon08}
Bondi, M., Ciliegi, P., Schinnerer, E., Smolcic, V., Jahnke, K.,
  Carilli, C., \& Zamorani, G.~2008, ApJ, 628, 1129

\bibitem[Bonzini et al.(2012)]{bon12}
  Bonzini, M., Mainieri, V., Padovani, P., Kellermann, K.~I., 
  Miller, N., Rosati, P., Tozzi, P., Vattakunnel, S., Balestra, I., 
  Brandt, W.~N., Luo, B, \& Xue, Y.~Q.~2012, ApJS, 203:15


\bibitem[Bridle et al.(1997)]{bri97}
  Bridle, A.~H., Backer, D.~C., Churchwell, E.~B., Haynes, M.~P.,
  Hewitt, J.~N., Hogg, D.~E., \& Lo, K.~Y. 1997, ``Report of the
  NRAO Large Proposal Committee'', posted as an attachment on \hfil\newline
 {\tt https://staff.nrao.edu/wiki/bin/view/NM/VLASSSciStaffForum}


\bibitem[Chang et al.(2004)]{cha04}
  Chang, T.~C., Refrigier, A., \& Helfand, D.~J.~2004, ApJ, 617, 794

\bibitem[Ciliegi et al.(2003)]{cil03}
  Ciliegi, P., Zamorani, G., Hasinger, G., Lehmann, I., 
  Szokoly, G., \& Wilson, G.~2003, A\&A, 398, 901



\bibitem[Condon(1984)]{con84}
  Condon, J.~J.~1984, \apj, 287, 461

\bibitem[Condon(1989)]{con89}
  Condon, J.~J.~1989, \apj, 338, 13


\bibitem[Condon(1997)]{con97}
  Condon, J.~J.~1997, PASP, 109, 166

\bibitem[Condon(2014)]{con14}
  Condon, J.~J.~2014, ``VLASS Notes, 2014 September 23'', posted as an attachment on \hfil\newline
{\tt https://staff.nrao.edu/wiki/bin/view/NM/VLASSSciStaffForum}



\bibitem[Condon et al.(1975)]{con75}
  Condon, J.~J., Balonek, T.~J., \& Jauncey, D.~L.~1975, AJ, 80, 887

\bibitem[Condon et al.(2002)]{con02}
  Condon, J.~J., Cotton, W.~D., \& Broderick, J.~J.~2002, AJ, 124, 675

\bibitem[Condon et al.(1998)]{con98}
  Condon, J.~.J., Cotton, W.~D., Greisen, E.~W.~et al.~1998, \aj, 115, 1693

\bibitem[Condon et al.(2013)]{con13}
  Condon, J.~J., Kellermann, K.~I., Kimball, A.~E., Ivezic, Z., \&
  Perley, R.~A.~2013, ApJ, 768:37

\bibitem[Condon \& Kaplan(1998)]{con98pne}
  Condon, J.~J., \& Kaplan, D.~L.~1998, ApJS, 117, 361






\bibitem[Fomalont(1969)]{fom69}
  Fomalont, E.~B.~1969, ApJ, 157, 1027




\bibitem[Hodge et al.(2011)]{hod11}
  Hodge, J.~A., Becker, R.~H., White, R.~L., Richards, G.~T., \&
  Zeimann, G.~R.~2011, A.~J., 142:3

\bibitem[Hummel(1981)]{hum81}
  Hummel, E.~1981, A\&A, 93, 93


\bibitem[Lindsay et al.(2014)]{lin14}
   Lindsay, S.~N., Jarvis, M.~J., Santos, M.~G., Brown, M.~J.~I., 
   Croom, S.~M., Driver, S.~P., Hopkins, A.~M., Liske, J., 
   Loveday, J., Norberg, P., \& Roboham, A.~S.~G. 2014,
   MNRAS, 440, 1527

\bibitem[Magliocchetti et al.(2002)]{mag02}
  Magliocchetti, M., et al.~2002, MNRAS, 333, 100


\bibitem[Murphy et al.(2014a)]{mur140801}
  Murphy, E., Baum, S., et al. 2014, ``VLASS Project Description'' 
(August 1 version)

\bibitem[Murphy et al.(2014b)]{mur140808}
  Murphy, E., Baum, S., et al. 2014, 
``The Jansky-Very Large Array Sky Survey (VLASS)'' 
(August 8 version)



\bibitem[Norris et al.(2011)]{nor11}
  Norris, R.~P.~et al.~2011, PASA, 28, 215


\bibitem[Ray et al.(2012)]{ray12}
  Ray, P.~S. et al. 2012, arXiv:1205.3089

\bibitem[SSG(2015)]{ssg15}
SSG15, VLA Survey Science Group~2015, VLASS ``final'' proposal \\
{\tt https://safe.nrao.edu/wiki/pub/JVLA/VLASS/VLASS\_final.pdf}

\bibitem[Sutherland \& Saunders(1992)]{sut92}
  Sutherland, W., \& Sanders, W.~1992, MNRAS, 259, 413




\bibitem[White et al.(1997)]{whi97}
  White, R.~L., Becker, R.~H., Helfand, D.~J., \& Gregg, M.~D.~1997,
  ApJ, 475, 479


\bibitem[Wilman et al.(2008)]{wil08}
  Wilman, R.~J., Miller, L., Jarvis, M.~J.~et al.~ 2008, \mnras, 388, 1335



\end{thebibliography}
\end{document}